\documentclass[twocolumn,pre,nopacs,amsmath,amssymb,nofootinbib]{revtex4-1}

\usepackage{amssymb}
\usepackage{amsmath}

\usepackage{color}
\usepackage[pdftex]{graphicx}

\usepackage{empheq}

\usepackage{mathtools}

\usepackage{makecell}

\usepackage{enumerate}
\usepackage{mathtools}
\usepackage{stmaryrd}

\usepackage{enumitem}

\usepackage{textcomp}
\usepackage{gensymb}

\newcommand{\ex}[1]{\mathrm{e}^{#1}}

\newcommand{\dd}[0]{\mathrm{d}}
\newcommand{\ii}[0]{\mathrm{i}}

\newcommand{\zz}[0]{\mathbf{0}}

\newcommand{\rr}[0]{\boldsymbol{r}}

\newcommand{\xx}[0]{\boldsymbol{x}}
\newcommand{\qq}[0]{\boldsymbol{q}}
\newcommand{\vv}[0]{\boldsymbol{v}}

\newcommand{\yy}[0]{\boldsymbol{y}}

\newcommand{\kB}[0]{k_{\mathrm{B}}}

\definecolor{darkblue}{rgb}{0,0,0.6}
\definecolor{darkred}{rgb}{0.6,0,0}
\usepackage[colorlinks=true,urlcolor=darkblue,citecolor=darkblue,linkcolor=darkred]{hyperref}

\newcommand{\nocontentsline}[3]{}
\newcommand{\tocless}[2]{\bgroup\let\addcontentsline=\nocontentsline#1{#2}\egroup}

\begin{document}

\title{The Dean-Kawasaki equation and stochastic density functional theory}

\author{Pierre Illien}
\affiliation{Sorbonne Université, CNRS, Physical Chemistry of Electrolytes and Interfacial Nanosystems (PHENIX), 4 place Jussieu, Paris, France}

\date{\today}

\begin{abstract}

The Dean-Kawasaki (DK) equation, which is at the basis of stochastic density functional theory (SDFT), was proposed in the mid-nineties to describe the evolution of the density of interacting Brownian particles, which can represent a large number of systems such as colloidal suspensions, supercooled liquids, polymer melts, biological molecules, active or chemotactic particles, or ions in solution. This theoretical framework, which can be summarized as a mathematical reformulation of the coupled overdamped Langevin equations that govern the dynamics of the particles, has attracted a significant amount of attention during the past thirty years. In this review, I present the context in which this framework was introduced, and I recall the main assumptions and calculation techniques that are employed to derive the DK equation. Then, in the broader context of statistical mechanics, I show how SDFT is connected to other theories, such fluctuating hydrodynamics, macroscopic fluctuation theory, or mode-coupling theory. The mathematical questions that are raised by the DK equation are presented in a non-specialist language. In the last parts of the review, I show how the original result was extended in several directions, I present the different strategies and approximations that have been employed to solve the DK equation, both analytically and numerically. I finally list the different situations where SDFT was employed to describe the fluctuations of Brownian suspensions, from the physics of active matter to the description of charged particles and electrolytes.

\end{abstract}

\maketitle

\tableofcontents

$\ $

\section{Introduction}

\subsection{From a single Brownian particle...}

The erratic motion of a mesoscopic particle in a fluid, which originates from the random collisions between the solvent molecules and the particle, is usually referred to as Brownian motion~\cite{Brown1828a}. From a theoretical point of view, such a system is \emph{a priori} very complicated to study, as it couples the evolution of the particle with that of all the molecules that constitute the solvent. In pioneering works, Einstein~\cite{Einstein1905a} and Smoluchowski~\cite{vonSmoluchowski1906} proposed simplified descriptions of this erratic motion. In these models, the dynamics of the solvent particles are ignored, and the motion of the mesoscopic particle, instead of being explicitly described as the result of the multiple collisions with solvent particles, is modeled by a sequence of random elementary displacements. As a consequence of the central limit theorem, the distribution of the position of the particle is typically described by a Gaussian distribution, whose variance (the mean-square displacement) increases linearly with time. Such models are valid as long as the mesoscopic particle is much larger than the solvent molecules, but small enough for thermal fluctuations to overcome its weight, which corresponds to characteristic sizes from $\sim 1$ nm to $\sim 1$ \textmu m, and when the density of the particle is comparable to that of the solvent. This stochastic view of microscopic motion in a fluid environment is one of the cornerstone of soft matter and biological physics.

Relying on this stochastic perspective, Langevin~\cite{Langevin1908} intended to write the equations of motions of the particle, i.e. the equations satisfied by by its position $\rr$ and velocity $\vv$. He proposed to model the effect on the solvent through two contributions: (i) a contribution that accounts for the dissipation induced by the solvent, i.e. its resistance to any perturbation that the colloid may impose on the solvent because of some external forcing. Within linear response, the resulting force typically reads $-m\gamma \vv$, where $\gamma$ has the dimension of an inverse time; (ii) a contribution that accounts for the fluctuation of the solvent, that fluctuates on a timescale $\tau_c$ comparable to the duration of the mean free path of a solvent molecule (i.e. the time during which a solvent molecule travels without hitting another molecule). Assuming that $\tau_c$ is typically smaller than other relevant timescales of the problem (the timescale of dissipation and that of the typical diffusion of the colloid, i.e. the time taken by a colloid to diffuse over its own length), the fluctuation force is generally assumed to be $\delta$-correlated, where $\delta$ refers to Dirac's distribution. Its amplitude follows from the equipartition theorem, which ensures that $\frac{1}{2}m\langle v^2 \rangle = \frac{1}{2}\kB T$, where $\kB$ is the Boltzmann constant and $T$ the temperature.
These assumptions result in the following equation obeyed by $\vv$ (the position $\rr$ is simply obtained by integrating $\frac{\dd \rr}{\dd t} = \vv$), usually referred to as the Langevin equation:
\begin{equation}
	m\frac{\dd \vv}{\dd t} = -\gamma m \vv + \sqrt{2m\gamma \kB T} \boldsymbol{\eta}(t),
	\label{Langevin}
\end{equation}
where $\boldsymbol{\eta}(t)$ is a unit Gaussian white noise, such that $\langle \eta_{i}(t) \rangle = 0$ and $\langle \eta_i(t) \eta_j (t') \rangle = \delta_{ij}\delta(t-t')$.

In the limit of large frictions, i.e. when the fluid is very viscous (or, equivalently, when one only observes the system on durations much larger than the typical time $\gamma^{-1}$), the degrees of freedom associated with the velocity have all reached their stationary value, and the term accounting for inertia in the Langevin equation (the left-hand side of Eq. \eqref{Langevin}) becomes negligible. The equation simply becomes
\begin{equation}
	\frac{\dd \rr}{\dd t} =  \sqrt{2 D} \boldsymbol{\eta}(t),
\end{equation}
where $D =  \kB T/m\gamma$ is the bare diffusion coefficient. This equation is often referred to as the \emph{overdamped} Langevin equation.

\subsection{...to multiple interacting Langevin processes}

For a single, isolated particle, both the Langevin equation and its overdamped limit can be solved very straightforwardly -- this is a textbook example of a Gaussian stochastic process~\cite{Gardiner1985,vanKampen1981}. However, in many systems of biological or physical interest (e.g. biomolecules or organelles in the intracellular medium, colloidal suspensions, emulsions, polymeric solutions),  diffusion occurs in conditions which are much more complicated than that of an isolated mesoscopic particle. Indeed, the erratic motion of each particle is strongly affected by interactions with the other particles  in the system, for instance because of excluded-volume effects, or because the particles are charged. Diffusion in `real' systems therefore depends on the complex interplay between thermal fluctuations, due to the presence of a solvent, and pair interactions. In the simple example of a suspension made of $N$ identical particles interacting via a pair potential $V$, and in the overdamped limit, their positions $\rr^1,\dots,\rr^N$ obey:
\begin{equation}
	\label{overdamped_Langevin}
	\frac{\dd \rr^\alpha(t)}{\dd t} = - \mu \sum_{\beta=1}^{N} \nabla V(\rr^\alpha(t)-\rr^\beta(t))  + \sqrt{2 D} \boldsymbol{\eta}^\alpha(t),
\end{equation}
where $\mu = 1/(m\gamma)$ is the mobility of the particle, and is related to the bare diffusion coefficient through the fluctuation-dissipation theorem $D=\mu\kB T$, and where the noises $\boldsymbol{\eta}^\alpha(t)$ are uncorrelated one with another: $\langle \eta_i^\alpha (t) \eta_j^\beta(t') \rangle = \delta_{ij}\delta^{\alpha\beta}\delta(t-t')$ (throughout the review, the following convention will be used: Greek letters will denote the label of the particle, and Roman letters their Cartesian coordinates). In this framework, the dynamics of the suspension are therefore described by the set of $N$ coupled stochastic differential equations given by Eq.~\eqref{overdamped_Langevin}.

\subsection{Theoretical challenges and purpose of the Dean-Kawasaki approach}

From a numerical perspective, such coupled equations can be integrated quite straightforwardly through Brownian dynamics simulations (see Ref.~\cite{Allen1987,Kloeden1992} for the fundamentals of this method, and Refs.~\cite{Leimkuhler2013,Rackauckas2017,Sammuller2021} for recent refinements). However, predicting the behavior of such a suspension from an analytical perspective is a theoretical challenge which raises numerous difficulties, in spite of its importance to understand the underlying physics.  The $N$-body problem that is set out in the previous section, and the different strategies that can be employed to solve it (at least partially or under suitable approximations), is at the heart of the works that are reviewed in this manuscript. My aim is to present, in the most instructive and non-technical way, some of the different analytical techniques that have been proposed to study the dynamics of Brownian particles coupled by pair interactions, and which obey equations such as Eq. \eqref{overdamped_Langevin}. I will focus on the Dean-Kawasaki equation, also called more recently `stochastic density functional theory'.

Investigating the timeline of this topic reveals that its conceptual aspects are at the crossroad between different topics of theoretical physics (stochastic processes, statistical field theory, disordered systems, classical density functional theory) and of mathematics (probability theory, stochastic partial differential equations, numerical analysis). On top of its fundamental richness, this level of modeling finds its applications to predict and analyze the behaviour of a wide range of nonequilibrium systems, such as supercooled liquids, active matter, or driven electrolytes.

The starting point of the review will be the framework that was set out respectively by Kawasaki and Dean, which are closely related to each other even though they differ on the calculation strategies, and on the physical observables they describe. Their common goal was to obtain an evolution equation for the density of particles at a given point of the system. In Section~\ref{sec_fundamental}, I present their fundamental results, and give some details on the derivations and underlying assumptions of the so-called `Dean-Kawasaki' (DK) equation. In Section~\ref{sec_links}, I place these results in the more general context of theoretical statistical mechanics, and show how they can be related to alternative strategies that were employed earlier or later to describe the dynamics of interacting Brownian particles. A few years after its derivation, the DK equation has raised a number of questions of mathematical interest, regarding its well-posedness and possible regularization. Even though this review is aimed at the physical community, I attempt in Section~\ref{sec_math} to summarize briefly the different works that recenlty addressed the DK equation as an object of mathematical interest. In Section~\ref{sec_extensions}, I show how the original DK equation, which applies in principle to identical particles obeying simple overdamped dynamics, can be extended to account for more complex situations. Section~\ref{sec_solutions} is devoted to the different strategies that have been employed to solve the DK equation, both analytically and numerically. Finally, Section~\ref{sec_applications} reviews the different systems that have been studied thanks to the DK equation, and the results of physical, chemical and biological interest that were obtained.

\subsection{Terminology}

The first occurrence of the phrasing `stochastic density functional theory' is probably due to Archer and Rauscher~\cite{Archer2004a}, who use the adjective `stochastic' to emphasize that the DK equation should not be confused with `dynamical density functional theory' (DDFT): the connections between SDFT and DDFT will be discussed in Section~\ref{sec_ddft}. This denomination was subsequently adopted by different authors.  Importantly, the expression `stochastic density functional theory' was also introduced recently in the theoretical chemistry community, to denote a stochastic method to sample electronic structure of molecules~\cite{Fabian2019}: we emphasize that the framework described in the present manuscript is unrelated.

\section{Fundamental equations}
\label{sec_fundamental}

\subsection{Kawasaki's approach}
\label{sec_Kawasaki}

Studying the dynamics of Brownian suspensions which obey evolution equations such as the one given in Eq. \eqref{overdamped_Langevin} has been the object of many theoretical approaches. Among them, mode-coupling theory (MCT), which has been proposed in the context of supercooled liquids and glass transition~\cite{Gotze2009,Gotze1992}, has been particularly successful. Its idea goes as follows: starting from the $N$-body dynamics of the suspension, one derives formally the evolution equations for the two-point, two-time density correlations. As expected, these equations are unclosed, and give rise to an infinite hierarchy of equations obeyed by correlation functions of higher and higher order. The central idea of MCT is to close this hierarchy of equations at the two-body level, by providing a suitable approximation for the memory kernels that appear in the formal solution of the Smoluchowski dynamics. This `mode-coupling approximation' typically takes as input the static structure factor of the liquid, that can be evaluated through numerical simulations or through the usual approximations from the static theories of liquids (additional comments on MCT will be given in Section \ref{sec_MCT}).

In 1994, Kawasaki proposed an alternative to classical MCT, and introduced another way to close the hierarchy of equations \cite{Kawasaki1994}. His idea was to start from the Smoluchowski equation, obeyed by the $N$-body probability distribution $P_N(\rr^1,\dots,\rr^N;t)$  that can be deduced from Eq. \eqref{overdamped_Langevin}, and which reads $\partial_t P_N(\rr^1,\dots,\rr^N;t) = \Omega P_N$, with the operator 
\begin{equation}
	\Omega=\nabla^N\cdot \left[D\nabla^N +\mu \sum_{\alpha>\beta} \nabla^N V(\rr^\alpha-\rr^\beta)\right],
\end{equation}
where $\nabla^N$ is the $3N$-dimensional gradient operator. Instead of relying on the usual MCT strategy, which would consist in projecting the $N$-body dynamics onto collective variables, for instance using the Mori-Zwanzig formalism~\cite{Mori1965b,Zwanzig1961}, Kawasaki suggested to perform a local coarse-graining of the system. More precisely, the system is assumed to be divided into small cells, which contain a number of particles much larger than 1, but much smaller than the total number of particles: the system is then described at a truly mesoscopic level. The quantity of interest then becomes $\mathcal{P}(\{\hat \rho \},t)$, i.e. the probability to observe a given set of local densities $\{\hat \rho \} = \{\hat \rho_1,\dots,\hat\rho_{N_c} \} $ in each of the $N_c$ coarse-graining cells.

In order to derive an evolution equation for $\mathcal{P}(\{\hat \rho \},t)$, one needs to relate it to the $N$-body distribution function $P_N$. The coarse-graining procedure is then completed by an hypothesis of local equilibrium. More precisely, one assumes that the system is equilibrated at any time at the scale of each coarse-graining cell. Within this approximation, the $N$-body probability distribution $P_N$ is replaced by an equilibrium counterpart which is compatible with the coarse-grained density distribution functional: $P_N(\rr^1,\dots,\rr^N;t)\simeq  P_N^\text{eq}(\rr^1,\dots,\rr^N)\mathcal{P}(\{\hat \rho \},t)/ \mathcal{P}^\text{eq}(\{\hat \rho \})$. This hypothesis is expected to be valid for slow enough evolution.

In this procedure, which is performed on scales much larger than the interaction range, the interaction part of the operator $\Omega$ is to be replaced by an effective interaction potential connected to the direct pair correlation function of the fluid. Kawasaki finally gets an equation obeyed by the probability distribution functional $\mathcal{P}(\{\hat\rho\},t)$ of the density variable $\hat \rho$ that is derived from the coarse-graining procedure, which reads 
\begin{align}
	&\frac{\partial}{\partial t} \mathcal{P}(\{\hat\rho\},t) = -D\int \dd \xx \frac{\delta}{\delta \hat\rho(\xx)} \nabla \cdot\Big\{\hat \rho(\xx) \nonumber\\
	&\times \nabla \left[ \frac{\delta}{\delta\hat\rho(\xx)} + \frac{1}{\kB T} \frac{\delta F[\hat\rho]}{\delta\hat\rho(\xx)} \right] P(\{\hat\rho\},t) \Big\},
	\label{Kawasaki_eq}
\end{align}
with the functional 
\begin{eqnarray}
	&&F [\hat\rho] = \kB T \int \dd \xx\;  \hat\rho(\xx) \left[\ln \frac{\hat\rho(\xx)}{\rho_0}-1 \right] \nonumber\\
	&&+\frac{1}{2} \int \dd \xx \int\dd \xx' \; [\hat\rho(\xx)-\rho_0] c(\xx-\xx') [\hat\rho(\xx')-\rho_0], \nonumber\\
	\label{functional}
\end{eqnarray}
where $c$ is the direct pair correlation function, and where $\rho_0=N/\mathcal{V}$ is the overall density of particles, with $ \mathcal{V}$ the volume of the system. This set of equations, which actually has connections with earlier MCT studies~\cite{Munakata1977, Munakata1989, Bagchi1987}, is the main result presented by Kawasaki in Ref. \cite{Kawasaki1994}. Just like in MCT, this approach requires the \textit{a priori} knowledge of the static structure of the liquid. However, the equation obeyed by the probability distribution function  $\mathcal{P}$ is particulary difficult to analyze, as opposed to the typical MCT equations, which are integro-differential equations obeyed by correlation functions, that can usually be integrated numerically.

\begin{figure*}
	\begin{center}
		\includegraphics[width=2\columnwidth]{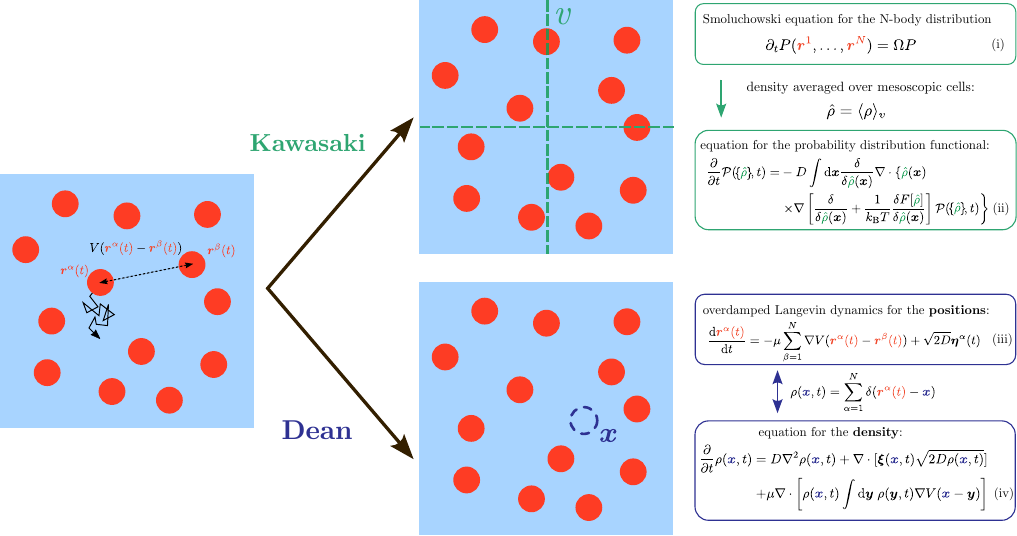}
	\end{center}
	\caption{Schematic representation of the derivations by Kawasaki \cite{Kawasaki1994} and Dean \cite{Dean1996}. See text for the definitions of the different quantities involved in Eqs. (i)-(iv). The blue background on the left panel represents the `solvent' in which the Brownian particles are embedded, and which causes their stochastic motion. The corresponding microscopic, overdamped dynamics is treated by a local coarse-graining procedure by Kawasaki (top panel) and by the introduction of a stochastic density by Dean (bottom panel). From the point of view of hydrodynamics, deriving the Dean-Kawasaki equation from the overdamped Langevin equations is equivalent to going from a Lagrangian description (which amounts to tracking and following individual particles over time, as in Eqs. (i) and (iii) on the Figure) to a Eulerian description (which amounts to observing a given point in the system and counting whether a particle is present or not, as in Eqs. (ii) and (iv)).}
	\label{sketch}
\end{figure*}

\subsection{Dean's derivation}
\label{sec_Dean}

In 1996, Dean also considered interacting Browian particles, and intended to derive the evolution equation of the density of particles at a given point of space~\cite{Dean1996}. The spirit of his approach was somewhat related to that of Kawasaki. However, the technical treatment of the overdamped Langevin dynamics is largely different, and I summarize it here.  Starting from the set of coupled Langevin equations [Eq. \eqref{overdamped_Langevin}], and considering some arbitrary test function $f$, Ito's lemma~\cite{Gardiner1985} yields, for any $\alpha\in\{1,\dots,N\}$:
\begin{eqnarray}
	%\label{ }
	&&\frac{\dd f(\rr^\alpha(t))}{\dd t} = D \nabla^2 f(\rr^\alpha(t)) + \sqrt{2D} \boldsymbol{\eta}^\alpha(t)\cdot \nabla f(\rr^\alpha(t)) \nonumber\\
	&& -\mu\left[ \sum_{\beta=1}^N  \nabla V(\rr^\alpha(t)-\rr^\beta(t))   \right] \cdot \nabla f(\rr^\alpha(t))  .
\end{eqnarray}
One then defines the density function of a single particle $\rho_\alpha(\xx,t) \equiv \delta (\rr^\alpha(t)-\xx)$
and deduces
\begin{eqnarray}
	\label{ }
&&	\frac{\dd f(\rr^\alpha(t))}{\dd t} =\nonumber\\
&& \int \dd \xx \; \rho_\alpha(\xx,t) \Bigg[ D \nabla^2 f(\xx)+  \sqrt{2D} \boldsymbol{\eta}^\alpha(t)\cdot \nabla f(\xx)  \nonumber\\
	&&  - \mu \nabla f(\xx)\cdot \sum_{\beta=1}^N  \nabla V(\xx-\rr^\beta(t))      \Bigg] .
\end{eqnarray}
Integrating by parts yields
\begin{eqnarray}
	\label{eq_f_IPP}
&&	\frac{\dd f(\rr^\alpha(t))}{\dd t} = \nonumber\\
&&\int \dd \xx \; f(\xx) \Bigg\{  D\nabla^2\rho_\alpha(\xx,t)- \sqrt{2D}\nabla\cdot[ \rho_\alpha(\xx,t)  \boldsymbol{\eta}^\alpha(t)]  \nonumber\\
&&	 +  \mu\nabla \cdot \left[ \rho_\alpha(\xx,t) \sum_{\beta=1}^N \nabla V (\xx-\rr^\beta(t))   \right]  \Bigg\}.\nonumber\\
\end{eqnarray}
From the definition of $ \rho_\alpha(\xx,t)$, it is clear that $f(\rr^\alpha(t)) = \int\dd\xx\;  \rho_\alpha(\xx,t) f(\xx)$. The derivative of this relation with respect to $t$ reads 
\begin{equation}
	\frac{\dd f(\rr^\alpha(t))}{\dd t} = \int \dd \xx \; \left(  \frac{\partial}{\partial t}\rho_\alpha(\xx,t)   \right)  f(\xx). 
	\label{eq_f_deriv}
\end{equation}
Comparing Eqs. \eqref{eq_f_deriv} and \eqref{eq_f_IPP}, and considering that these equalities hold for any test function $f$, one finds
\begin{eqnarray}
	\label{eq_EDP_rhoalpha}
	&&\frac{\partial}{\partial t}\rho_\alpha(\xx,t)  =  D\nabla^2\rho_\alpha(\xx,t) -   \sqrt{2D} \nabla\cdot[\rho_\alpha(\xx,t)  \boldsymbol{\eta}^\alpha(t)] \nonumber\\
	&&+\mu  \nabla \cdot \left[ \rho_\alpha(\xx,t) \sum_{\beta=1}^N \nabla V (\xx-\rr^\beta)   \right]  . 
\end{eqnarray}
The last step is to define the global density 
\begin{equation}
	\rho(\xx,t) = \sum_{\alpha=1}^N\delta (\rr^\alpha(t)-\xx)= \sum_{\alpha=1}^N \rho_\alpha(\xx,t), 	
\end{equation}
and summing Eq. \eqref{eq_EDP_rhoalpha} for $\alpha=1,\dots,N$ yields:
\begin{eqnarray}
&&	\frac{\partial}{\partial t}\rho(\xx,t) = D\nabla^2 \rho(\xx,t)+ \nabla \cdot [\boldsymbol{\xi}(\xx,t) \sqrt{2D\rho(\xx,t)}]  \nonumber\\
&&	+ \mu\nabla \cdot \left[ \rho(\xx,t) \int \dd \yy \; \rho(\yy,t)   \nabla V(\xx-\yy) \right] ,
\label{Dean_eq}
\end{eqnarray}
where one uses the relation $\nabla V(\xx-\rr^\beta) = \int \dd \yy \; \nabla V(\xx-\yy) \rho_\beta(\yy,t)$. Moreover, to derive Eq. \eqref{Dean_eq}, one uses the fact that the noise term $\Xi(\xx,t)\equiv - \sum_{\alpha=1}^N \nabla \cdot [\rho_\alpha(\xx,t)   \boldsymbol{\eta}^\alpha(t)] $ is Gaussian (this follows from the fact that, when computing its cumulants, one can average over the noises $\boldsymbol{\eta}^\alpha(t)$ independently), and has the same variance as  $\nabla \cdot [\sqrt{\rho(\xx,t)}\boldsymbol{\xi}(\xx,t)]$, $\boldsymbol{\xi}(\xx,t)$  being a space-dependent Gaussian random variable, which satisfies 
\begin{align}
	\langle \xi_i(\xx,t) \rangle &= 0, \\
	\langle \xi_i(\xx,t) \xi_j(\xx',t') \rangle &= \delta_{ij} \delta(t-t')\delta(\xx-\xx').
\end{align}
This can be proven by showing that $\langle \Xi(\xx,t) \Xi(\xx',t')\rangle = \delta(t-t')\sum_{\alpha=1}^N \nabla_{\xx} \cdot \nabla_{\xx'}(\rho_\alpha(\xx,t)\rho_\alpha(\xx',t'))$, and using that $\rho_\alpha(\xx,t)\rho_\alpha(\xx',t')=\delta(\xx-\xx')\rho_\alpha(\xx,t)$. Eq. \eqref{Dean_eq} is the main result from Ref.~\cite{Dean1996}, and is usually called Dean's equation.

Several comments follow: (i)~It is important to underline that the derivation of Eq. \eqref{Dean_eq} does not rely on any approximation, and that it is equivalent to the set of coupled Langevin equations given by Eq. \eqref{overdamped_Langevin} (this equivalence is summarized on the sketch shown on Fig.~\ref{sketch}). In other words, it is not associated with any coarse-graining procedure, in such a way that it preserves the notion of `particle entity', i.e. the property that individual particles only exist at one position in space at any given time (this was proven rigorously in Ref.~\cite{Bothe2023}); (ii)~However, the unknown of this equation is the stochastic density $\rho(\xx,t) = \sum_{\alpha=1}^N \delta (\rr^\alpha(t)-\xx)$, which is a sum of singular $\delta$-functions, and should therefore be seen as a density operator defined in the sense of distributions. Its physical meaning is therefore unclear, unless one performs ensemble averages or local spatial averages. Dean's equation should then be understood as being set in  a distribution space; (iii) Eq. \eqref{Dean_eq} has two nonlinearities in $\rho$. The interaction term is proportional to $\rho^2$, which directly stems from the pairwise interactions between particles. Moreover, the noise term scales as $\sqrt{\rho}$: Dean's equation is nonlinear even for noninteracting particles; (iv) The noise term in Eq. \eqref{Dean_eq} is multiplicative, i.e. its amplitude depends on the random variable $\rho$ itself. This raises a number of difficulties in the analysis of this equation, which will be discussed later in this review; (v) Finally, Eq. \eqref{Dean_eq} can be rewritten under the form
\begin{eqnarray}
	\frac{\partial}{\partial t}\rho(\xx,t) &=&  \mu\nabla \cdot \left[ \rho(\xx,t) \nabla \frac{\delta \mathcal{F}[\rho]}{\delta \rho(\xx,t)}  \right]  \nonumber\\
		&&+\nabla \cdot [\boldsymbol{\xi}(\xx,t) \sqrt{2D\rho(\xx,t)}],
\end{eqnarray}
with the functional 
\begin{eqnarray}
	\mathcal{F}[\rho] &=& \kB T \int \dd \xx \, \rho(\xx) \ln \rho(\xx)\nonumber\\
	&& + \frac{1}{2}\int \dd \xx \int \dd \xx' \rho(\xx) V(\xx-\xx')\rho(\xx').
\end{eqnarray}

This highlights the strong connection between the result derived by Kawasaki, and that obtained by Dean. While the latter derived a Langevin-like equation obeyed by the stochastic density $\rho$, the former derived the evolution equation of a probability density functional $\mathcal{P}$ associated with a locally coarse-grained density $\hat\rho$. To some extent, and as a guiding analogy, the result by Kawasaki can be understood as a `Fokker-Planck' counterpart to the `Langevin' equation derived by Dean (see for instance Refs.~\cite{Frusawa2000,Archer2004a} for a discussion of the relationship between Eqs. \eqref{Kawasaki_eq} and \eqref{Dean_eq}). Even though the analogy should remain limited, since the densities $\rho$ and $\hat \rho$ do not have the same physical meaning, and since the two derivations rely on rather different hypotheses, Eq. \eqref{Dean_eq} is sometimes called the `Dean-Kawasaki' equation in the literature -- this is the terminology that I will adopt in what follows.

\section{Relationships to other theories}
\label{sec_links}

In this Section, I show how the Dean-Kawasaki equation can be related to other classical theories from statistical mechanics, that either preceded or followed its derivation in the mid-nineties.

\subsection{Fluctuating hydrodynamics and macroscopic fluctuation theory}

The typical program of statistical mechanics consists in starting from the microscopic laws of evolution of the many particles that constitute the system, and in deducing a macroscopic description of the overall `fluid'. One typically ends up with hydrodynamic equations, that are obeyed by conserved fields (e.g. particle density, momentum or energy). These equations are  valid on sufficiently coarse timescales and lengthscales. Such hydrodynamic laws of evolution can be refined by computing the fluctuations around the deterministic evolution under suitable hypotheses. It typically relies on local equilibrium assumptions: more precisely, the system is assumed to reach microscopic equilibrium in a time much shorter than the typical times associated with macroscopic evolution. This framework, usually called fluctuating hydrodynamics, was initiated by Landau and Lifshitz~\cite{Landau1980}, and subsequently discussed and developed by many authors, such as Fox~\cite{Fox1978} or Spohn~\cite{Spohn1991}. By extension, the phrasing `fluctuating hydrodynamics' is often used to describe stochastic differential equations obeyed by density functionals, independently on how they are derived from microscopic principles \cite{Donev2010,Delong2013,Delong2014,DeLaTorre2015,Kim2017}.

Denoting by $\check \rho(\xx,t)$ a suitably coarse-grained particle density, the evolution of a diffusive system is given by a continuity equation:
\begin{equation}
	\partial_t \check \rho(\xx,t) = -\nabla\cdot \check{\boldsymbol{j}}(\xx,t),
	\label{FH1}
\end{equation} 
which is completed by the analogous of a constitutive equation, which relates the current $\check{\boldsymbol{j}}$ to the density $\check \rho$:
\begin{equation}
	\check{\boldsymbol{j}}(\xx,t) = - \check D(\check\rho(\xx,t)) \nabla \check\rho(\xx,t)+\sqrt{ \check \sigma(\check\rho(\xx,t))} \eta(\xx,t),
	\label{FH2}
\end{equation} 
where  $\eta(\xx,t)$ is a Gaussian noise of zero average and variance $\langle\eta(\xx,t)\eta(\xx',t') \rangle = \delta(\xx-\xx')\delta(t-t') $, and where $\check D$ and $\check \sigma$ are respectively the (collective) diffusion coefficient and the mobility. Under the assumption of local equilibrium, they are related through the relation $ 2\check{D}(\check{\rho})/\check{\sigma}(\check{\rho})=f''(\check{\rho})$, where $f$ is the free energy density \cite{Derrida2007}. Importantly, all the microscopic aspects of the dynamics (such as pairwise interactions) are encoded in the transport coefficients $\check{D}$ and $\check{\sigma}$.

Such description of the dynamics of the system is particularly convenient. Together with initial and boundary conditions, it allows to compute many quantities, such as stationary profiles, density correlations, the probability to observe a given macroscopic profile, relaxation towards equilibrium etc. In this context, macroscopic fluctuation theory (MFT) was proposed in the early 2000s by Bertini and collaborators~\cite{Bertini2001,Bertini2002,Bertini2015a}. MFT is a deterministic reformulation of fluctuating hydrodynamics: through a path-integral reformulation of the stochastic dynamics given in Eqs. \eqref{FH1}-\eqref{FH2}, one can calculate exactly large deviation functions of the density in generic driven diffusive systems. This framework has been successfully applied by various authors to get exact results on current fluctuations and tracer diffusion in paradigmatic one-dimensional models of statistical mechanics, in which the coefficients $\check{D}$ and $\check{\sigma}$ are known exactly~\cite{Derrida2009,Krapivsky2014c,Krapivsky2015,Poncet2021a,Grabsch2022,Mallick2022,Grabsch2024}.

At this stage, it is tempting to establish a relationship between Eqs. \eqref{FH1}-\eqref{FH2} on the one hand, and the DK equation [Eq. \eqref{Dean_eq}] on the other hand, since there exist very strong similarities between the two sets of equations (see Refs.~\cite{Duran-Olivencia2017,Chavanis2008} for discussions on the connection between both approaches). However, we must emphasize that the fluctuating hydrodynamics approach differs from that of SDFT, in the sense where the DK equation does not rely on any coarse-graining approximation, and is an exact reformulation of the microscopic dynamics. It is however interesting to draw a link between the two approaches, since the mathematical methods developed to study the large deviation of the fluctuating hydrodynamics equation can also be employed to obtain exact results on the DK equation. For instance, for a finite number of non-interacting particles, this was essentially the method followed by Velenich et al., in a paper~\cite{Velenich2008} that will be commented later (Section~\ref{sec_exact}).

Finally, in the particular situation where the particles interact via a mean-field-type interaction (for instance when the strength of the potential scales with the inverse number of particles), and in the limit of a very large number of particles ($N\to\infty$), the noise term in Eq. \eqref{Dean_eq} is subdominant. It can then be shown that the DK equation becomes equivalent to the (deterministic)  McKean-Vlasov equation~\cite{McKean1966}, which has been studied extensively in the mathematical literature. In this particular limit, a connection between the DK equation and MFT has been established by Bouchet et al.~\cite{Bouchet2016}, which allowed them to study the phase diagram of a mean-field model of coupled stochastic rotators (on mean-field approaches, see also Ref. \cite{Chavanis2014a}).

\subsection{Mode-coupling theory}
\label{sec_MCT}

Mode-coupling theory (MCT) was initially proposed to describe the dynamics of glass-forming liquids, in the seminal works by G\"otze~\cite{Gotze2009}. The quantity of interest in MCT is the intermediate scattering function $F(q,t)$, defined as
\begin{equation}
	F(q,t) = \frac{1}{N} \langle\tilde\rho(\qq,t)\tilde\rho(-\qq,0)\rangle,
\end{equation}
with the Fourier transform of the density $\tilde\rho(\qq,t)=\sum_{\alpha=1}^N \ex{-\ii \qq\cdot \rr^\alpha(t)}$ (throughout this review, we will use the following convention for Fourier transformation: $\tilde\psi(\qq) = \int \dd\rr\, \ex{-\ii\qq\rr} \psi(\rr)$, and its inverse: $\psi(\rr) = \frac{1}{(2\pi)^d} \int \dd \qq \,  \ex{\ii\qq\rr}\tilde\psi(\qq)$). An  equation of motion for $F(q,t)$ can be obtained by studying the variables of interest, i.e. the density modes, with the Mori-Zwanzig projection formalism~\cite{Mori1965b,Zwanzig1961}. This results in an exact equation for $F(q,t)$ which involves a memory kernel, that contains all the information about particle interactions, and that cannot be expressed simply in terms of the unknown $F$. The main approximations of MCT are (i) to express the memory kernel as a four-point correlation function, and (ii) to decouple four-point correlation functions as a product of two-point correlation functions. This typically yields a closed equation for the intermediate scattering function $F$, that only requires as inputs the static structure factor $S(q)=F(q,0)$ or equivalently the direct correlation function $c(q)=(1-1/S(q))/\rho_0$, which can be computed from numerical simulations or through the usual closures from the static theory of liquids~\cite{Hansen1986}. I will not go into further details on MCT and refer the reader to recent reviews~\cite{Reichman2005,Szamel2013, Janssen2018}.

This standard MCT was long considered as the most successful theory that is derived from microscopic principles and that may explain many features of the glass transitions as observed in experiments and numerical simulations. Unfortunately, in the low-temperature or high-density regime, standard MCT predicts that the system may become non-ergodic, in disagreement with all other observations. The decoupling approximation that is at the basis of MCT was therefore discussed, and it was suggested that adding higher-order correlation functions could resolve the issue of non-ergodicity~\cite{Mayer2006}, but it remained technically challenging.

In this context, approaches such as fluctuating hydrodynamics, and more specifically the DK equation, appeared as promising to rederive MCT-like equations, and potentially to improve the standard result~\cite{Mazenko2010, Mazenko2011,Das2013b}. Indeed, the intermediate scattering function $F(q,t)$ is defined in terms of the Fourier transform of the density $\rho$, which happens to be the quantity which obeys the DK equation. The idea was then to adopt a path-integral formulation of the DK equation, and to derive the associated action following the Martin-Siggia-Rose/Janssen–De Dominicis–Peliti  formalism~\cite{Martin1973,Janssen1976,DeDominicis1978a}. The perturbative derivation of standard MCT using this method was actually quite subtle, as fluctuation-dissipation relations need to be preserved~\cite{Miyazaki2005,Nishino2008,Das2020} -- an aspect which is closely related to the time-reversal symmetry of the action~\cite{Andreanov2006a}. Standard MCT was eventually successfully rederived from the DK equation by Kim et al.~\cite{Kim2014}, improving on a preliminary attempt~\cite{Kim2008}.  This derivation relies on a field-theoretical formulation of the Dean-Kawasaki equation, that is expanded around a Gaussian theory, i.e. not around the interaction-free case, that is non-Gaussian (this will be discussed in Section \ref{sec_exact}).  In summary, this series of works highlights the strong connections that exist between SDFT and MCT. As a final remark, note that MCT can also be connected to DDFT, that will be the object of the next section, in a less rigorous but more physical way~\cite{Archer2006a,teVrugt2020a}.

\subsection{Dynamical density functional theory (DDFT)}
\label{sec_ddft}

As emphasized in Section~\ref{sec_Dean}, Dean's equation is obeyed by a stochastic density, defined as a sum of $\delta$-functions, evaluated at fluctuating positions. A natural way to analyze the dynamics of this density would be to start by studying its average behavior, and it is tempting to perform an ensemble average of Eq. \eqref{Dean_eq}. Assuming that each realisation of the stochastic density field $\rho$ has a probability $P[\rho]$, and defining ensemble average as $\langle \cdot \rangle = \int \mathcal{D} \rho \;  \cdot P[\rho] / \int \mathcal{D} \rho \;   P[\rho]$, I define the ensemble-averaged density (i.e. averaged over the noise realizations) as $\bar \rho(\xx,t) = \langle \rho(\xx,t) \rangle$. From Eq. \eqref{Dean_eq}, one finds~\cite{Marconi1999}:
\begin{eqnarray}
	&&\frac{\partial}{\partial t}\bar\rho(\xx,t) = D\nabla^2 \bar\rho(\xx,t) \nonumber\\
	&&+  \mu\nabla \cdot \left[  \int \dd \yy \; \langle \rho(\xx,t) \rho(\yy,t)   \rangle \nabla V(\xx-\yy) \right].
	\label{unclosed_ddft}
\end{eqnarray}
In order to close this equation, one needs to express the two-point correlation function $ \langle \rho(\xx,t) \rho(\yy,t)   \rangle$ in terms of the one-point density $\bar\rho(\xx,t)$: again, this requires some closure approximation. Indeed, any exact evolution equation for the  two-point correlation function $ \langle \rho(\xx,t) \rho(\yy,t)   \rangle$ will involve some three-point correlation functions, and so on: this is the usual BBGKY-like hierarchy of equations that appears when describing interacting particles~\cite{Hansen1986,McQuarrie1976}.  

The simplest approximation would consist in writing $\langle \hat\rho(\xx,t) \hat\rho(\yy,t)   \rangle \simeq  \bar\rho(\xx,t) \bar \rho(\yy,t)  $. This mean-field approximation, which can be interesting in certain limits, would nonetheless be pathological for particles with strong repulsion, for instance. The idea of dynamical density functional theory (DDFT) is to approximate the two-point correlation function with the help of equilibrium free energy density functionals. More precisely, the excess part of the free energy, which contains information about the interparticle correlations,  may be used to `close' Eq. \eqref{unclosed_ddft}. This was initiated phenomenologically by Evans~\cite{Evans1979}, and a theoretical framework was developed later on. I follow here the ideas from Ref.~\cite{Marconi1999}, where Marconi and Tarazona proposed the `adiabatic approximation'. Consider an equilibrium system, described by a time-independent density function $\rho_\text{eq}$,  with the same interactions as in the dynamical system of interest. Such a system can be described by usual (static) density functional theory~\cite{Evans1979, Evans1992,Hansen1986}, with the free energy functional
\begin{equation}
	F_\text{eq}[\rho_\text{eq}] = \kB T \int \dd \xx \; \rho_\text{eq}(\xx)\left[\ln\frac{\rho_\text{eq}(\xx)}{\rho_0}+1\right] +  F_\text{exc}[\rho_\text{eq}],
\end{equation}
where $ F_\text{exc}[\rho_\text{eq}]$ is the excess free energy density functional, which contains all the information about particle interactions, and  where $\rho_\text{eq}(\xx)$ is the equilibrium, static counterpart to $\bar \rho(\xx,t)$.

Marconi and Tarazona then proposed the following approximation:
\begin{equation}
	\int \dd \yy \; \langle \hat\rho(\xx,t) \hat\rho(\yy,t)   \rangle \nabla V(\xx-\yy) \simeq \bar \rho(\xx,t) \nabla \frac{\delta  F_\text{exc}[\bar \rho(\xx,t)]}{\delta \bar\rho(\xx,t)}.
\end{equation}
This is obtained by relying on the fact that, at each time, one can find a fictitious external potential that equilibrates the system~\cite{Lovett1976} (i.e. that minimizes the grand potential, in the language of classical DFT). In other words, the DDFT approximation replaces the `true' non-equilibrium pair distribution function $ \langle \hat\rho(\xx,t) \hat\rho(\yy,t)   \rangle$ by the equilibrium one, and then uses the equilibrium density functional $ F_\text{exc}$ to express it. In summary, provided that the equilibrium $F_\text{exc}[\rho]$ is known explicitly (which is the case for many systems through accurate approximations that have been in the framework of classical DFT), one gets a closed equation for the (ensemble-averaged) one-body density:
\begin{equation}
	\frac{\partial}{\partial t}\bar \rho(\xx,t) = \mu \nabla\cdot  \left[\bar \rho(\xx,t) \nabla \frac{\delta F_\text{eq}[\bar \rho]}{\delta\bar{\rho}(\xx,t)}     \right].
	\label{DDFT_eq}
\end{equation}
Other theoretical ways can be followed to get this DDFT equation, for instance using projection operators~\cite{Yoshimori2005,Espanol2009}.

As opposed to classical, static DFT, which predicts the equilibrium configuration of a system of interacting particles typically through functional minimization, DDFT provides, within a set of approximations, the time evolution of a system to its equilibrium configuration. This approach quickly became successful to predict dynamical phenomena in suspensions of interacting Brownian particles, such as phase separation, nucleation, pattern formation, in systems ranging from polymers to passive and active colloidal fluids (see Ref.~\cite{teVrugt2020a} for a recent review on DDFT). More recently, this approach has been completed by power functional theory, a formal framework to treat dynamics in a variational way~\cite{Schmidt2013,Lutsko2021,Schmidt2022,DeLasHeras2023}. Finally, and to go back to the purpose of this review, I emphasize that the main drawback of DDFT is that it leads to deterministic equations, that are able to predict the time evolution of the system, but only its  average behavior. It does not give information on fluctuations around the average behavior, which is the core of SDFT and its main advantage.

We conclude this Section by discussing the relationships between the equation derived by Kawasaki in 1994 [Eq. \eqref{Kawasaki_eq}], that derived by Dean in 1996 [Eq. \eqref{Dean_eq}], and the DDFT equation written in this Section [Eq. \eqref{DDFT_eq}]. These three equations have apparent similarities: they all stem from the same microscopic dynamics ($N$ interacting Brownian particles), and they give the time evolution of a  `particle density'. However, the three densities $\hat{\rho}$, $\rho$ and $\bar{\rho}$, which are the variables involved in Eqs. \eqref{Kawasaki_eq}, \eqref{Dean_eq} and \eqref{DDFT_eq}, respectively have very different physical meanings -- the choice of notation highlights this difference. The first one is a spatially coarse-grained density; the second is a `proper' microscopic density, which is however defined in distribution space, stricly speaking; the third one is an ensemble-averaged density. These three equations have therefore different physical grounds, and the choice of describing a system of Brownian particles with one approach or the other should be done carefully, guided by the level of description that is to be adopted and the physical conclusions that are to be drawn. This misleading relatedness has been a source of confusion in the literature, which was eventually clarified by different authors, see in particular Refs.~\cite{Frusawa2000,Archer2004a,Kawasaki2006,Chavanis2011b,Chavanis2019}.

\section{Mathematical considerations}
\label{sec_math}

As shown by its history and recent developments, the Dean-Kawasaki equation has strong links with the physical world, and has motivated a lot of work in different physics communities, both on its theoretical aspects and on applications. However, the mathematical analysis of this equation has begun only recently. Several fundamental questions have been raised and addressed. Although this review is aimed at physicists, I found it interesting to review these recent mathematical results in a non-technical way.

The first problem that was addressed by mathematicians concerns the well-posedness of the Dean-Kawasaki equation, and more precisely the existence and uniqueness of its solutions. Interestingly, it was shown, first in the case of non-interacting particles~\cite{Konarovskyi2019}, that the DK equation is only well-posed for a discrete set of values of the diffusion coefficient $D$.  In other words, from a rigorous point of view, the DK equation only admits solutions for very specific values of parameters. This is rather surprising and puzzling given the interest of the DK equation and its predictive power when it is used in less rigorous ways. This result was later extended to the case of particles interacting through smooth enough potentials~\cite{Konarovskyi2020},  to the case where a specific initial condition is imposed ~\cite{Konarovskyi2024}, to non-local and singular interaction kernels~\cite{Wang2024}, and to the underdamped case \cite{Müller2025}.

The meaning of the density $\rho$ (sometimes called `empirical density' in the mathematical literature), defined in the original setting as a sum of delta functions, is rather unphysical, and the DK equation would only be defined rigorously in distributional space. An interesting alternative is to regularize the density (in the spirit of numerical methods such as smoothed-particle hydrodynamics \cite{Violeau2012}), and to define it as 
\begin{equation}
	\rho_\epsilon(\xx,t) = \sum_{\alpha=1}^N w_\epsilon(\xx-\rr^\alpha(t)),
\end{equation}
where $w_\epsilon(\yy)=\ex{-\yy^2/2\epsilon^2}/\sqrt{2\pi \epsilon^2}$ is a Gaussian kernel of variance $\epsilon^2$. The original DK equation may therefore be retrieved by taking the limit $\epsilon\to 0$. The equation satisfied by $\rho_\epsilon$ was derived and analyzed by Cornalba et al. both in one dimension~\cite{Cornalba2019,Cornalba2020} and higher dimensions~\cite{Cornalba2021} -- note that these works also include the case of inertial, underdamped dynamics. 

Finally, I mention that recent mathematical developments have focused on regularizing the DK equation by resorting to spatial discretization~\cite{Cornalba2023}, whose validity is checked by verifying that density fluctuations are correctly predicted. This provides a formal basis for some of the numerical schemes that are discussed in Section~\ref{sec_numerical}.

\section{Some extensions of the original result}
\label{sec_extensions}

The original DK equation, as stated in Eqs. \eqref{Kawasaki_eq} and \eqref{Dean_eq}, holds for identical Brownian particles, that obey overdamped dynamics, and which are immersed in a solvent which is described implicitly (in the sense that it only influences the dynamics of the particles through viscous damping and through its thermal fluctuations, and not through any fluid-mediated interactions). This `simple' result was subsequently extended to more general situations, by adding different ingredients to the original model. In this section, I list some of the different extensions of the original DK equation.

 Going beyond the overdamped limit, the effect of inertia was included in the DK equations in Refs.~\cite{Nakamura2009, Das2013}. The starting point of this calculation is the set of (underdamped) Langevin equations:
\begin{eqnarray}
	\frac{\dd \rr^\alpha}{\dd t}  &=& \vv^\alpha, \\
	m\frac{\dd \vv^\alpha}{\dd t} &=& -\gamma m \vv^\alpha -\sum_{\beta=1}^{N} \nabla V(\rr^\alpha-\rr^\beta) \nonumber\\
	&&+ \sqrt{2m\gamma \kB T} \boldsymbol{\eta}^\alpha(t).
\label{underdampedLangevin2}
\end{eqnarray}
It\^o calculus can be performed on both these equations to yield coupled evolution equations for the density of particles, defined as before $\rho(\xx,t)=\sum_{\alpha=1}^N\delta(\xx-\rr^\alpha(t))$, and the momentum density  $\boldsymbol{p}(\xx,t)=\sum_{\alpha=1}^N m\vv^\alpha(t)\delta(\xx-\rr^\alpha(t))$. Still in the situation where momentum of the particles matter, one can include the effect of the collisions between particles, aiming at application to granular materials~\cite{Lopez2004}.
	
Donev and collaborators studied the effect of hydrodynamic interactions, and derived extensions of the DK equation which include explicitly the hydrodynamic tensors that encode from momentum exchange between the particles~\cite{Donev2014, Pelaez2018}. In the presence of hydrodynamic interactions, the difficulty lies in the noise term of Eq. \eqref{Dean_eq}, which becomes multiplicative. Indeed, the scalar diffusion coefficient $D$ needs to be replaced by a diffusion tensor, which generally depends on the positions of all the particles and therefore on the density $\rho$, i.e. on the random variable itself. Note that, at the microscopic level (i.e. even before deriving the DK equation), integrating numerically the equations of motion for $\rr^1,\dots,\rr^N$ [Eq. \eqref{overdamped_Langevin}] with hydrodynamic interactions is challenging, and requires some advanced numerical methods \cite{Ermak1978}.

We finally list a few other recent extensions of the original DK equation:
(i)~The situation where the Brownian particles bear an orientational degree of freedom can be addressed by extending straightforwardly the original framework: this is important in the case where the Brownian particles represent force or charge dipoles~\cite{Cugliandolo2015b,Illien2024a}, or in the context of active matter, see Section~\ref{active_matter};
(ii)~In the situation where the suspension is made of several species of particles (that may differ through their sizes, their charge, their interaction potentials...), one can obtain sets of equations obeyed by the densities associated to each species~\cite{Demery2016,Poncet2017,Jardat2022,Benois2023}; 
(iii)~A generalization of the Dean-Kawasaki equation can be formally derived in the situation where the noise is non-Gaussian, and has non-zero cumulants of arbitrary order~\cite{Fodor2018b};
(iv)~The situation where the Brownian particles may undergo simple unimolecular `chemical reactions', in such a way that they switch randomly between different states, has been addressed recently~\cite{Spinney2024,Bressloff2024c};
(v)~Finally, Bressloff recently derived extensions of the DK equations with stochastic resetting~\cite{Bressloff2024}, or in the presence of a reflecting or partially absorbing boundary~\cite{Bressloff2024b}.

%with reactions~\cite{Spinney2024} 
%\cite{ Lefevre2007} 
%\cite{Wiese2016} 
%\cite{Kim2017}  

%\cite{Benitez2016} 
%(the question of Langevin equation with imaginary noise?) 

\section{Exact and approximate solutions to the Dean-Kawasaki equation}
\label{sec_solutions}

\subsection{Exact results}
\label{sec_exact}

In the limiting case where the Brownian particles do not interact with each other (i.e. $V= 0$), the DK equation reduces to 
\begin{equation}
	\partial_t\rho(\xx,t) =D \nabla^2 \rho(\xx,t)+ \nabla \cdot [\boldsymbol{\xi}(\xx,t) \sqrt{2D\rho(\xx,t)}].
\end{equation}
Importantly, even in the non-interacting case, the equation obeyed by the density $\rho(\xx,t)$ is non-trivial, since it is nonlinear and includes multiplicative noise. In particular, this shows that the statistics of the density is a priori non-Gaussian, even in the absence of interactions. Using a path-integral formalism~\cite{Martin1973,Janssen1976,DeDominicis1978a}, Velenich et al.~\cite{Velenich2008} reformulated the DK equation as a field theory, which contains an interaction term: it originates from the constraint that the density $\rho$ must remain positive (in contrast with a simple free field). Using both a direct calculation and a more elaborated approach with Feynman diagrams, the $n$-point correlation functions of the density field $\rho$ are computed recovering the physically expected result, namely that the density $\rho$ has Poissonian statistics. This is, to my knowledge, the only limiting case where the DK equation can been solved exactly -- albeit in a rather formal and unpractical way.  More recently, this result was extended by studying perturbatively the effect of an external quenched potential on the statistics of the noninteracting gas \cite{Kim2020}. Finally, we recently characterized analytically the non-Gaussian fluctuations in the case of interacting particles, in the limit of high-density and weak interactions between particles \cite{Illien2025}.

\subsection{Perturbative solutions}

As underlined in Section~\ref{sec_Dean}, the main difficulty in studying analytically the Dean equation is due to the two sources of non-linearities in Eq. \eqref{Dean_eq}, namely the noise term, which is of order  ${\rho}^{1/2}$, and the interaction term, which is of order $\rho^2$. Assuming that some ground state of the dynamics $\rho_*(\xx)$ is known, it appears natural to write the density as $\rho(\xx,t)=\rho_* (\xx)+ \sqrt{\rho_0} \phi(\xx,t)$, where $\phi(\xx,t)$ is small compared to $\rho_* (\xx)/ \sqrt{\rho_0}$. The prefactor $ \sqrt{\rho_0}$ is added for dimensional reasons -- this will be made clear in what follows. Several choices can be made for the ground state $\rho_*(\xx)$.

\subsubsection{Linearization around a constant, uniform state}
\label{sec_linconstant}

A natural choice for $\rho_*(\xx)$ is the constant, uniform value $\rho_0=N/\mathcal{V}$, where $\mathcal{V}$ is the volume of the system~\cite{Chavanis2008,Dean2014a, Demery2014}. Writing 
\begin{equation}
	\rho(\xx,t) = \rho_0 + \sqrt{\rho_0}\phi(\xx,t),
\end{equation}
Eq. \eqref{Dean_eq}  becomes, after having divided both sides by $\sqrt{\rho_0}$: 
\begin{align}
	%\label{LDE}
	&\partial_t \phi(\xx,t) = D\nabla^2 \phi(\xx,t) +\mu \rho_0  \nabla \cdot [(\phi*\nabla V) (\xx,t)] \nonumber\\
	&+\mu \sqrt{\rho_0} \nabla \cdot[\phi (\xx,t)(\phi*\nabla V)(\xx,t) ] \nonumber\\
	&+ \sqrt{2D}\nabla \cdot \left[ \boldsymbol{\xi}(\xx,t) \left( 1+\frac{\phi(\xx,t)}{\sqrt{\rho_0}}\right)^{1/2}  \right],
\end{align}
where one introduces the convolution operator $*$:	$(V *\phi) (\xx,t) \equiv \int \dd \yy \; V(\xx-\yy)\phi(\yy,t)$. In the limit where the perturbation from the constant uniform state is small ($\phi \ll \sqrt{\rho_0} $), one writes $( 1+{\phi(\xx,t)}/{\sqrt{\rho_0}})^{1/2}\simeq 1+{\phi(\xx,t)}/{2\sqrt{\rho_0}}$, and two terms may be neglected: the terms proportional to $\phi^2$, if one stays at linear order in $\phi$, and the multiplicative noise term, if one assumes that $\rho_0$ is large, i.e. if one linearizes around a \emph{dense} homogeneous state  -- rigorously, this perturbative scheme is valid in the joint limit where the density $\rho_0$ is very large and the strength of the interactions very small, the product of these two quantities being constant. This yields
\begin{align}
	\label{linDK}
	\partial_t \phi(\xx,t) = &D\nabla^2 \phi(\xx,t) + \rho_0 \mu \nabla^2 [(V *\phi) (\xx,t)] \nonumber\\
	&+ \sqrt{2D}\nabla \cdot \boldsymbol{\xi}(\xx,t).
\end{align}
%In Fourier space, using the following convention for the transformation:
%\begin{equation}
%	\tilde f(\qq) = \int \dd \rr \; f(\rr) \ex{-\ii \qq\cdot \rr}  \qquad \qquad ; \qquad \qquad f(\rr) = \int \frac{\dd \qq}{(2\pi)^d} \tilde f(\qq) \ex{\ii \qq\cdot \rr} ,
%\end{equation}
This linear equation for $\phi$ can be solved for in Fourier space, in which it reads
\begin{equation}
	\partial_t \tilde \phi(\qq,t) = -Dq^2\tilde \phi(\qq,t) - \mu \rho_0 q^2 \tilde V(\qq) \tilde \phi(\qq,t) + \sqrt{2D} \tilde{{\eta}}(\qq,t),
	\label{linDK_Fourier}
\end{equation}
where the (scalar) noise $\tilde{\eta}$ has zero average and variance $\langle \tilde \eta (\qq,t) \tilde \eta(\qq',t') \rangle= (2\pi)^d q^2 \delta(\qq+\qq')\delta(t-t')$.

Interestingly, with this linearized equation, it is easy to derive the pair correlation function, defined in real space in its translationally invariant form as $h(\rr) = [\langle \phi(\rr,t) \phi(\zz,t) \rangle- \delta(\rr)]/\rho_0$~\cite{Hansen1986}. The structure factor $S(\qq)$ can be deduced using its definition: $S(\qq)=1+\rho_0 \tilde{h}(\qq)$, and one gets
\begin{equation}
	S(\qq) = \left(1+\frac{\rho_0 \tilde V(\qq)}{\kB T}\right)^{-1},
	\label{S_RPA}
\end{equation}
which coincides with the result obtained within the random phase approximation~\cite{Andersen1970,Wheeler1971,Hansen1986}. This approximation is one of the classical closures that is used in the static theory of liquids. It consists in assuming that the direction correlation function $\tilde{c}(\qq)$ (related to the pair correlation function through the Ornstein-Zernike relation $\tilde{c}(\qq)=\tilde{h}(\qq)/[1+\rho_0 \tilde{h}(\qq)]$ \cite{Hansen1986}) is simply related to the pair potential through 
\begin{equation}
	\tilde{c}(\qq)=-\frac{\tilde{V}(\qq)}{\kB T},
\end{equation}
which is assumed to hold for any $\qq$. It was proposed in the context of long-range interaction (such as Coulombian) and was successfully applied to study the structure of liquids of softcore particles, such as in the Gaussian core model~\cite{Likos2001,Lang2000,Louis2000}. The linearized DK equation \eqref{linDK} can therefore be seen as a dynamical extension of the static random phase approximation. This linearized version has been used in many of the applications that will be presented in Section \ref{sec_applications}.

It is clear from Eq. \eqref{linDK} that the field $\phi$ will have Gaussian fluctuations (the linearization gets rid of all the non-Gaussianities that can be measured in a more thorough treatment of the original  DK equation \cite{Illien2025}). Eq. \eqref{linDK} can therefore serve as the basis of a simple Gaussian and dynamical theory of Brownian suspensions~\cite{Kruger2017a}, in which stress correlations and viscosity can be computed explicitly~\cite{Kruger2018}, at least within this level of description where the inner degrees of freedom of the solvent are ignored. For instance, the two-point, two-time correlation function of the perturbation $\phi$ can be computed in Fourier space, and simply reads 
\begin{align}
	&\langle \tilde\phi(\qq,t) \tilde\phi(\qq',t') \rangle = \nonumber \\ 
	& \frac{(2\pi)^d \delta(\qq+\qq')}{ 1+ \rho_0 \tilde V(\qq)/\kB T}
	  \exp\left\{-D q^2\left[1+\frac{ \rho_0  \tilde V(\qq)}{\kB T}\right]|t-t'|\right\}.
\end{align}
As a final remark, I emphasize that, to solve Eq. \eqref{linDK}, one transforms it into Fourier space, which relies on the assumption that the interaction potential $V(\rr)$ admits a Fourier transform. This is actually a very strong limitation when one tries to apply this procedure to `realistic' potentials, for instance with short-range repulsion, which typically leads to functional dependencies which are non-integrable. The resulting divergent interactions in reciprocal space could potentially be addressed by perturbative methods that have recently been put forward~\cite{Jin2024}.

\subsubsection{Linearization around a metastable state}

An alternative to the linearization around a constant uniform state, is the linearization around some metastable state of the dynamics $\rho_*$, i.e. a state which is such that:
\begin{equation}
\lim_{t\to\infty} \left. \frac{\delta F}{\delta \rho(\xx,t)} \right|_{\rho = \rho_*}=\mu,
\end{equation}
where $\mu$ is the chemical potential at which the system is maintained. A linear equation satisfied by the perturbation around $\rho_*$ can be obtained in a similar fashion to the calculation presented in Section \ref{sec_linconstant}. Frusawa presented the general idea of this calculation in Ref.~\cite{Frusawa2019a}, and subsequently applied it to study the relaxation of a metastable state of densely packed hard spheres~\cite{Frusawa2021}.

\subsection{Numerical solutions}
\label{sec_numerical}

From the DK equation, one can aim at computing either the average value of the solution, or second-order moments, such as two-point, two-time density correlation functions (or, in other words, dynamical structure factors). A naive numerical way to solve Eq. \eqref{Dean_eq} would consist in spatial discretization and time integration. However, if not chosen carefully, the interplay between these two schemes may result in breaking of the balance between the dissipation and fluctuation terms, and introduce spurious correlations and unphysical result. For this reason, the numerical integration of fluctuating hydrodynamics equations is subtle, and has motivated a lot of work in the computational physics literature.

Some methods are not specific to the Dean-Kawasaki equation, but provide good examples of the specific schemes for temporal integration that may be employed to obtain meaningful results. For instance, a fluctuating version of the standard lattice Boltzmann method, i.e. a discretization of the Boltzmann equation for the collisional dynamics, to which a fluctuating term is added, has been proposed \cite{Adhikari2005}. When it comes to spatially discretized schemes, analogous to finite-element methods, the main difficulty is that it typically induces unphysical and therefore undesired correlations. This is circumvented by converting the stochastic partial differential equations into ordinary differential equations obeyed by the cell number densities, i.e. density fields integrated over the volumes of the integration cells. Mathematically, this consists in converting the volume integrals of divergence term into surface integrals using the divergence theorem. Even though the resulting differential equations may resemble the ones studied in usual finite-element methods, the variables have very different meaning. This scheme, sometimes called finite-volume method, has  been employed by various authors to deal with the issues of spatial discretizations~\cite{Donev2010,Delong2013,Delong2014,Russo2021,Mendes2021}. The temporal integrators typically mix implicit methods (for diffusion) and explicit methods (for advection), and involve stochastic equivalent of Runge-Kutta integrators. The finite-volume method has also been refined to study the effect of reactions \cite{Kim2017}. Finally, as an alternative, advanced finite-element methods have also been designed and used in this context: for instance, the strategy in Ref.~\cite{Martinez-Lera2024} is to propose a transformation that subtracts the correlation artefact originating from the discretization procedure, in order to correct the final numerical result.

%Refs.~\cite{Delong2013,Delong2014,Donev2010}, which are not specific to the DK equation, provide good examples of the specific schemes for temporal integration that may be employed to obtain meaningful results. Alternatively, the finite volume method (which consists in converting the volume integrals of divergence term into surface integrals using the divergence theorem) has been employed by various authors to deal with the issues of spatial discretizations~\cite{Donev2010,Russo2021,Mendes2021}. 

Another challenge in the numerical resolution of fluctuating hydrodynamics equations lies in the fact that the density, which is subject to noise, must remain positive. When simulating a homogeneous and dense system, this issue might be safely ignored, but it becomes predominant when the system might display liquid-gas coexistence, for instance. This issue was recently addressed by proposing refined discretization schemes to numerically integrate the DK equation while preserving the positivity of the solution~\cite{Magaletti2022}. Technically, this requires the combination between several numerical schemes, that fulfill the different physical properties such as positivity and fluctuation-dissipation balance.

%\cite{Delong2014} description of numerical schemes suitable to integrate equations of "fluctuating hydrodynamics" (that include DK?)

%\cite{Azarnykh2016} numerical resolution, discussion on Eulerian vs. Lagrangian strategies

%\cite{Embacher2018, Li2019} numerical strategy to integrate the stochastic PDE

Finally, and more recently, the mathematical community has devoted some effort to analyzing rigorously possible discretization schemes, both for the original DK equation~\cite{Cornalba2023a}, and its regularized version~\cite{Cornalba2023b}.

%///////////////////////

%\cite{Kim2017} `analysis of finite element discretisations in the context of reaction-diffusion (agent-based) systems models'

%\cite{Cornalba2023a}: `We show that structure-preserving discretisations of the Dean–Kawasaki equation may approximate the density fluctuations of N non-interacting diffusing particles to arbitrary order in $N^{-1}$ (in suitable weak metrics). In other words, the Dean–Kawasaki equation may be interpreted as a “recipe” for accurate and efficient numerical simulations of the density fluctuations of independent diffusing particles.'

%\cite{Djurdjevac2024}

%\cite{Russo2021} +~\cite{Donev2010} : `finite-volume schemes for stochastic gradient flow equations' (creuser finite volume schemes ?)

%\cite{Cornalba2023b}`convergence analysis of discountinuous Galerkin scheme – and modelling – for the regularised inertial Dean–Kawasaki model'

\section{Applications}
\label{sec_applications}

In this final Section, I briefly review the different applications of the DK equation -- the bibliographical review does not aim at being fully exhaustive, but rather at giving a faithful overview of the wealth of situations where SDFT is relevant.

\subsection{Supercooled liquids}

In close link with the context in which Kawasaki proposed Eq. \eqref{Kawasaki_eq}, the stochastic equation \eqref{Dean_eq} was applied to study correlations and diffusion in supercooled liquids~\cite{Gupta2011,Bidhoodi2016}, and to address the question of the ergodicity breaking predicted by standard MCT in the low-temperature regime (we refer the reader interested in this topic to Section \ref{sec_MCT}, which contains more details and references).

%\subsection{The ergodic/non-ergodic transition}

%\cite{Iwata2006}

%\cite{Das2013b,Das2020}

%
%\subsection{Fluctuation relations and theorems}
%
%\cite{Harada2009,Yoshimori2012}

\subsection{Active matter}
\label{active_matter}

Active matter refers to non-equilibrium systems whose constitutive agents continuously convert the energy available from sources in their environment into mechanical work, in order to swim, self-propel or form complex structures. This line of research has a significant experimental part (synthesis of articial microswimmers, observation of emerging collective phenomena among biological agents...), and has also raised an important list of questions to be addressed by theoretical physicists. Among them, deriving laws for active matter systems starting from microscopic principles has attracted a lot of attention \cite{Ramaswamy2010,Marchetti2013}.

In this context, the DK equation appeared as a promising tool to describe interacting particles. Technically, the difficulty lies on the fact that active particles usually bear an additional degree of freedom (typically a head-tail orientation, that gives the direction of propulsion) to which the translational degrees of freedom, denoted earlier by $\rr_1,\dots,\rr_N$, are coupled. The DK equation was first derived for run-an-tumble particles in 1D~\cite{Tailleur2008}, in which the orientational degree of freedom takes discrete values. Its average version (which is actually closer to a DDFT description) was used to study the stability of such a system, and the possibility of motility-induced phase separation.  More recently, proper stochastic equations for 1D active spins were solved numerically, highlighting the importance of the multiplicative noise term on the flocking transition~\cite{OLaighleis2018}.  The DK equation for other sorts of active particles (namely active nematics, active Brownian particles (ABPs) and active Ornstein-Uhlenbeck particles (AOUPs)) were derived rigorously~\cite{Bertin2013,Solon2015, Martin2021,Dinelli2024}, and used to compute the pair correlations between ABPs in the dilute regime~\cite{Poncet2021b}. In the meantime, different authors tried to make active matter enter the usual classification of stochastic models in the context of critical phenomena (model A, B, H...)~\cite{Hohenberg1977,Chaikin}, and proposed an active version of model B, whose relationship to a DK-like approach was recently discussed~\cite{Tjhung2018}. The DK description of active fluids was also employed to measure  entropy production~\cite{GrandPre2021}, to etablish links between dissipation, phase transitions particle correlations~\cite{Tociu2019,Fodor2020,Rassolov2022}, and to characterize their structure~\cite{Tociu2022}. Finally, a DK equation for active chiral particles (i.e. whose activity results in a forced rotation instead of a forced translation) was derived recently~\cite{Kuroda2023}.

\subsection{Other nonequilibrium systems}

The DK equation has also been applied to other types of non-equilibrium Brownian suspensions, which do not exactly fall in the `active matter' landscape depicted in the previous section. For instance,  systems made of two species of particles which are driven by external fields in opposite direction and that display a laning transition, were studied in Ref.~\cite{Poncet2017}. This is an example where the original DK equation can be extended to mixtures of multiple species of particles, and allows one to compute the density correlations between the different species (within the linear approximation from Section \ref{sec_linconstant}). Similarly, the DK equation has been used to study mixtures of particles connected to different thermostats ~\cite{Jardat2022, Damman2024}, in the range of parameters where such mixtures do not phase separate (such phase separation may be studied by alternative methods~\cite{Grosberg2015}). Finally, I also mention the emerging topic of mixtures of particles with non-reciprocal interactions, which have also been studied within the DK framework~\cite{Ghimenti2024, Dinelli2023, Benois2023}.

\subsection{Chemotactic particles}

In many situations of biological interest, the agents that constitute the system interact via chemical signals, which are often emitted by the agents themselves. In the language adopted in this review, this means that the Brownian particles are submitted to drift forces, which are proportional to gradients of an auxiliary chemical field. This field is either created or consumed by the particles, depending on whether they play the role of source or sink, respectively. This problem was first studied in the framework of SDFT by Chavanis, who set up the problem and computed the fluctuations within the linearized approximation~\cite{Chavanis2008}, retrieved usual mean-field model and introduced the effect of inertia and delayed interactions~\cite{Chavanis2010} and studied some metastable states of the system~\cite{Chavanis2014}.

Beyond the linearized approximation, the analysis of the DK equation becomes much more complicated, as stated previously. However, in the context of chemotactic particles, the nonlinearities of the DK equation were treated perturbatively in the framework of the dynamical renormalization group~\cite{Forster1977,Medina1989}. This was achieved by Golestanian et al., who added logistic growth as a feature of the model, and studied thoroughly phase transitions in this system, and the associated critical exponents~\cite{Gelimson2015, Mahdisoltani2021, BenAliZinati2021, Mahdisoltani2023}. With the same method, the role of demographic noise was studied recently~\cite{VanDerKolk2023}. Similar nonlinear equations were also studied using the method of stochastic quantization~\cite{Samanta2019,Samanta2020}.

%\subsection{Fluctuation-induced effects, Casimir effect}

\subsection{Charged particles and electrolytes}

Among the different systems that may be studied using SDFT, electrolytes, and more generally charged systems, have attracted a lot of attention. One of the reasons for this is that the long-range Coulombic interactions through which ions or charged particles interact are sufficiently smooth and well-behaved to allow explicit calculations in Fourier domain. Another reason is that such systems have been central in the classical theories of liquids~\cite{Hansen1986, Kjellander2019a}, in such a way that many analytical results are known on their structure and dynamics, which allow to `benchmark' the results from the DK/SDFT approach.

In this context, Dean and Démery computed from the linearized DK equation the conductivity of dilute and strong electrolytes (retrieving the classical results by Debye-H\"uckel-Onsager) as well as the density correlations between different ionic species~\cite{Demery2016}. This framework was more recently extended to compute the temporal response of electrolytes when submitted to an external field~\cite{Bonneau2023,Bonneau2025}, and the conductivity beyond the small-field limit~\cite{Berthoumieux2024}. Within the same level of description, Frusawa studied fluctuations of electrolytes near a charged plate~\cite{Frusawa2019}, and Hoang Ngoc Minh et al. studied hyperuniformity that emerges when observing ionic fluctuations in finite volumes ~\cite{HoangNgoc2023}. Okamoto recently attempted to go beyond the linearized DK equation by performing a systematic diagrammatic expansion, in a rather formal way~\cite{Okamoto2022}. Fluctuations in the charge density for a bulk electrolyte were also studied by numerical resolution of the equations ~\cite{Poitevin2016}.

Going beyond the limit of dilute electrolytes, SDFT was combined with truncated Coulomb potentials to account for short-range repulsion between the ions, and was used in order to compute conductivity~\cite{Avni2022a, Avni2022} and viscosity~\cite{Robin2024} beyond the dilute limit --  the validity of such approximations was recently discussed~\cite{Bernard2023}. Alternatively, in order to describe dense electrolytes, Frusawa proposed a hybrid approach which combines SDFT with the usual equilibrium DFT approach, which typically includes density functionals that account faithfully for short-range repulsion~\cite{Frusawa2022}.

Finally, SDFT was also used to study Casimir forces that may emerge in electrolytes when geometric constraints are imposed on their fluctuations.  Although Casimir interactions are present in a wide range of quantum and classical systems~\cite{Kardar1999}, in the present case, these fluctuation-induced interactions are related to the long-range nature of the Coulombic forces. The DK equation, which is an intrinsically fluctuating description of the dynamics, appears as the right tool to study them. These effects were studied in different setups: net neutral plates containing Brownian charges ~\cite{Dean2014a,Lu2015,Dean2016}, and more recently, electrolytes submitted to a constant electric field in a slab bounded by media of different dielectric permittivities~\cite{Mahdisoltani2021a,Mahdisoltani2021b,Du2024,Du2025}. In all these situations, the linearized DK equation provides explicit, analytical estimates of the fluctuation-induced forces.

\subsection{Tracer particles}

The DK equation was also used to study the diffusion of a tagged particle (or tracer particle) within the suspension. More precisely, among the $N$ particles that constitute the system, one of them (the particle labeled $1$, with no loss of generality) is assumed to be tagged, whereas the remaining $N-1$ constitute a bath to which the tracer is coupled. Technically, the dynamics of the bath can still be described by a DK equation, by defining a bath density $\rho_\text{b}(\xx,t)=\sum_{\alpha=2}^N\delta(\xx-\rr^\alpha(t))$, that excludes the tracer particle. The position of the tracer particle $\rr^1(t)$ obeys a simple overdamped Langevin equation. One ends up with  a set of two coupled equations: one for $\rho_\text{b}$, and one for $\rr^1$~\cite{Demery2014}. While the equation for $\rho_\text{b}$ can be linearized following the method given in Section \ref{sec_linconstant}, the coupling between $\rho_\text{b}$ and $\rr^1$ remains generally nonlinear. The problem can then be studied perturbatively, assuming a weak coupling between the tracer and the bath~\cite{Demery2011}. 

This method was applied to study the statistics of the position of the tracer, as well as its correlations with the bath of particles, in different settings: in the case where the tracer is driven by some external force in a bath of passive particles~\cite{Demery2014, Demery2015, Demery2019} (a situation related to active microrheology experiments), in the case where the tracer is self-propelled~\cite{Martin2018} or the bath is made of self-propelled particles~\cite{Feng2023}, when the tracer is a tagged ion in an electrolyte \cite{Bernard2023}, in the situation where the tracer is coupled to a mixture of particles connected to different thermostats~\cite{Jardat2022} or with non-reciprocal interactions~\cite{Benois2023}, in the presence of confinement~\cite{Wang2023}, and finally when the tracer is odd-diffusive \cite{Muzzeddu2025a}. The long-distance decay tracer-bath correlations were also studied within the framework of the Dean-Kawasaki equation \cite{Venturelli2025, Venturelli2025a}.

\subsection{One-dimensional diffusive systems}

One-dimensional systems of diffusing and interacting particles plays a special role in statistical mechanics. First, from a technical point of view, this particular dimensionality allows the derivation of a wealth of exact results, by relying on mappings between different classes of models, and on specific mathematical methods (matrix ansatz, Bethe ansatz, integrable systems, random matrix theory etc.). Second, in the particular situation where the pair interactions between the particles are sufficiently hard to prevent them from bypassing each other, macroscopic observables (such as the current of particles) and properties associated with tracer particles exhibit anomalous scalings, typically subdiffusive, which are the signature of the very strong geometric constraints imposed on the system. This situation is generally referred to as `single-file diffusion'.

In a series of papers, Ooshida et al. employed a DK approach to characterize two-point correlations and cooperativity effects in single-file diffusion with hardcore repulsion~\cite{Ooshida2011,Ooshida2013,Ooshida2015,Ooshida2018} (these studies  are actually closer to a one-dimensional application of MCT~\cite{Abel2009}). These results also gave insight into higher-dimensional systems~\cite{Ooshida2016,Ooshida2016a}.

More recently, different authors relied on the DK formalism to study one-dimensional gases with longer-ranged interactions. I mention the `active Dyson gas', referring to run-and-tumble particles with logarithmic interactions and an external confinement, for which the stationary density was computed \cite{Touzo2023}; ranked diffusion (particles on a line that undergo a drift proportional to their rank), for which the equation for the density was mapped onto a Burgers equation, that allowed the computation of the steady density and the joint distribution of positions~\cite{LeDoussal2022,Flack2023}; the Riesz gas, where particles interact with a potential $V(\rr) \propto |\rr|^{-s}$ ($s>0$), or the Dyson gas, where particles interact with a logarithmic potential, for which the fluctuations of the integrated current and of the position of a tagged particle were computed explicitly in quenched and annealed settings~\cite{Dandekar2023, Dandekar2024}.

%\subsection{Oscillators and synchronization}
%
%\cite{Grossmann2016}
%
%\cite{Buendia2024}

\subsection{Machine learning}

Interestingly, the DK equation has recently been  used in the context of machine learning. Training a neural network to perform some tasks, such as speech or image recognition, consists in finding optimal values for the numerous parameters at stake, to minimize error and maximize accuracy of the predictions. The parameters in the neural network may be seen as particles, and the cost function, with respect to which the problem should be optimized, may be seen as the interaction between particles. This formal mapping allows one to study the training of the network as the evolution of the particles within this potential, and to rely on the results from nonequilibrium statistical mechanics and interacting particle systems~\cite{Rotskoff2022}.

\subsection{First-passage problems}

Finally, viewing the density of particles $\rho$ as a stochastic process, one can compute its first-passage properties. For instance, this point of view was adopted to study nucleation phenomena in colloidal suspensions~\cite{Lutsko2012}. More recently, Liu et al. considered the Kramers problem associated to a system obeying a DK-like equation, and computed the mean first-passage time to a potential barrier~\cite{Liu2024}.

\section{Conclusion and perspectives}

This review has examined the Dean-Kawasaki equation, also referred to as Stochastic Density Functional Theory (SDFT), highlighting its origins, fundamental principles, and broad applicability. Initially developed to describe the dynamics of interacting Brownian particles, SDFT has since expanded to become a versatile tool in statistical mechanics. I have explored how this framework connects with other theories, such as fluctuating hydrodynamics and mode-coupling theory, underscoring its potential to provide a unified approach to various complex systems. Additionally, the article has addressed several extensions of the original Dean-Kawasaki equation, including considerations of hydrodynamic interactions, inertial effects, and the dynamics of active matter. These advances demonstrate the equation’s robustness in modeling diverse nonequilibrium phenomena, from supercooled liquids to active and driven systems. By capturing both the deterministic and stochastic aspects of particle behavior, SDFT offers a comprehensive framework for understanding and predicting the intricate dynamics of complex fluids. As research continues, the application of SDFT is likely to expand further, providing deeper insights into the behavior of increasingly complex systems in physics, chemistry, and beyond.

However, significant challenges remain in the domain. One of the primary difficulties is the treatment of nonlinearities and multiplicative noise within the Dean-Kawasaki equation, which complicates both analytical and numerical solutions. Moreover, ensuring the well-posedness and stability of solutions, particularly in systems with strong interactions or inhomogeneities, is an ongoing concern. The need for more efficient computational methods that can handle the intricate dynamics and high dimensionality of real-world systems is also pressing. Additionally, extending SDFT to account for more complex interactions presents a formidable challenge. Addressing these issues will be crucial for advancing the theory and broadening its applicability to new areas of research.

%At this stage, the goal would be to refine the analytical treatment of the Dean-Kawasaki equation Eq. \eqref{DE}. It would be interesting to go beyond the linearization procedure, which is quite limiting, in the sense that it is only valid for interaction potentials which are sufficiently soft. MCT is of course an alternative, but the complexity of the resulting equations generally prevents any analytical insight. 

%So, starting from Eq. \eqref{DE}, what could be the approximations (beyond the obvious linearization) that would result in a \emph{stochastic} evolution equation for the density $\hat\rho$, in which the pair interactions would be encoded in a non-trivial but explicit way (for instance relying on the knowledge of the direct pair correlation functions from alternative approaches, such as MSA?)? Work in progress!

\section*{Acknowledgments}

I warmly thank Antoine Carof, Benjamin Rotenberg, Sophie Herrmann, and Davide Venturelli for their in-depth reading of the manuscript. The writing of the present review was motivated by numerous discussions and collaborations over the past few years, who enlightened my understanding of the technical and physical aspects of the Dean-Kawasaki equation. I wish to acknowledge 
Ramin Golestanian, 
Vincent D\'emery,
Olivier B\'enichou,
Sophie Marbach,
Marie Jardat,
Olivier Bernard,
Vincent Dahirel,
Aur\'elien Grabsch,
Th\^e Hoang Ngoc Minh,
Alexis Poncet, 
Pierre Rizkallah,
Guillaume Jeanmairet,
and
Daniel Borgis.

Finally, after its pre-publication, this review has benefited from the comments of  Joseph Brader, Max von Renesse, Johannes Zimmer, Bongsoo Kim, \'Elie Rapha\"el, Pierre-Henri Chavanis and Cesare Nardini.

%\bibliographystyle{/Users/pierreillien/work/docs/mystyle3.bst}
%\bibliography{review_DK_references.bib}

\begin{thebibliography}{209}%
	\makeatletter
	\providecommand \@ifxundefined [1]{%
		\@ifx{#1\undefined}
	}%
	\providecommand \@ifnum [1]{%
		\ifnum #1\expandafter \@firstoftwo
		\else \expandafter \@secondoftwo
		\fi
	}%
	\providecommand \@ifx [1]{%
		\ifx #1\expandafter \@firstoftwo
		\else \expandafter \@secondoftwo
		\fi
	}%
	\providecommand \natexlab [1]{#1}%
	\providecommand \enquote  [1]{``#1''}%
	\providecommand \bibnamefont  [1]{#1}%
	\providecommand \bibfnamefont [1]{#1}%
	\providecommand \citenamefont [1]{#1}%
	\providecommand \href@noop [0]{\@secondoftwo}%
	\providecommand \href [0]{\begingroup \@sanitize@url \@href}%
	\providecommand \@href[1]{\@@startlink{#1}\@@href}%
	\providecommand \@@href[1]{\endgroup#1\@@endlink}%
	\providecommand \@sanitize@url [0]{\catcode `\\12\catcode `\$12\catcode
		`\&12\catcode `\#12\catcode `\^12\catcode `\_12\catcode `\%12\relax}%
	\providecommand \@@startlink[1]{}%
	\providecommand \@@endlink[0]{}%
	\providecommand \url  [0]{\begingroup\@sanitize@url \@url }%
	\providecommand \@url [1]{\endgroup\@href {#1}{\urlprefix }}%
	\providecommand \urlprefix  [0]{URL }%
	\providecommand \Eprint [0]{\href }%
	\providecommand \doibase [0]{https://doi.org/}%
	\providecommand \selectlanguage [0]{\@gobble}%
	\providecommand \bibinfo  [0]{\@secondoftwo}%
	\providecommand \bibfield  [0]{\@secondoftwo}%
	\providecommand \translation [1]{[#1]}%
	\providecommand \BibitemOpen [0]{}%
	\providecommand \bibitemStop [0]{}%
	\providecommand \bibitemNoStop [0]{.\EOS\space}%
	\providecommand \EOS [0]{\spacefactor3000\relax}%
	\providecommand \BibitemShut  [1]{\csname bibitem#1\endcsname}%
	\let\auto@bib@innerbib\@empty
	%</preamble>
	\bibitem [{\citenamefont {Brown}(1828)}]{Brown1828a}%
	\BibitemOpen
	\bibfield  {author} {\bibinfo {author} {\bibfnamefont {R.}~\bibnamefont
			{Brown}},\ }\bibfield  {title} {\bibinfo {title} {{{XXVII}}. {{A}} brief
			account of microscopical observations made in the months of {{June}},
			{{July}} and {{August}} 1827, on the particles contained in the pollen of
			plants; and on the general existence of active molecules in organic and
			inorganic bodies},\ }\href {https://doi.org/10.1080/14786442808674769}
	{\bibfield  {journal} {\bibinfo  {journal} {The Philosophical Magazine}\
		}\textbf {\bibinfo {volume} {4}},\ \bibinfo {pages} {161} (\bibinfo {year}
		{1828})}\BibitemShut {NoStop}%
	\bibitem [{\citenamefont {Einstein}(1905)}]{Einstein1905a}%
	\BibitemOpen
	\bibfield  {author} {\bibinfo {author} {\bibfnamefont {A.}~\bibnamefont
			{Einstein}},\ }\bibfield  {title} {\bibinfo {title} {{\"U}ber die von der
			molekularkinetischen {{Theorie}} der {{W{\"a}rme}} geforderte {{Bewegung}}
			von in ruhenden {{Fl{\"u}ssigkeiten}} suspendierten {{Teilchen}}},\ }\href
	{https://doi.org/10.1002/andp.19053220806} {\bibfield  {journal} {\bibinfo
			{journal} {Annalen der Physik}\ }\textbf {\bibinfo {volume} {17}},\ \bibinfo
		{pages} {549} (\bibinfo {year} {1905})}\BibitemShut {NoStop}%
	\bibitem [{\citenamefont {{von Smoluchowski}}(1906)}]{vonSmoluchowski1906}%
	\BibitemOpen
	\bibfield  {author} {\bibinfo {author} {\bibfnamefont {M.}~\bibnamefont {{von
					Smoluchowski}}},\ }\bibfield  {title} {\bibinfo {title} {Zur kinetischen
			{{Theorie}} der {{Brownschen Molekularbewegung}} und der {{Suspensionen}}},\
	}\href {https://doi.org/10.1002/andp.19063261405} {\bibfield  {journal}
		{\bibinfo  {journal} {Ann. Phys.}\ }\textbf {\bibinfo {volume} {21}},\
		\bibinfo {pages} {759} (\bibinfo {year} {1906})}\BibitemShut {NoStop}%
	\bibitem [{\citenamefont {Langevin}(1908)}]{Langevin1908}%
	\BibitemOpen
	\bibfield  {author} {\bibinfo {author} {\bibfnamefont {P.}~\bibnamefont
			{Langevin}},\ }\bibfield  {title} {\bibinfo {title} {Sur la th{\'e}orie du
			mouvement brownien},\ }\href {https://doi.org/10.1119/1.18725} {\bibfield
		{journal} {\bibinfo  {journal} {Comptes rendus de l'Acad{\'e}mie des Sciences
				(Paris)}\ }\textbf {\bibinfo {volume} {146}},\ \bibinfo {pages} {530}
		(\bibinfo {year} {1908})}\BibitemShut {NoStop}%
	\bibitem [{\citenamefont {Gardiner}(1985)}]{Gardiner1985}%
	\BibitemOpen
	\bibfield  {author} {\bibinfo {author} {\bibfnamefont {C.~W.}\ \bibnamefont
			{Gardiner}},\ }\href@noop {} {\emph {\bibinfo {title} {Handbook of
				{{Stochastic Methods}}}}}\ (\bibinfo  {publisher} {Springer},\ \bibinfo
	{year} {1985})\BibitemShut {NoStop}%
	\bibitem [{\citenamefont {{van Kampen}}(1981)}]{vanKampen1981}%
	\BibitemOpen
	\bibfield  {author} {\bibinfo {author} {\bibfnamefont {N.~G.}\ \bibnamefont
			{{van Kampen}}},\ }\href@noop {} {\emph {\bibinfo {title} {Stochastic
				{{Processes}} in {{Physics}} and {{Chemistry}}}}}\ (\bibinfo  {publisher}
	{North-Holland},\ \bibinfo {address} {Amsterdam},\ \bibinfo {year}
	{1981})\BibitemShut {NoStop}%
	\bibitem [{\citenamefont {Allen}\ and\ \citenamefont
		{Tildesley}(1987)}]{Allen1987}%
	\BibitemOpen
	\bibfield  {author} {\bibinfo {author} {\bibfnamefont {M.~P.}\ \bibnamefont
			{Allen}}\ and\ \bibinfo {author} {\bibfnamefont {D.~J.}\ \bibnamefont
			{Tildesley}},\ }\href@noop {} {\emph {\bibinfo {title} {Computer
				{{Simulation}} of {{Liquids}}}}}\ (\bibinfo  {publisher} {Oxford University
		Press},\ \bibinfo {year} {1987})\BibitemShut {NoStop}%
	\bibitem [{\citenamefont {Kloeden}\ and\ \citenamefont
		{Platen}(1992)}]{Kloeden1992}%
	\BibitemOpen
	\bibfield  {author} {\bibinfo {author} {\bibfnamefont {P.~E.}\ \bibnamefont
			{Kloeden}}\ and\ \bibinfo {author} {\bibfnamefont {E.}~\bibnamefont
			{Platen}},\ }\href {https://doi.org/10.1007/978-3-662-12616-5} {\emph
		{\bibinfo {title} {Numerical {{Solution}} of {{Stochastic Differential
						Equations}}}}}\ (\bibinfo  {publisher} {Springer Berlin Heidelberg},\
	\bibinfo {address} {Berlin, Heidelberg},\ \bibinfo {year} {1992})\BibitemShut
	{NoStop}%
	\bibitem [{\citenamefont {Leimkuhler}\ and\ \citenamefont
		{Matthews}(2013)}]{Leimkuhler2013}%
	\BibitemOpen
	\bibfield  {author} {\bibinfo {author} {\bibfnamefont {B.}~\bibnamefont
			{Leimkuhler}}\ and\ \bibinfo {author} {\bibfnamefont {C.}~\bibnamefont
			{Matthews}},\ }\bibfield  {title} {\bibinfo {title} {Rational construction of
			stochastic numerical methods for molecular sampling},\ }\href
	{https://doi.org/10.1093/amrx/abs010} {\bibfield  {journal} {\bibinfo
			{journal} {Applied Mathematics Research eXpress}\ }\textbf {\bibinfo {volume}
			{2013}},\ \bibinfo {pages} {34} (\bibinfo {year} {2013})}\BibitemShut
	{NoStop}%
	\bibitem [{\citenamefont {Rackauckas}\ and\ \citenamefont
		{Nie}(2017)}]{Rackauckas2017}%
	\BibitemOpen
	\bibfield  {author} {\bibinfo {author} {\bibfnamefont {C.}~\bibnamefont
			{Rackauckas}}\ and\ \bibinfo {author} {\bibfnamefont {Q.}~\bibnamefont
			{Nie}},\ }\bibfield  {title} {\bibinfo {title} {Adaptive methods for
			stochastic differential equations via natural embeddings and rejection
			sampling with memory},\ }\href {https://doi.org/10.3934/dcdsb.2017133}
	{\bibfield  {journal} {\bibinfo  {journal} {Discrete \& Continuous Dynamical
				Systems - B}\ }\textbf {\bibinfo {volume} {22}},\ \bibinfo {pages} {2731}
		(\bibinfo {year} {2017})}\BibitemShut {NoStop}%
	\bibitem [{\citenamefont {Samm{\"u}ller}\ and\ \citenamefont
		{Schmidt}(2021)}]{Sammuller2021}%
	\BibitemOpen
	\bibfield  {author} {\bibinfo {author} {\bibfnamefont {F.}~\bibnamefont
			{Samm{\"u}ller}}\ and\ \bibinfo {author} {\bibfnamefont {M.}~\bibnamefont
			{Schmidt}},\ }\bibfield  {title} {\bibinfo {title} {Adaptive {{Brownian
					Dynamics}}},\ }\href {https://doi.org/10.1063/5.0062396} {\bibfield
		{journal} {\bibinfo  {journal} {The Journal of Chemical Physics}\ }\textbf
		{\bibinfo {volume} {155}},\ \bibinfo {pages} {134107} (\bibinfo {year}
		{2021})}\BibitemShut {NoStop}%
	\bibitem [{\citenamefont {Archer}\ and\ \citenamefont
		{Rauscher}(2004)}]{Archer2004a}%
	\BibitemOpen
	\bibfield  {author} {\bibinfo {author} {\bibfnamefont {A.~J.}\ \bibnamefont
			{Archer}}\ and\ \bibinfo {author} {\bibfnamefont {M.}~\bibnamefont
			{Rauscher}},\ }\bibfield  {title} {\bibinfo {title} {Dynamical density
			functional theory for interacting {{Brownian}} particles: {{Stochastic}} or
			deterministic?},\ }\href {https://doi.org/10.1088/0305-4470/37/40/001}
	{\bibfield  {journal} {\bibinfo  {journal} {Journal of Physics A:
				Mathematical and General}\ }\textbf {\bibinfo {volume} {37}},\ \bibinfo
		{pages} {9325} (\bibinfo {year} {2004})}\BibitemShut {NoStop}%
	\bibitem [{\citenamefont {Fabian}\ \emph {et~al.}(2019)\citenamefont {Fabian},
		\citenamefont {Shpiro}, \citenamefont {Rabani}, \citenamefont {Neuhauser},\
		and\ \citenamefont {Baer}}]{Fabian2019}%
	\BibitemOpen
	\bibfield  {author} {\bibinfo {author} {\bibfnamefont {M.~D.}\ \bibnamefont
			{Fabian}}, \bibinfo {author} {\bibfnamefont {B.}~\bibnamefont {Shpiro}},
		\bibinfo {author} {\bibfnamefont {E.}~\bibnamefont {Rabani}}, \bibinfo
		{author} {\bibfnamefont {D.}~\bibnamefont {Neuhauser}},\ and\ \bibinfo
		{author} {\bibfnamefont {R.}~\bibnamefont {Baer}},\ }\bibfield  {title}
	{\bibinfo {title} {Stochastic density functional theory},\ }\href
	{https://doi.org/10.1002/wcms.1412} {\bibfield  {journal} {\bibinfo
			{journal} {WIREs Computational Molecular Science}\ }\textbf {\bibinfo
			{volume} {9}},\ \bibinfo {pages} {e1412} (\bibinfo {year}
		{2019})}\BibitemShut {NoStop}%
	\bibitem [{\citenamefont {G{\"o}tze}(2009)}]{Gotze2009}%
	\BibitemOpen
	\bibfield  {author} {\bibinfo {author} {\bibfnamefont {W.}~\bibnamefont
			{G{\"o}tze}},\ }\href@noop {} {\emph {\bibinfo {title} {Complex {{Dynamics}}
				of {{Glass-Forming Liquids}}}}}\ (\bibinfo  {publisher} {Oxford University
		Press},\ \bibinfo {year} {2009})\BibitemShut {NoStop}%
	\bibitem [{\citenamefont {Gotze}\ and\ \citenamefont
		{Sjogren}(1992)}]{Gotze1992}%
	\BibitemOpen
	\bibfield  {author} {\bibinfo {author} {\bibfnamefont {W.}~\bibnamefont
			{Gotze}}\ and\ \bibinfo {author} {\bibfnamefont {L.}~\bibnamefont
			{Sjogren}},\ }\bibfield  {title} {\bibinfo {title} {Relaxation processes in
			supercooled liquids},\ }\href {https://doi.org/10.1088/0034-4885/55/3/001}
	{\bibfield  {journal} {\bibinfo  {journal} {Reports on Progress in Physics}\
		}\textbf {\bibinfo {volume} {55}},\ \bibinfo {pages} {241} (\bibinfo {year}
		{1992})}\BibitemShut {NoStop}%
	\bibitem [{\citenamefont {Kawasaki}(1994)}]{Kawasaki1994}%
	\BibitemOpen
	\bibfield  {author} {\bibinfo {author} {\bibfnamefont {K.}~\bibnamefont
			{Kawasaki}},\ }\bibfield  {title} {\bibinfo {title} {Stochastic model of slow
			dynamics in supercooled liquids and dense colloidal suspensions},\ }\href
	{https://doi.org/10.1016/0378-4371(94)90533-9} {\bibfield  {journal}
		{\bibinfo  {journal} {Physica A: Statistical Mechanics and its Applications}\
		}\textbf {\bibinfo {volume} {208}},\ \bibinfo {pages} {35} (\bibinfo {year}
		{1994})}\BibitemShut {NoStop}%
	\bibitem [{\citenamefont {Mori}(1965)}]{Mori1965b}%
	\BibitemOpen
	\bibfield  {author} {\bibinfo {author} {\bibfnamefont {H.}~\bibnamefont
			{Mori}},\ }\bibfield  {title} {\bibinfo {title} {Transport, {{Collective
					Motion}}, and {{Brownian Motion}}},\ }\href
	{https://doi.org/10.1143/PTP.33.423} {\bibfield  {journal} {\bibinfo
			{journal} {Progress of Theoretical Physics}\ }\textbf {\bibinfo {volume}
			{33}},\ \bibinfo {pages} {423} (\bibinfo {year} {1965})}\BibitemShut
	{NoStop}%
	\bibitem [{\citenamefont {Zwanzig}(1961)}]{Zwanzig1961}%
	\BibitemOpen
	\bibfield  {author} {\bibinfo {author} {\bibfnamefont {R.}~\bibnamefont
			{Zwanzig}},\ }\bibfield  {title} {\bibinfo {title} {Memory {{Effects}} in
			{{Irreversible Thermodynamics}}},\ }\href
	{https://doi.org/10.1103/PhysRev.124.983} {\bibfield  {journal} {\bibinfo
			{journal} {Physical Review}\ }\textbf {\bibinfo {volume} {124}},\ \bibinfo
		{pages} {983} (\bibinfo {year} {1961})}\BibitemShut {NoStop}%
	\bibitem [{\citenamefont {Munakata}(1977)}]{Munakata1977}%
	\BibitemOpen
	\bibfield  {author} {\bibinfo {author} {\bibfnamefont {T.}~\bibnamefont
			{Munakata}},\ }\bibfield  {title} {\bibinfo {title} {Liquid {{Instability}}
			and {{Freezing}} - {{Reductive Perturbation Approach}}},\ }\href
	{https://doi.org/10.1143/JPSJ.43.1723} {\bibfield  {journal} {\bibinfo
			{journal} {J. Phys. Soc. Jpn.}\ }\textbf {\bibinfo {volume} {43}},\ \bibinfo
		{pages} {1723} (\bibinfo {year} {1977})}\BibitemShut {NoStop}%
	\bibitem [{\citenamefont {Munakata}(1989)}]{Munakata1989}%
	\BibitemOpen
	\bibfield  {author} {\bibinfo {author} {\bibfnamefont {T.}~\bibnamefont
			{Munakata}},\ }\bibfield  {title} {\bibinfo {title} {A {{Dynamical
					Extension}} of {{Density Functional Theory}}},\ }\href
	{https://doi.org/10.1143/JPSJ.58.2434} {\bibfield  {journal} {\bibinfo
			{journal} {J. Phys. Soc. Jpn.}\ }\textbf {\bibinfo {volume} {58}},\ \bibinfo
		{pages} {2434} (\bibinfo {year} {1989})}\BibitemShut {NoStop}%
	\bibitem [{\citenamefont {Bagchi}(1987)}]{Bagchi1987}%
	\BibitemOpen
	\bibfield  {author} {\bibinfo {author} {\bibfnamefont {B.}~\bibnamefont
			{Bagchi}},\ }\bibfield  {title} {\bibinfo {title} {Stability of a supercooled
			liquid to periodic density waves and dynamics of freezing},\ }\href
	{https://doi.org/10.1016/0378-4371(87)90252-4} {\bibfield  {journal}
		{\bibinfo  {journal} {Physica A: Statistical Mechanics and its Applications}\
		}\textbf {\bibinfo {volume} {145}},\ \bibinfo {pages} {273} (\bibinfo {year}
		{1987})}\BibitemShut {NoStop}%
	\bibitem [{\citenamefont {Dean}(1996)}]{Dean1996}%
	\BibitemOpen
	\bibfield  {author} {\bibinfo {author} {\bibfnamefont {D.~S.}\ \bibnamefont
			{Dean}},\ }\bibfield  {title} {\bibinfo {title} {Langevin equation for the
			density of a system of interacting {{Langevin}} processes},\ }\href
	{https://doi.org/10.1088/0305-4470/29/24/001} {\bibfield  {journal} {\bibinfo
			{journal} {J. Phys. A: Math. Gen.}\ }\textbf {\bibinfo {volume} {29}},\
		\bibinfo {pages} {L613} (\bibinfo {year} {1996})}\BibitemShut {NoStop}%
	\bibitem [{\citenamefont {Bothe}\ \emph {et~al.}(2023)\citenamefont {Bothe},
		\citenamefont {Cocconi}, \citenamefont {Zhen},\ and\ \citenamefont
		{Pruessner}}]{Bothe2023}%
	\BibitemOpen
	\bibfield  {author} {\bibinfo {author} {\bibfnamefont {M.}~\bibnamefont
			{Bothe}}, \bibinfo {author} {\bibfnamefont {L.}~\bibnamefont {Cocconi}},
		\bibinfo {author} {\bibfnamefont {Z.}~\bibnamefont {Zhen}},\ and\ \bibinfo
		{author} {\bibfnamefont {G.}~\bibnamefont {Pruessner}},\ }\bibfield  {title}
	{\bibinfo {title} {Particle entity in the {{Doi}}--{{Peliti}} and response
			field formalisms},\ }\href {https://doi.org/10.1088/1751-8121/acc498}
	{\bibfield  {journal} {\bibinfo  {journal} {Journal of Physics A:
				Mathematical and Theoretical}\ }\textbf {\bibinfo {volume} {56}},\ \bibinfo
		{pages} {175002} (\bibinfo {year} {2023})}\BibitemShut {NoStop}%
	\bibitem [{\citenamefont {Frusawa}\ and\ \citenamefont
		{Hayakawa}(2000)}]{Frusawa2000}%
	\BibitemOpen
	\bibfield  {author} {\bibinfo {author} {\bibfnamefont {H.}~\bibnamefont
			{Frusawa}}\ and\ \bibinfo {author} {\bibfnamefont {R.}~\bibnamefont
			{Hayakawa}},\ }\bibfield  {title} {\bibinfo {title} {On the controversy over
			the stochastic density functional equations},\ }\href
	{https://doi.org/10.1088/0305-4470/33/15/101} {\bibfield  {journal} {\bibinfo
			{journal} {Journal of Physics A: Mathematical and General}\ }\textbf
		{\bibinfo {volume} {33}},\ \bibinfo {pages} {L155} (\bibinfo {year}
		{2000})}\BibitemShut {NoStop}%
	\bibitem [{\citenamefont {Landau}\ \emph {et~al.}(1980)\citenamefont {Landau},
		\citenamefont {Lifshitz},\ and\ \citenamefont {Pitaevskii}}]{Landau1980}%
	\BibitemOpen
	\bibfield  {author} {\bibinfo {author} {\bibfnamefont {L.~D.}\ \bibnamefont
			{Landau}}, \bibinfo {author} {\bibfnamefont {E.~M.}\ \bibnamefont
			{Lifshitz}},\ and\ \bibinfo {author} {\bibfnamefont {L.~P.}\ \bibnamefont
			{Pitaevskii}},\ }\href@noop {} {\emph {\bibinfo {title} {Course of
				Theoretical Physics. {{Vol}}. 9: {{Statistical}} Physics ({{Part}} 2)}}},\
	\bibinfo {edition} {2nd}\ ed.\ (\bibinfo  {publisher} {Pergamon Press},\
	\bibinfo {address} {Oxford},\ \bibinfo {year} {1980})\BibitemShut {NoStop}%
	\bibitem [{\citenamefont {Fox}(1978)}]{Fox1978}%
	\BibitemOpen
	\bibfield  {author} {\bibinfo {author} {\bibfnamefont {R.~F.}\ \bibnamefont
			{Fox}},\ }\bibfield  {title} {\bibinfo {title} {Gaussian stochastic processes
			in physics},\ }\href {https://doi.org/10.1016/0370-1573(78)90145-X}
	{\bibfield  {journal} {\bibinfo  {journal} {Physics Reports}\ }\textbf
		{\bibinfo {volume} {48}},\ \bibinfo {pages} {179} (\bibinfo {year}
		{1978})}\BibitemShut {NoStop}%
	\bibitem [{\citenamefont {Spohn}(1991)}]{Spohn1991}%
	\BibitemOpen
	\bibfield  {author} {\bibinfo {author} {\bibfnamefont {H.}~\bibnamefont
			{Spohn}},\ }\href@noop {} {\emph {\bibinfo {title} {Large-{{Scale Dynamics}}
				of {{Interacting Particles}}}}}\ (\bibinfo  {publisher} {Springer},\ \bibinfo
	{year} {1991})\BibitemShut {NoStop}%
	\bibitem [{\citenamefont {Donev}\ \emph {et~al.}(2010)\citenamefont {Donev},
		\citenamefont {{Vanden-Eijnden}}, \citenamefont {Garcia},\ and\ \citenamefont
		{Bell}}]{Donev2010}%
	\BibitemOpen
	\bibfield  {author} {\bibinfo {author} {\bibfnamefont {A.}~\bibnamefont
			{Donev}}, \bibinfo {author} {\bibfnamefont {E.}~\bibnamefont
			{{Vanden-Eijnden}}}, \bibinfo {author} {\bibfnamefont {A.}~\bibnamefont
			{Garcia}},\ and\ \bibinfo {author} {\bibfnamefont {J.}~\bibnamefont {Bell}},\
	}\bibfield  {title} {\bibinfo {title} {On the accuracy of finite-volume
			schemes for fluctuating hydrodynamics},\ }\href
	{https://doi.org/10.2140/camcos.2010.5.149} {\bibfield  {journal} {\bibinfo
			{journal} {Communications in Applied Mathematics and Computational Science}\
		}\textbf {\bibinfo {volume} {5}},\ \bibinfo {pages} {149} (\bibinfo {year}
		{2010})}\BibitemShut {NoStop}%
	\bibitem [{\citenamefont {Delong}\ \emph {et~al.}(2013)\citenamefont {Delong},
		\citenamefont {Griffith}, \citenamefont {{Vanden-Eijnden}},\ and\
		\citenamefont {Donev}}]{Delong2013}%
	\BibitemOpen
	\bibfield  {author} {\bibinfo {author} {\bibfnamefont {S.}~\bibnamefont
			{Delong}}, \bibinfo {author} {\bibfnamefont {B.~E.}\ \bibnamefont
			{Griffith}}, \bibinfo {author} {\bibfnamefont {E.}~\bibnamefont
			{{Vanden-Eijnden}}},\ and\ \bibinfo {author} {\bibfnamefont {A.}~\bibnamefont
			{Donev}},\ }\bibfield  {title} {\bibinfo {title} {Temporal integrators for
			fluctuating hydrodynamics},\ }\href
	{https://doi.org/10.1103/PhysRevE.87.033302} {\bibfield  {journal} {\bibinfo
			{journal} {Physical Review E}\ }\textbf {\bibinfo {volume} {87}},\ \bibinfo
		{pages} {033302} (\bibinfo {year} {2013})}\BibitemShut {NoStop}%
	\bibitem [{\citenamefont {Delong}\ \emph {et~al.}(2014)\citenamefont {Delong},
		\citenamefont {Sun}, \citenamefont {Griffith}, \citenamefont
		{{Vanden-Eijnden}},\ and\ \citenamefont {Donev}}]{Delong2014}%
	\BibitemOpen
	\bibfield  {author} {\bibinfo {author} {\bibfnamefont {S.}~\bibnamefont
			{Delong}}, \bibinfo {author} {\bibfnamefont {Y.}~\bibnamefont {Sun}},
		\bibinfo {author} {\bibfnamefont {B.~E.}\ \bibnamefont {Griffith}}, \bibinfo
		{author} {\bibfnamefont {E.}~\bibnamefont {{Vanden-Eijnden}}},\ and\ \bibinfo
		{author} {\bibfnamefont {A.}~\bibnamefont {Donev}},\ }\bibfield  {title}
	{\bibinfo {title} {Multiscale temporal integrators for fluctuating
			hydrodynamics},\ }\href {https://doi.org/10.1103/PhysRevE.90.063312}
	{\bibfield  {journal} {\bibinfo  {journal} {Physical Review E}\ }\textbf
		{\bibinfo {volume} {90}},\ \bibinfo {pages} {063312} (\bibinfo {year}
		{2014})}\BibitemShut {NoStop}%
	\bibitem [{\citenamefont {De~La~Torre}\ \emph {et~al.}(2015)\citenamefont
		{De~La~Torre}, \citenamefont {Espa{\~n}ol},\ and\ \citenamefont
		{Donev}}]{DeLaTorre2015}%
	\BibitemOpen
	\bibfield  {author} {\bibinfo {author} {\bibfnamefont {J.~A.}\ \bibnamefont
			{De~La~Torre}}, \bibinfo {author} {\bibfnamefont {P.}~\bibnamefont
			{Espa{\~n}ol}},\ and\ \bibinfo {author} {\bibfnamefont {A.}~\bibnamefont
			{Donev}},\ }\bibfield  {title} {\bibinfo {title} {Finite element
			discretization of non-linear diffusion equations with thermal fluctuations},\
	}\href {https://doi.org/10.1063/1.4913746} {\bibfield  {journal} {\bibinfo
			{journal} {The Journal of Chemical Physics}\ }\textbf {\bibinfo {volume}
			{142}},\ \bibinfo {pages} {094115} (\bibinfo {year} {2015})}\BibitemShut
	{NoStop}%
	\bibitem [{\citenamefont {Kim}\ \emph {et~al.}(2017)\citenamefont {Kim},
		\citenamefont {Nonaka}, \citenamefont {Bell}, \citenamefont {Garcia},\ and\
		\citenamefont {Donev}}]{Kim2017}%
	\BibitemOpen
	\bibfield  {author} {\bibinfo {author} {\bibfnamefont {C.}~\bibnamefont
			{Kim}}, \bibinfo {author} {\bibfnamefont {A.}~\bibnamefont {Nonaka}},
		\bibinfo {author} {\bibfnamefont {J.~B.}\ \bibnamefont {Bell}}, \bibinfo
		{author} {\bibfnamefont {A.~L.}\ \bibnamefont {Garcia}},\ and\ \bibinfo
		{author} {\bibfnamefont {A.}~\bibnamefont {Donev}},\ }\bibfield  {title}
	{\bibinfo {title} {Stochastic simulation of reaction-diffusion systems: {{A}}
			fluctuating-hydrodynamics approach},\ }\href
	{https://doi.org/10.1063/1.4978775} {\bibfield  {journal} {\bibinfo
			{journal} {The Journal of Chemical Physics}\ }\textbf {\bibinfo {volume}
			{146}},\ \bibinfo {pages} {124110} (\bibinfo {year} {2017})}\BibitemShut
	{NoStop}%
	\bibitem [{\citenamefont {Derrida}(2007)}]{Derrida2007}%
	\BibitemOpen
	\bibfield  {author} {\bibinfo {author} {\bibfnamefont {B.}~\bibnamefont
			{Derrida}},\ }\bibfield  {title} {\bibinfo {title} {Non-equilibrium steady
			states: {{Fluctuations}} and large deviations of the density and of the
			current},\ }\href {https://doi.org/10.1088/1742-5468/2007/07/P07023}
	{\bibfield  {journal} {\bibinfo  {journal} {Journal of Statistical Mechanics:
				Theory and Experiment}\ }\textbf {\bibinfo {volume} {2007}},\ \bibinfo
		{pages} {P07023} (\bibinfo {year} {2007})}\BibitemShut {NoStop}%
	\bibitem [{\citenamefont {Bertini}\ \emph {et~al.}(2001)\citenamefont
		{Bertini}, \citenamefont {De~Sole}, \citenamefont {Gabrielli}, \citenamefont
		{{Jona-Lasinio}},\ and\ \citenamefont {Landim}}]{Bertini2001}%
	\BibitemOpen
	\bibfield  {author} {\bibinfo {author} {\bibfnamefont {L.}~\bibnamefont
			{Bertini}}, \bibinfo {author} {\bibfnamefont {A.}~\bibnamefont {De~Sole}},
		\bibinfo {author} {\bibfnamefont {D.}~\bibnamefont {Gabrielli}}, \bibinfo
		{author} {\bibfnamefont {G.}~\bibnamefont {{Jona-Lasinio}}},\ and\ \bibinfo
		{author} {\bibfnamefont {C.}~\bibnamefont {Landim}},\ }\bibfield  {title}
	{\bibinfo {title} {Fluctuations in {{Stationary Nonequilibrium States}} of
			{{Irreversible Processes}}},\ }\href
	{https://doi.org/10.1103/PhysRevLett.87.040601} {\bibfield  {journal}
		{\bibinfo  {journal} {Physical Review Letters}\ }\textbf {\bibinfo {volume}
			{87}},\ \bibinfo {pages} {040601} (\bibinfo {year} {2001})}\BibitemShut
	{NoStop}%
	\bibitem [{\citenamefont {Bertini}\ \emph {et~al.}(2002)\citenamefont
		{Bertini}, \citenamefont {De~Sole}, \citenamefont {Gabrielli}, \citenamefont
		{{Jona-Lasinio}},\ and\ \citenamefont {Landim}}]{Bertini2002}%
	\BibitemOpen
	\bibfield  {author} {\bibinfo {author} {\bibfnamefont {L.}~\bibnamefont
			{Bertini}}, \bibinfo {author} {\bibfnamefont {A.}~\bibnamefont {De~Sole}},
		\bibinfo {author} {\bibfnamefont {D.}~\bibnamefont {Gabrielli}}, \bibinfo
		{author} {\bibfnamefont {G.}~\bibnamefont {{Jona-Lasinio}}},\ and\ \bibinfo
		{author} {\bibfnamefont {C.}~\bibnamefont {Landim}},\ }\bibfield  {title}
	{\bibinfo {title} {Macroscopic {{Fluctuation Theory}} for {{Stationary
					Non-Equilibrium States}}},\ }\href {https://doi.org/10.1023/A:1014525911391}
	{\bibfield  {journal} {\bibinfo  {journal} {J. Stat. Phys.}\ }\textbf
		{\bibinfo {volume} {107}},\ \bibinfo {pages} {635} (\bibinfo {year}
		{2002})}\BibitemShut {NoStop}%
	\bibitem [{\citenamefont {Bertini}\ \emph {et~al.}(2015)\citenamefont
		{Bertini}, \citenamefont {De~Sole}, \citenamefont {Gabrielli}, \citenamefont
		{{Jona-Lasinio}},\ and\ \citenamefont {Landim}}]{Bertini2015a}%
	\BibitemOpen
	\bibfield  {author} {\bibinfo {author} {\bibfnamefont {L.}~\bibnamefont
			{Bertini}}, \bibinfo {author} {\bibfnamefont {A.}~\bibnamefont {De~Sole}},
		\bibinfo {author} {\bibfnamefont {D.}~\bibnamefont {Gabrielli}}, \bibinfo
		{author} {\bibfnamefont {G.}~\bibnamefont {{Jona-Lasinio}}},\ and\ \bibinfo
		{author} {\bibfnamefont {C.}~\bibnamefont {Landim}},\ }\bibfield  {title}
	{\bibinfo {title} {Macroscopic fluctuation theory},\ }\href
	{https://doi.org/10.1103/RevModPhys.87.593} {\bibfield  {journal} {\bibinfo
			{journal} {Reviews of Modern Physics}\ }\textbf {\bibinfo {volume} {87}},\
		\bibinfo {pages} {593} (\bibinfo {year} {2015})}\BibitemShut {NoStop}%
	\bibitem [{\citenamefont {Derrida}\ and\ \citenamefont
		{Gerschenfeld}(2009)}]{Derrida2009}%
	\BibitemOpen
	\bibfield  {author} {\bibinfo {author} {\bibfnamefont {B.}~\bibnamefont
			{Derrida}}\ and\ \bibinfo {author} {\bibfnamefont {A.}~\bibnamefont
			{Gerschenfeld}},\ }\bibfield  {title} {\bibinfo {title} {Current
			{{Fluctuations}} in {{One Dimensional Diffusive Systems}} with a {{Step
					Initial Density Profile}}},\ }\href
	{https://doi.org/10.1007/s10955-009-9830-1} {\bibfield  {journal} {\bibinfo
			{journal} {Journal of Statistical Physics}\ }\textbf {\bibinfo {volume}
			{137}},\ \bibinfo {pages} {978} (\bibinfo {year} {2009})}\BibitemShut
	{NoStop}%
	\bibitem [{\citenamefont {Krapivsky}\ \emph {et~al.}(2014)\citenamefont
		{Krapivsky}, \citenamefont {Mallick},\ and\ \citenamefont
		{Sadhu}}]{Krapivsky2014c}%
	\BibitemOpen
	\bibfield  {author} {\bibinfo {author} {\bibfnamefont {P.~L.}\ \bibnamefont
			{Krapivsky}}, \bibinfo {author} {\bibfnamefont {K.}~\bibnamefont {Mallick}},\
		and\ \bibinfo {author} {\bibfnamefont {T.}~\bibnamefont {Sadhu}},\ }\bibfield
	{title} {\bibinfo {title} {Large {{Deviations}} in {{Single-File
					Diffusion}}},\ }\href {https://doi.org/10.1103/PhysRevLett.113.078101}
	{\bibfield  {journal} {\bibinfo  {journal} {Physical Review Letters}\
		}\textbf {\bibinfo {volume} {113}},\ \bibinfo {pages} {078101} (\bibinfo
		{year} {2014})}\BibitemShut {NoStop}%
	\bibitem [{\citenamefont {Krapivsky}\ \emph {et~al.}(2015)\citenamefont
		{Krapivsky}, \citenamefont {Mallick},\ and\ \citenamefont
		{Sadhu}}]{Krapivsky2015}%
	\BibitemOpen
	\bibfield  {author} {\bibinfo {author} {\bibfnamefont {P.~L.}\ \bibnamefont
			{Krapivsky}}, \bibinfo {author} {\bibfnamefont {K.}~\bibnamefont {Mallick}},\
		and\ \bibinfo {author} {\bibfnamefont {T.}~\bibnamefont {Sadhu}},\ }\bibfield
	{title} {\bibinfo {title} {Tagged {{Particle}} in {{Single-File
					Diffusion}}},\ }\href {https://doi.org/10.1007/s10955-015-1291-0} {\bibfield
		{journal} {\bibinfo  {journal} {Journal of Statistical Physics}\ }\textbf
		{\bibinfo {volume} {160}},\ \bibinfo {pages} {885} (\bibinfo {year}
		{2015})}\BibitemShut {NoStop}%
	\bibitem [{\citenamefont {Poncet}\ \emph
		{et~al.}(2021{\natexlab{a}})\citenamefont {Poncet}, \citenamefont {Grabsch},
		\citenamefont {Illien},\ and\ \citenamefont {B{\'e}nichou}}]{Poncet2021a}%
	\BibitemOpen
	\bibfield  {author} {\bibinfo {author} {\bibfnamefont {A.}~\bibnamefont
			{Poncet}}, \bibinfo {author} {\bibfnamefont {A.}~\bibnamefont {Grabsch}},
		\bibinfo {author} {\bibfnamefont {P.}~\bibnamefont {Illien}},\ and\ \bibinfo
		{author} {\bibfnamefont {O.}~\bibnamefont {B{\'e}nichou}},\ }\bibfield
	{title} {\bibinfo {title} {Generalized {{Correlation Profiles}} in
			{{Single-File Systems}}},\ }\href
	{https://doi.org/10.1103/PhysRevLett.127.220601} {\bibfield  {journal}
		{\bibinfo  {journal} {Physical Review Letters}\ }\textbf {\bibinfo {volume}
			{127}},\ \bibinfo {pages} {220601} (\bibinfo {year}
		{2021}{\natexlab{a}})}\BibitemShut {NoStop}%
	\bibitem [{\citenamefont {Grabsch}\ \emph {et~al.}(2022)\citenamefont
		{Grabsch}, \citenamefont {Poncet}, \citenamefont {Rizkallah}, \citenamefont
		{Illien},\ and\ \citenamefont {B{\'e}nichou}}]{Grabsch2022}%
	\BibitemOpen
	\bibfield  {author} {\bibinfo {author} {\bibfnamefont {A.}~\bibnamefont
			{Grabsch}}, \bibinfo {author} {\bibfnamefont {A.}~\bibnamefont {Poncet}},
		\bibinfo {author} {\bibfnamefont {P.}~\bibnamefont {Rizkallah}}, \bibinfo
		{author} {\bibfnamefont {P.}~\bibnamefont {Illien}},\ and\ \bibinfo {author}
		{\bibfnamefont {O.}~\bibnamefont {B{\'e}nichou}},\ }\bibfield  {title}
	{\bibinfo {title} {Exact closure and solution for spatial correlations in
			single-file diffusion},\ }\href {https://doi.org/10.1126/sciadv.abm5043}
	{\bibfield  {journal} {\bibinfo  {journal} {Science Advances}\ }\textbf
		{\bibinfo {volume} {8}},\ \bibinfo {pages} {eabm5043} (\bibinfo {year}
		{2022})}\BibitemShut {NoStop}%
	\bibitem [{\citenamefont {Mallick}\ \emph {et~al.}(2022)\citenamefont
		{Mallick}, \citenamefont {Moriya},\ and\ \citenamefont
		{Sasamoto}}]{Mallick2022}%
	\BibitemOpen
	\bibfield  {author} {\bibinfo {author} {\bibfnamefont {K.}~\bibnamefont
			{Mallick}}, \bibinfo {author} {\bibfnamefont {H.}~\bibnamefont {Moriya}},\
		and\ \bibinfo {author} {\bibfnamefont {T.}~\bibnamefont {Sasamoto}},\
	}\bibfield  {title} {\bibinfo {title} {Exact solution of the macroscopic
			fluctuation theory for the symmetric exclusion process},\ }\href
	{https://doi.org/10.1103/PhysRevLett.129.040601} {\bibfield  {journal}
		{\bibinfo  {journal} {Physical Review Letters}\ }\textbf {\bibinfo {volume}
			{129}},\ \bibinfo {pages} {40601} (\bibinfo {year} {2022})}\BibitemShut
	{NoStop}%
	\bibitem [{\citenamefont {Grabsch}\ and\ \citenamefont
		{B{\'e}nichou}(2024)}]{Grabsch2024}%
	\BibitemOpen
	\bibfield  {author} {\bibinfo {author} {\bibfnamefont {A.}~\bibnamefont
			{Grabsch}}\ and\ \bibinfo {author} {\bibfnamefont {O.}~\bibnamefont
			{B{\'e}nichou}},\ }\bibfield  {title} {\bibinfo {title} {Tracer {{Diffusion}}
			beyond {{Gaussian Behavior}}: {{Explicit Results}} for {{General Single-File
					Systems}}},\ }\href {https://doi.org/10.1103/PhysRevLett.132.217101}
	{\bibfield  {journal} {\bibinfo  {journal} {Physical Review Letters}\
		}\textbf {\bibinfo {volume} {132}},\ \bibinfo {pages} {217101} (\bibinfo
		{year} {2024})}\BibitemShut {NoStop}%
	\bibitem [{\citenamefont {{Dur{\'a}n-Olivencia}}\ \emph
		{et~al.}(2017)\citenamefont {{Dur{\'a}n-Olivencia}}, \citenamefont
		{Yatsyshin}, \citenamefont {Goddard},\ and\ \citenamefont
		{Kalliadasis}}]{Duran-Olivencia2017}%
	\BibitemOpen
	\bibfield  {author} {\bibinfo {author} {\bibfnamefont {M.~A.}\ \bibnamefont
			{{Dur{\'a}n-Olivencia}}}, \bibinfo {author} {\bibfnamefont {P.}~\bibnamefont
			{Yatsyshin}}, \bibinfo {author} {\bibfnamefont {B.~D.}\ \bibnamefont
			{Goddard}},\ and\ \bibinfo {author} {\bibfnamefont {S.}~\bibnamefont
			{Kalliadasis}},\ }\bibfield  {title} {\bibinfo {title} {General framework for
			fluctuating dynamic density functional theory},\ }\href
	{https://doi.org/10.1088/1367-2630/aa9041} {\bibfield  {journal} {\bibinfo
			{journal} {New Journal of Physics}\ }\textbf {\bibinfo {volume} {19}},\
		\bibinfo {pages} {123022} (\bibinfo {year} {2017})}\BibitemShut {NoStop}%
	\bibitem [{\citenamefont {Chavanis}(2008)}]{Chavanis2008}%
	\BibitemOpen
	\bibfield  {author} {\bibinfo {author} {\bibfnamefont {P.-H.}\ \bibnamefont
			{Chavanis}},\ }\bibfield  {title} {\bibinfo {title} {Hamiltonian and
			{{Brownian}} systems with long-range interactions: {{V}}. {{Stochastic}}
			kinetic equations and theory of fluctuations},\ }\href
	{https://doi.org/10.1016/j.physa.2008.06.016} {\bibfield  {journal} {\bibinfo
			{journal} {Physica A: Statistical Mechanics and its Applications}\ }\textbf
		{\bibinfo {volume} {387}},\ \bibinfo {pages} {5716} (\bibinfo {year}
		{2008})}\BibitemShut {NoStop}%
	\bibitem [{\citenamefont {Velenich}\ \emph {et~al.}(2008)\citenamefont
		{Velenich}, \citenamefont {Chamon}, \citenamefont {Cugliandolo},\ and\
		\citenamefont {Kreimer}}]{Velenich2008}%
	\BibitemOpen
	\bibfield  {author} {\bibinfo {author} {\bibfnamefont {A.}~\bibnamefont
			{Velenich}}, \bibinfo {author} {\bibfnamefont {C.}~\bibnamefont {Chamon}},
		\bibinfo {author} {\bibfnamefont {L.~F.}\ \bibnamefont {Cugliandolo}},\ and\
		\bibinfo {author} {\bibfnamefont {D.}~\bibnamefont {Kreimer}},\ }\bibfield
	{title} {\bibinfo {title} {On the {{Brownian}} gas: A field theory with a
			{{Poissonian}} ground state},\ }\href
	{https://doi.org/10.1088/1751-8113/41/23/235002} {\bibfield  {journal}
		{\bibinfo  {journal} {Journal of Physics A: Mathematical and Theoretical}\
		}\textbf {\bibinfo {volume} {41}},\ \bibinfo {pages} {235002} (\bibinfo
		{year} {2008})}\BibitemShut {NoStop}%
	\bibitem [{\citenamefont {McKean}(1966)}]{McKean1966}%
	\BibitemOpen
	\bibfield  {author} {\bibinfo {author} {\bibfnamefont {H.~P.}\ \bibnamefont
			{McKean}},\ }\bibfield  {title} {\bibinfo {title} {A class of {{Markov}}
			processes associated with nonlinear parabolic equations},\ }\href
	{https://doi.org/10.1073/pnas.56.6.1907} {\bibfield  {journal} {\bibinfo
			{journal} {Proceedings of the National Academy of Sciences}\ }\textbf
		{\bibinfo {volume} {56}},\ \bibinfo {pages} {1907} (\bibinfo {year}
		{1966})}\BibitemShut {NoStop}%
	\bibitem [{\citenamefont {Bouchet}\ \emph {et~al.}(2016)\citenamefont
		{Bouchet}, \citenamefont {Gawedzki},\ and\ \citenamefont
		{Nardini}}]{Bouchet2016}%
	\BibitemOpen
	\bibfield  {author} {\bibinfo {author} {\bibfnamefont {F.}~\bibnamefont
			{Bouchet}}, \bibinfo {author} {\bibfnamefont {K.}~\bibnamefont {Gawedzki}},\
		and\ \bibinfo {author} {\bibfnamefont {C.}~\bibnamefont {Nardini}},\
	}\bibfield  {title} {\bibinfo {title} {Perturbative {{Calculation}} of
			{{Quasi-Potential}} in {{Non-equilibrium Diffusions}}: {{A Mean-Field
					Example}}},\ }\href {https://doi.org/10.1007/s10955-016-1503-2} {\bibfield
		{journal} {\bibinfo  {journal} {Journal of Statistical Physics}\ }\textbf
		{\bibinfo {volume} {163}},\ \bibinfo {pages} {1157} (\bibinfo {year}
		{2016})}\BibitemShut {NoStop}%
	\bibitem [{\citenamefont {Chavanis}(2014)}]{Chavanis2014a}%
	\BibitemOpen
	\bibfield  {author} {\bibinfo {author} {\bibfnamefont {P.-H.}\ \bibnamefont
			{Chavanis}},\ }\bibfield  {title} {\bibinfo {title} {The {{Brownian}} mean
			field model},\ }\href {https://doi.org/10.1140/epjb/e2014-40586-6} {\bibfield
		{journal} {\bibinfo  {journal} {The European Physical Journal B}\ }\textbf
		{\bibinfo {volume} {87}},\ \bibinfo {pages} {120} (\bibinfo {year}
		{2014})}\BibitemShut {NoStop}%
	\bibitem [{\citenamefont {Hansen}\ and\ \citenamefont
		{McDonald}(1986)}]{Hansen1986}%
	\BibitemOpen
	\bibfield  {author} {\bibinfo {author} {\bibfnamefont {J.~P.}\ \bibnamefont
			{Hansen}}\ and\ \bibinfo {author} {\bibfnamefont {I.~R.}\ \bibnamefont
			{McDonald}},\ }\href@noop {} {\emph {\bibinfo {title} {Theory of Simple
				Liquids}}},\ \bibinfo {edition} {2nd}\ ed.\ (\bibinfo  {publisher} {Academic
		Press},\ \bibinfo {year} {1986})\BibitemShut {NoStop}%
	\bibitem [{\citenamefont {Reichman}\ and\ \citenamefont
		{Charbonneau}(2005)}]{Reichman2005}%
	\BibitemOpen
	\bibfield  {author} {\bibinfo {author} {\bibfnamefont {D.~R.}\ \bibnamefont
			{Reichman}}\ and\ \bibinfo {author} {\bibfnamefont {P.}~\bibnamefont
			{Charbonneau}},\ }\bibfield  {title} {\bibinfo {title} {Mode-coupling
			theory},\ }\href {https://doi.org/10.1088/1742-5468/2005/05/P05013}
	{\bibfield  {journal} {\bibinfo  {journal} {Journal of Statistical Mechanics:
				Theory and Experiment}\ }\textbf {\bibinfo {volume} {2005}},\ \bibinfo
		{pages} {P05013} (\bibinfo {year} {2005})}\BibitemShut {NoStop}%
	\bibitem [{\citenamefont {Szamel}(2013)}]{Szamel2013}%
	\BibitemOpen
	\bibfield  {author} {\bibinfo {author} {\bibfnamefont {G.}~\bibnamefont
			{Szamel}},\ }\bibfield  {title} {\bibinfo {title} {Mode-coupling theory and
			beyond: {{A}} diagrammatic approach},\ }\href
	{https://doi.org/10.1093/ptep/pts036} {\bibfield  {journal} {\bibinfo
			{journal} {Progress of Theoretical and Experimental Physics}\ }\textbf
		{\bibinfo {volume} {2013}},\ \bibinfo {pages} {012J01} (\bibinfo {year}
		{2013})}\BibitemShut {NoStop}%
	\bibitem [{\citenamefont {Janssen}(2018)}]{Janssen2018}%
	\BibitemOpen
	\bibfield  {author} {\bibinfo {author} {\bibfnamefont {L.~M.~C.}\
			\bibnamefont {Janssen}},\ }\bibfield  {title} {\bibinfo {title}
		{Mode-{{Coupling Theory}} of the {{Glass Transition}}: {{A Primer}}},\ }\href
	{https://doi.org/10.3389/fphy.2018.00097} {\bibfield  {journal} {\bibinfo
			{journal} {Frontiers in Physics}\ }\textbf {\bibinfo {volume} {6}},\ \bibinfo
		{pages} {97} (\bibinfo {year} {2018})}\BibitemShut {NoStop}%
	\bibitem [{\citenamefont {Mayer}\ \emph {et~al.}(2006)\citenamefont {Mayer},
		\citenamefont {Miyazaki},\ and\ \citenamefont {Reichman}}]{Mayer2006}%
	\BibitemOpen
	\bibfield  {author} {\bibinfo {author} {\bibfnamefont {P.}~\bibnamefont
			{Mayer}}, \bibinfo {author} {\bibfnamefont {K.}~\bibnamefont {Miyazaki}},\
		and\ \bibinfo {author} {\bibfnamefont {D.~R.}\ \bibnamefont {Reichman}},\
	}\bibfield  {title} {\bibinfo {title} {Cooperativity beyond {{Caging}}:
			{{Generalized Mode-Coupling Theory}}},\ }\href
	{https://doi.org/10.1103/PhysRevLett.97.095702} {\bibfield  {journal}
		{\bibinfo  {journal} {Physical Review Letters}\ }\textbf {\bibinfo {volume}
			{97}},\ \bibinfo {pages} {095702} (\bibinfo {year} {2006})}\BibitemShut
	{NoStop}%
	\bibitem [{\citenamefont {Mazenko}(2010)}]{Mazenko2010}%
	\BibitemOpen
	\bibfield  {author} {\bibinfo {author} {\bibfnamefont {G.~F.}\ \bibnamefont
			{Mazenko}},\ }\bibfield  {title} {\bibinfo {title} {Fundamental theory of
			statistical particle dynamics},\ }\href
	{https://doi.org/10.1103/PhysRevE.81.061102} {\bibfield  {journal} {\bibinfo
			{journal} {Physical Review E}\ }\textbf {\bibinfo {volume} {81}},\ \bibinfo
		{pages} {061102} (\bibinfo {year} {2010})}\BibitemShut {NoStop}%
	\bibitem [{\citenamefont {Mazenko}(2011)}]{Mazenko2011}%
	\BibitemOpen
	\bibfield  {author} {\bibinfo {author} {\bibfnamefont {G.~F.}\ \bibnamefont
			{Mazenko}},\ }\bibfield  {title} {\bibinfo {title} {Smoluchowski dynamics and
			the ergodic-nonergodic transition},\ }\href
	{https://doi.org/10.1103/PhysRevE.83.041125} {\bibfield  {journal} {\bibinfo
			{journal} {Physical Review E}\ }\textbf {\bibinfo {volume} {83}},\ \bibinfo
		{pages} {041125} (\bibinfo {year} {2011})}\BibitemShut {NoStop}%
	\bibitem [{\citenamefont {Das}\ and\ \citenamefont {Mazenko}(2013)}]{Das2013b}%
	\BibitemOpen
	\bibfield  {author} {\bibinfo {author} {\bibfnamefont {S.~P.}\ \bibnamefont
			{Das}}\ and\ \bibinfo {author} {\bibfnamefont {G.~F.}\ \bibnamefont
			{Mazenko}},\ }\bibfield  {title} {\bibinfo {title} {Newtonian {{Kinetic
					Theory}} and the {{Ergodic-Nonergodic Transition}}},\ }\href
	{https://doi.org/10.1007/s10955-013-0755-3} {\bibfield  {journal} {\bibinfo
			{journal} {Journal of Statistical Physics}\ }\textbf {\bibinfo {volume}
			{152}},\ \bibinfo {pages} {159} (\bibinfo {year} {2013})}\BibitemShut
	{NoStop}%
	\bibitem [{\citenamefont {Martin}\ \emph {et~al.}(1973)\citenamefont {Martin},
		\citenamefont {Siggia},\ and\ \citenamefont {Rose}}]{Martin1973}%
	\BibitemOpen
	\bibfield  {author} {\bibinfo {author} {\bibfnamefont {P.~C.}\ \bibnamefont
			{Martin}}, \bibinfo {author} {\bibfnamefont {E.~D.}\ \bibnamefont {Siggia}},\
		and\ \bibinfo {author} {\bibfnamefont {H.~A.}\ \bibnamefont {Rose}},\
	}\bibfield  {title} {\bibinfo {title} {Statistical {{Dynamics}} of
			{{Classical Systems}}},\ }\href {https://doi.org/10.1103/PhysRevA.8.423}
	{\bibfield  {journal} {\bibinfo  {journal} {Physical Review A}\ }\textbf
		{\bibinfo {volume} {8}},\ \bibinfo {pages} {423} (\bibinfo {year}
		{1973})}\BibitemShut {NoStop}%
	\bibitem [{\citenamefont {Janssen}(1976)}]{Janssen1976}%
	\BibitemOpen
	\bibfield  {author} {\bibinfo {author} {\bibfnamefont {H.-K.}\ \bibnamefont
			{Janssen}},\ }\bibfield  {title} {\bibinfo {title} {On a {{Lagrangean}} for
			classical field dynamics and renormalization group calculations of dynamical
			critical properties},\ }\href {https://doi.org/10.1007/BF01316547} {\bibfield
		{journal} {\bibinfo  {journal} {Zeitschrift f{\"u}r Physik B Condensed
				Matter and Quanta}\ }\textbf {\bibinfo {volume} {23}},\ \bibinfo {pages}
		{377} (\bibinfo {year} {1976})}\BibitemShut {NoStop}%
	\bibitem [{\citenamefont {De~Dominicis}\ and\ \citenamefont
		{Peliti}(1978)}]{DeDominicis1978a}%
	\BibitemOpen
	\bibfield  {author} {\bibinfo {author} {\bibfnamefont {C.}~\bibnamefont
			{De~Dominicis}}\ and\ \bibinfo {author} {\bibfnamefont {L.}~\bibnamefont
			{Peliti}},\ }\bibfield  {title} {\bibinfo {title} {Field-theory
			renormalization and critical dynamics above {{T}} c : {{Helium}},
			antiferromagnets, and liquid-gas systems},\ }\href
	{https://doi.org/10.1103/PhysRevB.18.353} {\bibfield  {journal} {\bibinfo
			{journal} {Physical Review B}\ }\textbf {\bibinfo {volume} {18}},\ \bibinfo
		{pages} {353} (\bibinfo {year} {1978})}\BibitemShut {NoStop}%
	\bibitem [{\citenamefont {Miyazaki}\ and\ \citenamefont
		{Reichman}(2005)}]{Miyazaki2005}%
	\BibitemOpen
	\bibfield  {author} {\bibinfo {author} {\bibfnamefont {K.}~\bibnamefont
			{Miyazaki}}\ and\ \bibinfo {author} {\bibfnamefont {D.~R.}\ \bibnamefont
			{Reichman}},\ }\bibfield  {title} {\bibinfo {title} {Mode-coupling theory and
			the fluctuation--dissipation theorem for nonlinear {{Langevin}} equations
			with multiplicative noise},\ }\href
	{https://doi.org/10.1088/0305-4470/38/20/L03} {\bibfield  {journal} {\bibinfo
			{journal} {Journal of Physics A: Mathematical and General}\ }\textbf
		{\bibinfo {volume} {38}},\ \bibinfo {pages} {L343} (\bibinfo {year}
		{2005})}\BibitemShut {NoStop}%
	\bibitem [{\citenamefont {Nishino}\ and\ \citenamefont
		{Hayakawa}(2008)}]{Nishino2008}%
	\BibitemOpen
	\bibfield  {author} {\bibinfo {author} {\bibfnamefont {T.~H.}\ \bibnamefont
			{Nishino}}\ and\ \bibinfo {author} {\bibfnamefont {H.}~\bibnamefont
			{Hayakawa}},\ }\bibfield  {title} {\bibinfo {title}
		{Fluctuation-dissipation-relation-preserving field theory of the glass
			transition in terms of fluctuating hydrodynamics},\ }\href
	{https://doi.org/10.1103/PhysRevE.78.061502} {\bibfield  {journal} {\bibinfo
			{journal} {Physical Review E}\ }\textbf {\bibinfo {volume} {78}},\ \bibinfo
		{pages} {061502} (\bibinfo {year} {2008})}\BibitemShut {NoStop}%
	\bibitem [{\citenamefont {Das}(2020)}]{Das2020}%
	\BibitemOpen
	\bibfield  {author} {\bibinfo {author} {\bibfnamefont {S.~P.}\ \bibnamefont
			{Das}},\ }\bibfield  {title} {\bibinfo {title} {Dynamic transition in a
			{{Brownian}} fluid: Role of fluctuation--dissipation constraints},\ }\href
	{https://doi.org/10.1088/1742-5468/ab684c} {\bibfield  {journal} {\bibinfo
			{journal} {Journal of Statistical Mechanics: Theory and Experiment}\ }\textbf
		{\bibinfo {volume} {2020}},\ \bibinfo {pages} {023208} (\bibinfo {year}
		{2020})}\BibitemShut {NoStop}%
	\bibitem [{\citenamefont {Andreanov}\ \emph {et~al.}(2006)\citenamefont
		{Andreanov}, \citenamefont {Biroli},\ and\ \citenamefont
		{Lef{\`e}vre}}]{Andreanov2006a}%
	\BibitemOpen
	\bibfield  {author} {\bibinfo {author} {\bibfnamefont {A.}~\bibnamefont
			{Andreanov}}, \bibinfo {author} {\bibfnamefont {G.}~\bibnamefont {Biroli}},\
		and\ \bibinfo {author} {\bibfnamefont {A.}~\bibnamefont {Lef{\`e}vre}},\
	}\bibfield  {title} {\bibinfo {title} {Dynamical field theory for
			glass-forming liquids, self-consistent resummations and time-reversal
			symmetry},\ }\href {https://doi.org/10.1088/1742-5468/2006/07/P07008}
	{\bibfield  {journal} {\bibinfo  {journal} {Journal of Statistical Mechanics:
				Theory and Experiment}\ }\textbf {\bibinfo {volume} {2006}},\ \bibinfo
		{pages} {P07008} (\bibinfo {year} {2006})}\BibitemShut {NoStop}%
	\bibitem [{\citenamefont {Kim}\ \emph {et~al.}(2014)\citenamefont {Kim},
		\citenamefont {Kawasaki}, \citenamefont {Jacquin},\ and\ \citenamefont
		{Van~Wijland}}]{Kim2014}%
	\BibitemOpen
	\bibfield  {author} {\bibinfo {author} {\bibfnamefont {B.}~\bibnamefont
			{Kim}}, \bibinfo {author} {\bibfnamefont {K.}~\bibnamefont {Kawasaki}},
		\bibinfo {author} {\bibfnamefont {H.}~\bibnamefont {Jacquin}},\ and\ \bibinfo
		{author} {\bibfnamefont {F.}~\bibnamefont {Van~Wijland}},\ }\bibfield
	{title} {\bibinfo {title} {Equilibrium dynamics of the {{Dean-Kawasaki}}
			equation: {{Mode-coupling}} theory and its extension},\ }\href
	{https://doi.org/10.1103/PhysRevE.89.012150} {\bibfield  {journal} {\bibinfo
			{journal} {Physical Review E}\ }\textbf {\bibinfo {volume} {89}},\ \bibinfo
		{pages} {012150} (\bibinfo {year} {2014})}\BibitemShut {NoStop}%
	\bibitem [{\citenamefont {Kim}\ and\ \citenamefont {Kawasaki}(2008)}]{Kim2008}%
	\BibitemOpen
	\bibfield  {author} {\bibinfo {author} {\bibfnamefont {B.}~\bibnamefont
			{Kim}}\ and\ \bibinfo {author} {\bibfnamefont {K.}~\bibnamefont {Kawasaki}},\
	}\bibfield  {title} {\bibinfo {title} {A fluctuation-dissipation
			relationship-preserving field theory for interacting {{Brownian}} particles:
			One-loop theory and mode coupling theory},\ }\href
	{https://doi.org/10.1088/1742-5468/2008/02/P02004} {\bibfield  {journal}
		{\bibinfo  {journal} {Journal of Statistical Mechanics: Theory and
				Experiment}\ }\textbf {\bibinfo {volume} {2008}},\ \bibinfo {pages} {P02004}
		(\bibinfo {year} {2008})}\BibitemShut {NoStop}%
	\bibitem [{\citenamefont {Archer}(2006)}]{Archer2006a}%
	\BibitemOpen
	\bibfield  {author} {\bibinfo {author} {\bibfnamefont {A.~J.}\ \bibnamefont
			{Archer}},\ }\bibfield  {title} {\bibinfo {title} {Dynamical density
			functional theory for dense atomic liquids},\ }\href
	{https://doi.org/10.1088/0953-8984/18/24/004} {\bibfield  {journal} {\bibinfo
			{journal} {Journal of Physics Condensed Matter}\ }\textbf {\bibinfo {volume}
			{18}},\ \bibinfo {pages} {5617} (\bibinfo {year} {2006})}\BibitemShut
	{NoStop}%
	\bibitem [{\citenamefont {{te Vrugt}}\ \emph {et~al.}(2020)\citenamefont {{te
				Vrugt}}, \citenamefont {L{\"o}wen},\ and\ \citenamefont
		{Wittkowski}}]{teVrugt2020a}%
	\BibitemOpen
	\bibfield  {author} {\bibinfo {author} {\bibfnamefont {M.}~\bibnamefont {{te
					Vrugt}}}, \bibinfo {author} {\bibfnamefont {H.}~\bibnamefont {L{\"o}wen}},\
		and\ \bibinfo {author} {\bibfnamefont {R.}~\bibnamefont {Wittkowski}},\
	}\bibfield  {title} {\bibinfo {title} {Classical dynamical density functional
			theory: From fundamentals to applications},\ }\href
	{https://doi.org/10.1080/00018732.2020.1854965} {\bibfield  {journal}
		{\bibinfo  {journal} {Adv. Phys.}\ }\textbf {\bibinfo {volume} {69}},\
		\bibinfo {pages} {121} (\bibinfo {year} {2020})}\BibitemShut {NoStop}%
	\bibitem [{\citenamefont {Marconi}\ and\ \citenamefont
		{Tarazona}(1999)}]{Marconi1999}%
	\BibitemOpen
	\bibfield  {author} {\bibinfo {author} {\bibfnamefont {U.~M.~B.}\
			\bibnamefont {Marconi}}\ and\ \bibinfo {author} {\bibfnamefont
			{P.}~\bibnamefont {Tarazona}},\ }\bibfield  {title} {\bibinfo {title}
		{Dynamic density functional theory of fluids},\ }\href
	{https://doi.org/10.1063/1.478705} {\bibfield  {journal} {\bibinfo  {journal}
			{J. Chem. Phys.}\ }\textbf {\bibinfo {volume} {110}},\ \bibinfo {pages}
		{8032} (\bibinfo {year} {1999})}\BibitemShut {NoStop}%
	\bibitem [{\citenamefont {McQuarrie}(1976)}]{McQuarrie1976}%
	\BibitemOpen
	\bibfield  {author} {\bibinfo {author} {\bibfnamefont {D.~A.}\ \bibnamefont
			{McQuarrie}},\ }\href@noop {} {\emph {\bibinfo {title} {Statistical
				{{Mechanics}}}}}\ (\bibinfo  {publisher} {{Harper and Row}},\ \bibinfo {year}
	{1976})\BibitemShut {NoStop}%
	\bibitem [{\citenamefont {Evans}(1979)}]{Evans1979}%
	\BibitemOpen
	\bibfield  {author} {\bibinfo {author} {\bibfnamefont {R.}~\bibnamefont
			{Evans}},\ }\bibfield  {title} {\bibinfo {title} {The nature of the
			liquid-vapour interface and other topics in the statistical mechanics of
			non-uniform, classical fluids},\ }\href
	{https://doi.org/10.1080/00018737900101365} {\bibfield  {journal} {\bibinfo
			{journal} {Advances in Physics}\ }\textbf {\bibinfo {volume} {28}},\ \bibinfo
		{pages} {143} (\bibinfo {year} {1979})}\BibitemShut {NoStop}%
	\bibitem [{\citenamefont {Evans}(1992)}]{Evans1992}%
	\BibitemOpen
	\bibfield  {author} {\bibinfo {author} {\bibfnamefont {R.}~\bibnamefont
			{Evans}},\ }\bibfield  {title} {\bibinfo {title} {Density functionals in the
			theory of nonuniform fluids},\ }in\ \href@noop {} {\emph {\bibinfo
			{booktitle} {Fundamentals of Inhomogeneous Fluids}}}\ (\bibinfo  {publisher}
	{Marcel Dekker: New York},\ \bibinfo {year} {1992})\ pp.\ \bibinfo {pages}
	{85--176}\BibitemShut {NoStop}%
	\bibitem [{\citenamefont {Lovett}\ \emph {et~al.}(1976)\citenamefont {Lovett},
		\citenamefont {Mou},\ and\ \citenamefont {Buff}}]{Lovett1976}%
	\BibitemOpen
	\bibfield  {author} {\bibinfo {author} {\bibfnamefont {R.}~\bibnamefont
			{Lovett}}, \bibinfo {author} {\bibfnamefont {C.~Y.}\ \bibnamefont {Mou}},\
		and\ \bibinfo {author} {\bibfnamefont {F.~P.}\ \bibnamefont {Buff}},\
	}\bibfield  {title} {\bibinfo {title} {The structure of the liquid--vapor
			interface},\ }\href {https://doi.org/10.1063/1.433110} {\bibfield  {journal}
		{\bibinfo  {journal} {The Journal of Chemical Physics}\ }\textbf {\bibinfo
			{volume} {65}},\ \bibinfo {pages} {570} (\bibinfo {year} {1976})}\BibitemShut
	{NoStop}%
	\bibitem [{\citenamefont {Yoshimori}(2005)}]{Yoshimori2005}%
	\BibitemOpen
	\bibfield  {author} {\bibinfo {author} {\bibfnamefont {A.}~\bibnamefont
			{Yoshimori}},\ }\bibfield  {title} {\bibinfo {title} {Microscopic derivation
			of time-dependent density functional methods},\ }\href
	{https://doi.org/10.1103/PhysRevE.71.031203} {\bibfield  {journal} {\bibinfo
			{journal} {Physical Review E}\ }\textbf {\bibinfo {volume} {71}},\ \bibinfo
		{pages} {031203} (\bibinfo {year} {2005})}\BibitemShut {NoStop}%
	\bibitem [{\citenamefont {Espa{\~n}ol}\ and\ \citenamefont
		{L{\"o}wen}(2009)}]{Espanol2009}%
	\BibitemOpen
	\bibfield  {author} {\bibinfo {author} {\bibfnamefont {P.}~\bibnamefont
			{Espa{\~n}ol}}\ and\ \bibinfo {author} {\bibfnamefont {H.}~\bibnamefont
			{L{\"o}wen}},\ }\bibfield  {title} {\bibinfo {title} {Derivation of dynamical
			density functional theory using the projection operator technique},\ }\href
	{https://doi.org/10.1063/1.3266943} {\bibfield  {journal} {\bibinfo
			{journal} {The Journal of Chemical Physics}\ }\textbf {\bibinfo {volume}
			{131}},\ \bibinfo {pages} {244101} (\bibinfo {year} {2009})}\BibitemShut
	{NoStop}%
	\bibitem [{\citenamefont {Schmidt}\ and\ \citenamefont
		{Brader}(2013)}]{Schmidt2013}%
	\BibitemOpen
	\bibfield  {author} {\bibinfo {author} {\bibfnamefont {M.}~\bibnamefont
			{Schmidt}}\ and\ \bibinfo {author} {\bibfnamefont {J.~M.}\ \bibnamefont
			{Brader}},\ }\bibfield  {title} {\bibinfo {title} {Power functional theory
			for {{Brownian}} dynamics},\ }\href {https://doi.org/10.1063/1.4807586}
	{\bibfield  {journal} {\bibinfo  {journal} {J. Chem. Phys.}\ }\textbf
		{\bibinfo {volume} {138}},\ \bibinfo {pages} {214101} (\bibinfo {year}
		{2013})}\BibitemShut {NoStop}%
	\bibitem [{\citenamefont {Lutsko}\ and\ \citenamefont
		{Oettel}(2021)}]{Lutsko2021}%
	\BibitemOpen
	\bibfield  {author} {\bibinfo {author} {\bibfnamefont {J.~F.}\ \bibnamefont
			{Lutsko}}\ and\ \bibinfo {author} {\bibfnamefont {M.}~\bibnamefont
			{Oettel}},\ }\bibfield  {title} {\bibinfo {title} {Reconsidering power
			functional theory},\ }\href {https://doi.org/10.1063/5.0055288} {\bibfield
		{journal} {\bibinfo  {journal} {J. Chem. Phys.}\ }\textbf {\bibinfo {volume}
			{155}},\ \bibinfo {pages} {094901} (\bibinfo {year} {2021})}\BibitemShut
	{NoStop}%
	\bibitem [{\citenamefont {Schmidt}(2022)}]{Schmidt2022}%
	\BibitemOpen
	\bibfield  {author} {\bibinfo {author} {\bibfnamefont {M.}~\bibnamefont
			{Schmidt}},\ }\bibfield  {title} {\bibinfo {title} {Power functional theory
			for many-body dynamics},\ }\href
	{https://doi.org/10.1103/RevModPhys.94.015007} {\bibfield  {journal}
		{\bibinfo  {journal} {Reviews of Modern Physics}\ }\textbf {\bibinfo {volume}
			{94}},\ \bibinfo {pages} {015007} (\bibinfo {year} {2022})}\BibitemShut
	{NoStop}%
	\bibitem [{\citenamefont {De~Las~Heras}\ \emph {et~al.}(2023)\citenamefont
		{De~Las~Heras}, \citenamefont {Zimmermann}, \citenamefont {Samm{\"u}ller},
		\citenamefont {Hermann},\ and\ \citenamefont {Schmidt}}]{DeLasHeras2023}%
	\BibitemOpen
	\bibfield  {author} {\bibinfo {author} {\bibfnamefont {D.}~\bibnamefont
			{De~Las~Heras}}, \bibinfo {author} {\bibfnamefont {T.}~\bibnamefont
			{Zimmermann}}, \bibinfo {author} {\bibfnamefont {F.}~\bibnamefont
			{Samm{\"u}ller}}, \bibinfo {author} {\bibfnamefont {S.}~\bibnamefont
			{Hermann}},\ and\ \bibinfo {author} {\bibfnamefont {M.}~\bibnamefont
			{Schmidt}},\ }\bibfield  {title} {\bibinfo {title} {Perspective: {{How}} to
			overcome dynamical density functional theory},\ }\href
	{https://doi.org/10.1088/1361-648X/accb33} {\bibfield  {journal} {\bibinfo
			{journal} {Journal of Physics: Condensed Matter}\ }\textbf {\bibinfo {volume}
			{35}},\ \bibinfo {pages} {271501} (\bibinfo {year} {2023})}\BibitemShut
	{NoStop}%
	\bibitem [{\citenamefont {Kawasaki}(2006)}]{Kawasaki2006}%
	\BibitemOpen
	\bibfield  {author} {\bibinfo {author} {\bibfnamefont {K.}~\bibnamefont
			{Kawasaki}},\ }\bibfield  {title} {\bibinfo {title} {Interpolation of
			stochastic and deterministic reduced dynamics},\ }\href
	{https://doi.org/10.1016/j.physa.2005.08.009} {\bibfield  {journal} {\bibinfo
			{journal} {Physica A: Statistical Mechanics and its Applications}\ }\textbf
		{\bibinfo {volume} {362}},\ \bibinfo {pages} {249} (\bibinfo {year}
		{2006})}\BibitemShut {NoStop}%
	\bibitem [{\citenamefont {Chavanis}(2011)}]{Chavanis2011b}%
	\BibitemOpen
	\bibfield  {author} {\bibinfo {author} {\bibfnamefont {P.-H.}\ \bibnamefont
			{Chavanis}},\ }\bibfield  {title} {\bibinfo {title} {Brownian particles with
			long- and short-range interactions},\ }\href
	{https://doi.org/10.1016/j.physa.2010.12.018} {\bibfield  {journal} {\bibinfo
			{journal} {Physica A: Statistical Mechanics and its Applications}\ }\textbf
		{\bibinfo {volume} {390}},\ \bibinfo {pages} {1546} (\bibinfo {year}
		{2011})}\BibitemShut {NoStop}%
	\bibitem [{\citenamefont {Chavanis}(2019)}]{Chavanis2019}%
	\BibitemOpen
	\bibfield  {author} {\bibinfo {author} {\bibfnamefont {P.-H.}\ \bibnamefont
			{Chavanis}},\ }\bibfield  {title} {\bibinfo {title} {The {{Generalized
					Stochastic Smoluchowski Equation}}},\ }\href
	{https://doi.org/10.3390/e21101006} {\bibfield  {journal} {\bibinfo
			{journal} {Entropy. An International and Interdisciplinary Journal of Entropy
				and Information Studies}\ }\textbf {\bibinfo {volume} {21}},\ \bibinfo
		{pages} {1006} (\bibinfo {year} {2019})}\BibitemShut {NoStop}%
	\bibitem [{\citenamefont {Konarovskyi}\ \emph {et~al.}(2019)\citenamefont
		{Konarovskyi}, \citenamefont {Lehmann},\ and\ \citenamefont {{von
				Renesse}}}]{Konarovskyi2019}%
	\BibitemOpen
	\bibfield  {author} {\bibinfo {author} {\bibfnamefont {V.}~\bibnamefont
			{Konarovskyi}}, \bibinfo {author} {\bibfnamefont {T.}~\bibnamefont
			{Lehmann}},\ and\ \bibinfo {author} {\bibfnamefont {M.~K.}\ \bibnamefont
			{{von Renesse}}},\ }\bibfield  {title} {\bibinfo {title} {Dean-kawasaki
			dynamics: {{Ill-posedness}} vs. triviality},\ }\href
	{https://doi.org/10.1214/19-ECP208} {\bibfield  {journal} {\bibinfo
			{journal} {Electron. Commun. Probab.}\ }\textbf {\bibinfo {volume} {24}},\
		\bibinfo {pages} {1} (\bibinfo {year} {2019})}\BibitemShut {NoStop}%
	\bibitem [{\citenamefont {Konarovskyi}\ \emph {et~al.}(2020)\citenamefont
		{Konarovskyi}, \citenamefont {Lehmann},\ and\ \citenamefont
		{Von~Renesse}}]{Konarovskyi2020}%
	\BibitemOpen
	\bibfield  {author} {\bibinfo {author} {\bibfnamefont {V.}~\bibnamefont
			{Konarovskyi}}, \bibinfo {author} {\bibfnamefont {T.}~\bibnamefont
			{Lehmann}},\ and\ \bibinfo {author} {\bibfnamefont {M.}~\bibnamefont
			{Von~Renesse}},\ }\bibfield  {title} {\bibinfo {title} {On
			{{Dean}}--{{Kawasaki Dynamics}} with {{Smooth Drift Potential}}},\ }\href
	{https://doi.org/10.1007/s10955-019-02449-3} {\bibfield  {journal} {\bibinfo
			{journal} {Journal of Statistical Physics}\ }\textbf {\bibinfo {volume}
			{178}},\ \bibinfo {pages} {666} (\bibinfo {year} {2020})}\BibitemShut
	{NoStop}%
	\bibitem [{\citenamefont {Konarovskyi}\ and\ \citenamefont
		{Müller}(2024)}]{Konarovskyi2024}%
	\BibitemOpen
	\bibfield  {author} {\bibinfo {author} {\bibfnamefont {V.}~\bibnamefont
			{Konarovskyi}}\ and\ \bibinfo {author} {\bibfnamefont {F.}~\bibnamefont
			{Müller}},\ }\bibfield  {title} {\bibinfo {title} {Dean–{Kawasaki}
			equation with initial condition in the space of positive distributions},\
	}\href {https://doi.org/10.1007/s00028-024-01018-w} {\bibfield  {journal}
		{\bibinfo  {journal} {Journal of Evolution Equations}\ }\textbf {\bibinfo
			{volume} {24}},\ \bibinfo {pages} {92} (\bibinfo {year} {2024})}\BibitemShut
	{NoStop}%
	\bibitem [{\citenamefont {Wang}\ \emph {et~al.}(2024)\citenamefont {Wang},
		\citenamefont {Wu},\ and\ \citenamefont {Zhang}}]{Wang2024}%
	\BibitemOpen
	\bibfield  {author} {\bibinfo {author} {\bibfnamefont {L.}~\bibnamefont
			{Wang}}, \bibinfo {author} {\bibfnamefont {Z.}~\bibnamefont {Wu}},\ and\
		\bibinfo {author} {\bibfnamefont {R.}~\bibnamefont {Zhang}},\ }\href
	{https://arxiv.org/abs/2207.12774} {\bibinfo {title} {Dean-{K}awasaki
			equation with singular interactions and applications to dynamical
			{I}sing-{K}ac model}} (\bibinfo {year} {2024}),\ \Eprint
	{https://arxiv.org/abs/2207.12774} {arXiv:2207.12774} \BibitemShut {NoStop}%
	\bibitem [{\citenamefont {M{\"u}ller}\ \emph {et~al.}(2025)\citenamefont
		{M{\"u}ller}, \citenamefont {von Renesse},\ and\ \citenamefont
		{Zimmer}}]{Müller2025}%
	\BibitemOpen
	\bibfield  {author} {\bibinfo {author} {\bibfnamefont {F.}~\bibnamefont
			{M{\"u}ller}}, \bibinfo {author} {\bibfnamefont {M.}~\bibnamefont {von
				Renesse}},\ and\ \bibinfo {author} {\bibfnamefont {J.}~\bibnamefont
			{Zimmer}},\ }\href {https://doi.org/10.48550/arXiv.2411.14334} {\bibinfo
		{title} {Well-{{Posedness}} for {{Dean-Kawasaki Models}} of
			{{Vlasov-Fokker-Planck Type}}}} (\bibinfo {year} {2025}),\ \Eprint
	{https://arxiv.org/abs/2411.14334} {arXiv:2411.14334 [math]} \BibitemShut
	{NoStop}%
	\bibitem [{\citenamefont {Violeau}(2012)}]{Violeau2012}%
	\BibitemOpen
	\bibfield  {author} {\bibinfo {author} {\bibfnamefont {D.}~\bibnamefont
			{Violeau}},\ }\href
	{https://doi.org/10.1093/acprof:oso/9780199655526.001.0001} {\emph {\bibinfo
			{title} {Fluid {{Mechanics}} and the {{SPH Method}}: {{Theory}} and
				{{Applications}}}}}\ (\bibinfo  {publisher} {Oxford University Press},\
	\bibinfo {year} {2012})\BibitemShut {NoStop}%
	\bibitem [{\citenamefont {Cornalba}\ \emph {et~al.}(2019)\citenamefont
		{Cornalba}, \citenamefont {Shardlow},\ and\ \citenamefont
		{Zimmer}}]{Cornalba2019}%
	\BibitemOpen
	\bibfield  {author} {\bibinfo {author} {\bibfnamefont {F.}~\bibnamefont
			{Cornalba}}, \bibinfo {author} {\bibfnamefont {T.}~\bibnamefont {Shardlow}},\
		and\ \bibinfo {author} {\bibfnamefont {J.}~\bibnamefont {Zimmer}},\
	}\bibfield  {title} {\bibinfo {title} {A {{Regularized Dean--Kawasaki
					Model}}: {{Derivation}} and {{Analysis}}},\ }\href
	{https://doi.org/10.1137/18M1172697} {\bibfield  {journal} {\bibinfo
			{journal} {SIAM Journal on Mathematical Analysis}\ }\textbf {\bibinfo
			{volume} {51}},\ \bibinfo {pages} {1137} (\bibinfo {year}
		{2019})}\BibitemShut {NoStop}%
	\bibitem [{\citenamefont {Cornalba}\ \emph {et~al.}(2020)\citenamefont
		{Cornalba}, \citenamefont {Shardlow},\ and\ \citenamefont
		{Zimmer}}]{Cornalba2020}%
	\BibitemOpen
	\bibfield  {author} {\bibinfo {author} {\bibfnamefont {F.}~\bibnamefont
			{Cornalba}}, \bibinfo {author} {\bibfnamefont {T.}~\bibnamefont {Shardlow}},\
		and\ \bibinfo {author} {\bibfnamefont {J.}~\bibnamefont {Zimmer}},\
	}\bibfield  {title} {\bibinfo {title} {From weakly interacting particles to a
			regularised {{Dean}}--{{Kawasaki}} model},\ }\href
	{https://doi.org/10.1088/1361-6544/ab5174} {\bibfield  {journal} {\bibinfo
			{journal} {Nonlinearity}\ }\textbf {\bibinfo {volume} {33}},\ \bibinfo
		{pages} {864} (\bibinfo {year} {2020})}\BibitemShut {NoStop}%
	\bibitem [{\citenamefont {Cornalba}\ \emph {et~al.}(2021)\citenamefont
		{Cornalba}, \citenamefont {Shardlow},\ and\ \citenamefont
		{Zimmer}}]{Cornalba2021}%
	\BibitemOpen
	\bibfield  {author} {\bibinfo {author} {\bibfnamefont {F.}~\bibnamefont
			{Cornalba}}, \bibinfo {author} {\bibfnamefont {T.}~\bibnamefont {Shardlow}},\
		and\ \bibinfo {author} {\bibfnamefont {J.}~\bibnamefont {Zimmer}},\
	}\bibfield  {title} {\bibinfo {title} {Well-posedness for a regularised
			inertial {{Dean}}--{{Kawasaki}} model for slender particles in several space
			dimensions},\ }\href {https://doi.org/10.1016/j.jde.2021.02.048} {\bibfield
		{journal} {\bibinfo  {journal} {Journal of Differential Equations}\ }\textbf
		{\bibinfo {volume} {284}},\ \bibinfo {pages} {253} (\bibinfo {year}
		{2021})}\BibitemShut {NoStop}%
	\bibitem [{\citenamefont {Cornalba}\ \emph {et~al.}(2023)\citenamefont
		{Cornalba}, \citenamefont {Fischer}, \citenamefont {Ingmanns},\ and\
		\citenamefont {Raithel}}]{Cornalba2023}%
	\BibitemOpen
	\bibfield  {author} {\bibinfo {author} {\bibfnamefont {F.}~\bibnamefont
			{Cornalba}}, \bibinfo {author} {\bibfnamefont {J.}~\bibnamefont {Fischer}},
		\bibinfo {author} {\bibfnamefont {J.}~\bibnamefont {Ingmanns}},\ and\
		\bibinfo {author} {\bibfnamefont {C.}~\bibnamefont {Raithel}},\ }\href
	{https://arxiv.org/abs/2303.00429} {\bibinfo {title} {Density fluctuations in
			weakly interacting particle systems via the {D}ean-{K}awasaki equation}}
	(\bibinfo {year} {2023}),\ \Eprint {https://arxiv.org/abs/2303.00429}
	{arXiv:2303.00429} \BibitemShut {NoStop}%
	\bibitem [{\citenamefont {Nakamura}\ and\ \citenamefont
		{Yoshimori}(2009)}]{Nakamura2009}%
	\BibitemOpen
	\bibfield  {author} {\bibinfo {author} {\bibfnamefont {T.}~\bibnamefont
			{Nakamura}}\ and\ \bibinfo {author} {\bibfnamefont {A.}~\bibnamefont
			{Yoshimori}},\ }\bibfield  {title} {\bibinfo {title} {Derivation of the
			nonlinear fluctuating hydrodynamic equation from the underdamped {{Langevin}}
			equation},\ }\href {https://doi.org/10.1088/1751-8113/42/6/065001} {\bibfield
		{journal} {\bibinfo  {journal} {Journal of Physics A: Mathematical and
				Theoretical}\ }\textbf {\bibinfo {volume} {42}},\ \bibinfo {pages} {065001}
		(\bibinfo {year} {2009})}\BibitemShut {NoStop}%
	\bibitem [{\citenamefont {Das}\ and\ \citenamefont
		{Yoshimori}(2013)}]{Das2013}%
	\BibitemOpen
	\bibfield  {author} {\bibinfo {author} {\bibfnamefont {S.~P.}\ \bibnamefont
			{Das}}\ and\ \bibinfo {author} {\bibfnamefont {A.}~\bibnamefont
			{Yoshimori}},\ }\bibfield  {title} {\bibinfo {title} {Coarse-grained forms
			for equations describing the microscopic motion of particles in a fluid},\
	}\href {https://doi.org/10.1103/PhysRevE.88.043008} {\bibfield  {journal}
		{\bibinfo  {journal} {Physical Review E}\ }\textbf {\bibinfo {volume} {88}},\
		\bibinfo {pages} {043008} (\bibinfo {year} {2013})}\BibitemShut {NoStop}%
	\bibitem [{\citenamefont {L{\'o}pez}\ and\ \citenamefont
		{Puglisi}(2004)}]{Lopez2004}%
	\BibitemOpen
	\bibfield  {author} {\bibinfo {author} {\bibfnamefont {C.}~\bibnamefont
			{L{\'o}pez}}\ and\ \bibinfo {author} {\bibfnamefont {A.}~\bibnamefont
			{Puglisi}},\ }\bibfield  {title} {\bibinfo {title} {Continuum description of
			finite-size particles advected by external flows: {{The}} effect of
			collisions},\ }\href {https://doi.org/10.1103/PhysRevE.69.046306} {\bibfield
		{journal} {\bibinfo  {journal} {Physical Review E}\ }\textbf {\bibinfo
			{volume} {69}},\ \bibinfo {pages} {046306} (\bibinfo {year}
		{2004})}\BibitemShut {NoStop}%
	\bibitem [{\citenamefont {Donev}\ and\ \citenamefont
		{{Vanden-Eijnden}}(2014)}]{Donev2014}%
	\BibitemOpen
	\bibfield  {author} {\bibinfo {author} {\bibfnamefont {A.}~\bibnamefont
			{Donev}}\ and\ \bibinfo {author} {\bibfnamefont {E.}~\bibnamefont
			{{Vanden-Eijnden}}},\ }\bibfield  {title} {\bibinfo {title} {Dynamic density
			functional theory with hydrodynamic interactions and fluctuations},\ }\href
	{https://doi.org/10.1063/1.4883520} {\bibfield  {journal} {\bibinfo
			{journal} {The Journal of Chemical Physics}\ }\textbf {\bibinfo {volume}
			{140}},\ \bibinfo {pages} {234115} (\bibinfo {year} {2014})}\BibitemShut
	{NoStop}%
	\bibitem [{\citenamefont {Pel{\'a}ez}\ \emph {et~al.}(2018)\citenamefont
		{Pel{\'a}ez}, \citenamefont {Usabiaga}, \citenamefont {Panzuela},
		\citenamefont {Xiao}, \citenamefont {{Delgado-Buscalioni}},\ and\
		\citenamefont {Donev}}]{Pelaez2018}%
	\BibitemOpen
	\bibfield  {author} {\bibinfo {author} {\bibfnamefont {R.~P.}\ \bibnamefont
			{Pel{\'a}ez}}, \bibinfo {author} {\bibfnamefont {F.~B.}\ \bibnamefont
			{Usabiaga}}, \bibinfo {author} {\bibfnamefont {S.}~\bibnamefont {Panzuela}},
		\bibinfo {author} {\bibfnamefont {Q.}~\bibnamefont {Xiao}}, \bibinfo {author}
		{\bibfnamefont {R.}~\bibnamefont {{Delgado-Buscalioni}}},\ and\ \bibinfo
		{author} {\bibfnamefont {A.}~\bibnamefont {Donev}},\ }\bibfield  {title}
	{\bibinfo {title} {Hydrodynamic fluctuations in quasi-two dimensional
			diffusion},\ }\href {https://doi.org/10.1088/1742-5468/aac2fb} {\bibfield
		{journal} {\bibinfo  {journal} {Journal of Statistical Mechanics: Theory and
				Experiment}\ }\textbf {\bibinfo {volume} {2018}},\ \bibinfo {pages} {063207}
		(\bibinfo {year} {2018})}\BibitemShut {NoStop}%
	\bibitem [{\citenamefont {Ermak}\ and\ \citenamefont
		{McCammon}(1978)}]{Ermak1978}%
	\BibitemOpen
	\bibfield  {author} {\bibinfo {author} {\bibfnamefont {D.~L.}\ \bibnamefont
			{Ermak}}\ and\ \bibinfo {author} {\bibfnamefont {J.~A.}\ \bibnamefont
			{McCammon}},\ }\bibfield  {title} {\bibinfo {title} {Brownian dynamics with
			hydrodynamic interactions},\ }\href {https://doi.org/10.1063/1.436761}
	{\bibfield  {journal} {\bibinfo  {journal} {The Journal of Chemical Physics}\
		}\textbf {\bibinfo {volume} {69}},\ \bibinfo {pages} {1352} (\bibinfo {year}
		{1978})}\BibitemShut {NoStop}%
	\bibitem [{\citenamefont {Cugliandolo}\ \emph {et~al.}(2015)\citenamefont
		{Cugliandolo}, \citenamefont {D{\'e}jardin}, \citenamefont {Lozano},\ and\
		\citenamefont {Van~Wijland}}]{Cugliandolo2015b}%
	\BibitemOpen
	\bibfield  {author} {\bibinfo {author} {\bibfnamefont {L.~F.}\ \bibnamefont
			{Cugliandolo}}, \bibinfo {author} {\bibfnamefont {P.-M.}\ \bibnamefont
			{D{\'e}jardin}}, \bibinfo {author} {\bibfnamefont {G.~S.}\ \bibnamefont
			{Lozano}},\ and\ \bibinfo {author} {\bibfnamefont {F.}~\bibnamefont
			{Van~Wijland}},\ }\bibfield  {title} {\bibinfo {title} {Stochastic dynamics
			of collective modes for {{Brownian}} dipoles},\ }\href
	{https://doi.org/10.1103/PhysRevE.91.032139} {\bibfield  {journal} {\bibinfo
			{journal} {Physical Review E}\ }\textbf {\bibinfo {volume} {91}},\ \bibinfo
		{pages} {032139} (\bibinfo {year} {2015})}\BibitemShut {NoStop}%
	\bibitem [{\citenamefont {Illien}\ \emph {et~al.}(2024)\citenamefont {Illien},
		\citenamefont {Carof},\ and\ \citenamefont {Rotenberg}}]{Illien2024a}%
	\BibitemOpen
	\bibfield  {author} {\bibinfo {author} {\bibfnamefont {P.}~\bibnamefont
			{Illien}}, \bibinfo {author} {\bibfnamefont {A.}~\bibnamefont {Carof}},\ and\
		\bibinfo {author} {\bibfnamefont {B.}~\bibnamefont {Rotenberg}},\ }\href
	{https://arxiv.org/abs/2407.17232} {\bibinfo {title} {Stochastic density
			functional theory for ions in a polar solvent}} (\bibinfo {year} {2024}),\
	\Eprint {https://arxiv.org/abs/2407.17232} {arXiv:2407.17232} \BibitemShut
	{NoStop}%
	\bibitem [{\citenamefont {D{\'e}mery}\ and\ \citenamefont
		{Dean}(2016)}]{Demery2016}%
	\BibitemOpen
	\bibfield  {author} {\bibinfo {author} {\bibfnamefont {V.}~\bibnamefont
			{D{\'e}mery}}\ and\ \bibinfo {author} {\bibfnamefont {D.~S.}\ \bibnamefont
			{Dean}},\ }\bibfield  {title} {\bibinfo {title} {The conductivity of strong
			electrolytes from stochastic density functional theory},\ }\href
	{https://doi.org/10.1088/1742-5468/2016/02/023106} {\bibfield  {journal}
		{\bibinfo  {journal} {J. Stat. Mech.}\ ,\ \bibinfo {pages} {023106}}
		(\bibinfo {year} {2016})}\BibitemShut {NoStop}%
	\bibitem [{\citenamefont {Poncet}\ \emph {et~al.}(2017)\citenamefont {Poncet},
		\citenamefont {B{\'e}nichou}, \citenamefont {D{\'e}mery},\ and\ \citenamefont
		{Oshanin}}]{Poncet2017}%
	\BibitemOpen
	\bibfield  {author} {\bibinfo {author} {\bibfnamefont {A.}~\bibnamefont
			{Poncet}}, \bibinfo {author} {\bibfnamefont {O.}~\bibnamefont
			{B{\'e}nichou}}, \bibinfo {author} {\bibfnamefont {V.}~\bibnamefont
			{D{\'e}mery}},\ and\ \bibinfo {author} {\bibfnamefont {G.}~\bibnamefont
			{Oshanin}},\ }\bibfield  {title} {\bibinfo {title} {Universal long ranged
			correlations in driven binary mixtures},\ }\href
	{https://doi.org/10.1103/PhysRevLett.118.118002} {\bibfield  {journal}
		{\bibinfo  {journal} {Phys. Rev. Lett.}\ }\textbf {\bibinfo {volume} {118}},\
		\bibinfo {pages} {118002} (\bibinfo {year} {2017})}\BibitemShut {NoStop}%
	\bibitem [{\citenamefont {Jardat}\ \emph {et~al.}(2022)\citenamefont {Jardat},
		\citenamefont {Dahirel},\ and\ \citenamefont {Illien}}]{Jardat2022}%
	\BibitemOpen
	\bibfield  {author} {\bibinfo {author} {\bibfnamefont {M.}~\bibnamefont
			{Jardat}}, \bibinfo {author} {\bibfnamefont {V.}~\bibnamefont {Dahirel}},\
		and\ \bibinfo {author} {\bibfnamefont {P.}~\bibnamefont {Illien}},\
	}\bibfield  {title} {\bibinfo {title} {Diffusion of a tracer in a dense
			mixture of soft particles connected to different thermostats},\ }\href
	{https://doi.org/10.1103/PhysRevE.106.064608} {\bibfield  {journal} {\bibinfo
			{journal} {Phys. Rev. E}\ }\textbf {\bibinfo {volume} {106}},\ \bibinfo
		{pages} {064608} (\bibinfo {year} {2022})}\BibitemShut {NoStop}%
	\bibitem [{\citenamefont {Benois}\ \emph {et~al.}(2023)\citenamefont {Benois},
		\citenamefont {Jardat}, \citenamefont {Dahirel}, \citenamefont {D{\'e}mery},
		\citenamefont {{Agudo-Canalejo}}, \citenamefont {Golestanian},\ and\
		\citenamefont {Illien}}]{Benois2023}%
	\BibitemOpen
	\bibfield  {author} {\bibinfo {author} {\bibfnamefont {A.}~\bibnamefont
			{Benois}}, \bibinfo {author} {\bibfnamefont {M.}~\bibnamefont {Jardat}},
		\bibinfo {author} {\bibfnamefont {V.}~\bibnamefont {Dahirel}}, \bibinfo
		{author} {\bibfnamefont {V.}~\bibnamefont {D{\'e}mery}}, \bibinfo {author}
		{\bibfnamefont {J.}~\bibnamefont {{Agudo-Canalejo}}}, \bibinfo {author}
		{\bibfnamefont {R.}~\bibnamefont {Golestanian}},\ and\ \bibinfo {author}
		{\bibfnamefont {P.}~\bibnamefont {Illien}},\ }\bibfield  {title} {\bibinfo
		{title} {Enhanced diffusion of tracer particles in nonreciprocal mixtures},\
	}\href {https://doi.org/10.1103/PhysRevE.108.054606} {\bibfield  {journal}
		{\bibinfo  {journal} {Physical Review E}\ }\textbf {\bibinfo {volume}
			{108}},\ \bibinfo {pages} {054606} (\bibinfo {year} {2023})}\BibitemShut
	{NoStop}%
	\bibitem [{\citenamefont {Fodor}\ \emph {et~al.}(2018)\citenamefont {Fodor},
		\citenamefont {Hayakawa}, \citenamefont {Tailleur},\ and\ \citenamefont
		{Van~Wijland}}]{Fodor2018b}%
	\BibitemOpen
	\bibfield  {author} {\bibinfo {author} {\bibfnamefont {{\'E}.}~\bibnamefont
			{Fodor}}, \bibinfo {author} {\bibfnamefont {H.}~\bibnamefont {Hayakawa}},
		\bibinfo {author} {\bibfnamefont {J.}~\bibnamefont {Tailleur}},\ and\
		\bibinfo {author} {\bibfnamefont {F.}~\bibnamefont {Van~Wijland}},\
	}\bibfield  {title} {\bibinfo {title} {Non-{{Gaussian}} noise without memory
			in active matter},\ }\href {https://doi.org/10.1103/PhysRevE.98.062610}
	{\bibfield  {journal} {\bibinfo  {journal} {Physical Review E}\ }\textbf
		{\bibinfo {volume} {98}},\ \bibinfo {pages} {062610} (\bibinfo {year}
		{2018})}\BibitemShut {NoStop}%
	\bibitem [{\citenamefont {Spinney}\ and\ \citenamefont
		{Morris}(2024)}]{Spinney2024}%
	\BibitemOpen
	\bibfield  {author} {\bibinfo {author} {\bibfnamefont {R.~E.}\ \bibnamefont
			{Spinney}}\ and\ \bibinfo {author} {\bibfnamefont {R.~G.}\ \bibnamefont
			{Morris}},\ }\href {https://arxiv.org/abs/2404.02487} {\bibinfo {title} {A
			{D}ean-{K}awasaki equation for reaction diffusion systems driven by {P}oisson
			noise}} (\bibinfo {year} {2024}),\ \Eprint {https://arxiv.org/abs/2404.02487}
	{arXiv:2404.02487} \BibitemShut {NoStop}%
	\bibitem [{\citenamefont {Bressloff}(2024{\natexlab{a}})}]{Bressloff2024c}%
	\BibitemOpen
	\bibfield  {author} {\bibinfo {author} {\bibfnamefont {P.~C.}\ \bibnamefont
			{Bressloff}},\ }\bibfield  {title} {\bibinfo {title} {Global density
			equations for a population of actively switching particles},\ }\href
	{https://doi.org/10.1088/1751-8121/ad2431} {\bibfield  {journal} {\bibinfo
			{journal} {Journal of Physics A: Mathematical and Theoretical}\ }\textbf
		{\bibinfo {volume} {57}},\ \bibinfo {pages} {085001} (\bibinfo {year}
		{2024}{\natexlab{a}})}\BibitemShut {NoStop}%
	\bibitem [{\citenamefont {Bressloff}(2024{\natexlab{b}})}]{Bressloff2024}%
	\BibitemOpen
	\bibfield  {author} {\bibinfo {author} {\bibfnamefont {P.~C.}\ \bibnamefont
			{Bressloff}},\ }\bibfield  {title} {\bibinfo {title} {Global density
			equations for interacting particle systems with stochastic resetting:
			{{From}} overdamped {{Brownian}} motion to phase synchronization},\ }\href
	{https://doi.org/10.1063/5.0196626} {\bibfield  {journal} {\bibinfo
			{journal} {Chaos: An Interdisciplinary Journal of Nonlinear Science}\
		}\textbf {\bibinfo {volume} {34}},\ \bibinfo {pages} {043101} (\bibinfo
		{year} {2024}{\natexlab{b}})}\BibitemShut {NoStop}%
	\bibitem [{\citenamefont {Bressloff}(2024{\natexlab{c}})}]{Bressloff2024b}%
	\BibitemOpen
	\bibfield  {author} {\bibinfo {author} {\bibfnamefont {P.~C.}\ \bibnamefont
			{Bressloff}},\ }\bibfield  {title} {\bibinfo {title} {A generalized
			{Dean}–{Kawasaki} equation for an interacting {Brownian} gas in a partially
			absorbing medium},\ }\href {https://doi.org/10.1098/rspa.2023.0915}
	{\bibfield  {journal} {\bibinfo  {journal} {Proceedings of the Royal Society
				A: Mathematical, Physical and Engineering Sciences}\ }\textbf {\bibinfo
			{volume} {480}},\ \bibinfo {pages} {20230915} (\bibinfo {year}
		{2024}{\natexlab{c}})}\BibitemShut {NoStop}%
	\bibitem [{\citenamefont {Kim}\ \emph {et~al.}(2020)\citenamefont {Kim},
		\citenamefont {Fuchs},\ and\ \citenamefont {Krakoviack}}]{Kim2020}%
	\BibitemOpen
	\bibfield  {author} {\bibinfo {author} {\bibfnamefont {B.}~\bibnamefont
			{Kim}}, \bibinfo {author} {\bibfnamefont {M.}~\bibnamefont {Fuchs}},\ and\
		\bibinfo {author} {\bibfnamefont {V.}~\bibnamefont {Krakoviack}},\ }\bibfield
	{title} {\bibinfo {title} {Dynamics of a noninteracting colloidal fluid in a
			quenched {{Gaussian}} random potential: A time-reversal-symmetry-preserving
			field-theoretic approach},\ }\href {https://doi.org/10.1088/1742-5468/ab632e}
	{\bibfield  {journal} {\bibinfo  {journal} {J. Stat. Mech.}\ }\textbf
		{\bibinfo {volume} {2020}},\ \bibinfo {pages} {023301} (\bibinfo {year}
		{2020})}\BibitemShut {NoStop}%
	\bibitem [{\citenamefont {Illien}\ and\ \citenamefont
		{Carof}(2025)}]{Illien2025}%
	\BibitemOpen
	\bibfield  {author} {\bibinfo {author} {\bibfnamefont {P.}~\bibnamefont
			{Illien}}\ and\ \bibinfo {author} {\bibfnamefont {A.}~\bibnamefont {Carof}},\
	}\href {https://doi.org/10.48550/arXiv.2501.16206} {\bibinfo {title}
		{Non-{{Gaussian}} density fluctuations in the {{Dean-Kawasaki}} equation}}
	(\bibinfo {year} {2025}),\ \Eprint {https://arxiv.org/abs/2501.16206}
	{arXiv:2501.16206} \BibitemShut {NoStop}%
	\bibitem [{\citenamefont {Dean}\ and\ \citenamefont
		{Podgornik}(2014)}]{Dean2014a}%
	\BibitemOpen
	\bibfield  {author} {\bibinfo {author} {\bibfnamefont {D.~S.}\ \bibnamefont
			{Dean}}\ and\ \bibinfo {author} {\bibfnamefont {R.}~\bibnamefont
			{Podgornik}},\ }\bibfield  {title} {\bibinfo {title} {Relaxation of the
			thermal {{Casimir}} force between net neutral plates containing {{Brownian}}
			charges},\ }\href {https://doi.org/10.1103/PhysRevE.89.032117} {\bibfield
		{journal} {\bibinfo  {journal} {Physical Review E}\ }\textbf {\bibinfo
			{volume} {89}},\ \bibinfo {pages} {032117} (\bibinfo {year}
		{2014})}\BibitemShut {NoStop}%
	\bibitem [{\citenamefont {D{\'e}mery}\ \emph {et~al.}(2014)\citenamefont
		{D{\'e}mery}, \citenamefont {B{\'e}nichou},\ and\ \citenamefont
		{Jacquin}}]{Demery2014}%
	\BibitemOpen
	\bibfield  {author} {\bibinfo {author} {\bibfnamefont {V.}~\bibnamefont
			{D{\'e}mery}}, \bibinfo {author} {\bibfnamefont {O.}~\bibnamefont
			{B{\'e}nichou}},\ and\ \bibinfo {author} {\bibfnamefont {H.}~\bibnamefont
			{Jacquin}},\ }\bibfield  {title} {\bibinfo {title} {Generalized {{Langevin}}
			equations for a driven tracer in dense soft colloids: Construction and
			applications},\ }\href {https://doi.org/10.1088/1367-2630/16/5/053032}
	{\bibfield  {journal} {\bibinfo  {journal} {New Journal of Physics}\ }\textbf
		{\bibinfo {volume} {16}},\ \bibinfo {pages} {053032} (\bibinfo {year}
		{2014})}\BibitemShut {NoStop}%
	\bibitem [{\citenamefont {Andersen}\ and\ \citenamefont
		{Chandler}(1970)}]{Andersen1970}%
	\BibitemOpen
	\bibfield  {author} {\bibinfo {author} {\bibfnamefont {H.~C.}\ \bibnamefont
			{Andersen}}\ and\ \bibinfo {author} {\bibfnamefont {D.}~\bibnamefont
			{Chandler}},\ }\bibfield  {title} {\bibinfo {title} {Mode {{Expansion}} in
			{{Equilibrium Statistical Mechanics}}. {{I}}. {{General Theory}} and
			{{Application}} to the {{Classical Electron Gas}}},\ }\href
	{https://doi.org/10.1063/1.1674024} {\bibfield  {journal} {\bibinfo
			{journal} {The Journal of Chemical Physics}\ }\textbf {\bibinfo {volume}
			{53}},\ \bibinfo {pages} {547} (\bibinfo {year} {1970})}\BibitemShut
	{NoStop}%
	\bibitem [{\citenamefont {Wheeler}\ and\ \citenamefont
		{Chandler}(1971)}]{Wheeler1971}%
	\BibitemOpen
	\bibfield  {author} {\bibinfo {author} {\bibfnamefont {J.~C.}\ \bibnamefont
			{Wheeler}}\ and\ \bibinfo {author} {\bibfnamefont {D.}~\bibnamefont
			{Chandler}},\ }\bibfield  {title} {\bibinfo {title} {Catastrophe in the
			{{Random-Phase Approximation}}: {{Critique}} of a {{Theory}} of {{Phase
					Transitions}}},\ }\href {https://doi.org/10.1063/1.1676291} {\bibfield
		{journal} {\bibinfo  {journal} {The Journal of Chemical Physics}\ }\textbf
		{\bibinfo {volume} {55}},\ \bibinfo {pages} {1645} (\bibinfo {year}
		{1971})}\BibitemShut {NoStop}%
	\bibitem [{\citenamefont {Likos}\ \emph {et~al.}(2001)\citenamefont {Likos},
		\citenamefont {Lang}, \citenamefont {Watzlawek},\ and\ \citenamefont
		{L{\"o}wen}}]{Likos2001}%
	\BibitemOpen
	\bibfield  {author} {\bibinfo {author} {\bibfnamefont {C.~N.}\ \bibnamefont
			{Likos}}, \bibinfo {author} {\bibfnamefont {A.}~\bibnamefont {Lang}},
		\bibinfo {author} {\bibfnamefont {M.}~\bibnamefont {Watzlawek}},\ and\
		\bibinfo {author} {\bibfnamefont {H.}~\bibnamefont {L{\"o}wen}},\ }\bibfield
	{title} {\bibinfo {title} {Criterion for determining clustering versus
			reentrant melting behavior for bounded interaction potentials},\ }\href
	{https://doi.org/10.1103/PhysRevE.63.031206} {\bibfield  {journal} {\bibinfo
			{journal} {Physical Review E}\ }\textbf {\bibinfo {volume} {63}},\ \bibinfo
		{pages} {031206} (\bibinfo {year} {2001})}\BibitemShut {NoStop}%
	\bibitem [{\citenamefont {Lang}\ \emph {et~al.}(2000)\citenamefont {Lang},
		\citenamefont {Likos}, \citenamefont {Watzlawek},\ and\ \citenamefont
		{L{\"o}wen}}]{Lang2000}%
	\BibitemOpen
	\bibfield  {author} {\bibinfo {author} {\bibfnamefont {A.}~\bibnamefont
			{Lang}}, \bibinfo {author} {\bibfnamefont {C.~N.}\ \bibnamefont {Likos}},
		\bibinfo {author} {\bibfnamefont {M.}~\bibnamefont {Watzlawek}},\ and\
		\bibinfo {author} {\bibfnamefont {H.}~\bibnamefont {L{\"o}wen}},\ }\bibfield
	{title} {\bibinfo {title} {Fluid and solid phases of the {{Gaussian}} core
			model},\ }\href {https://doi.org/10.1088/0953-8984/12/24/302} {\bibfield
		{journal} {\bibinfo  {journal} {Journal of Physics: Condensed Matter}\
		}\textbf {\bibinfo {volume} {12}},\ \bibinfo {pages} {5087} (\bibinfo {year}
		{2000})}\BibitemShut {NoStop}%
	\bibitem [{\citenamefont {Louis}\ \emph {et~al.}(2000)\citenamefont {Louis},
		\citenamefont {Bolhuis},\ and\ \citenamefont {Hansen}}]{Louis2000}%
	\BibitemOpen
	\bibfield  {author} {\bibinfo {author} {\bibfnamefont {A.~A.}\ \bibnamefont
			{Louis}}, \bibinfo {author} {\bibfnamefont {P.~G.}\ \bibnamefont {Bolhuis}},\
		and\ \bibinfo {author} {\bibfnamefont {J.~P.}\ \bibnamefont {Hansen}},\
	}\bibfield  {title} {\bibinfo {title} {Mean-field fluid behavior of the
			{{Gaussian}} core model},\ }\href {https://doi.org/10.1103/PhysRevE.62.7961}
	{\bibfield  {journal} {\bibinfo  {journal} {Physical Review E}\ }\textbf
		{\bibinfo {volume} {62}},\ \bibinfo {pages} {7961} (\bibinfo {year}
		{2000})}\BibitemShut {NoStop}%
	\bibitem [{\citenamefont {Kr{\"u}ger}\ and\ \citenamefont
		{Dean}(2017)}]{Kruger2017a}%
	\BibitemOpen
	\bibfield  {author} {\bibinfo {author} {\bibfnamefont {M.}~\bibnamefont
			{Kr{\"u}ger}}\ and\ \bibinfo {author} {\bibfnamefont {D.~S.}\ \bibnamefont
			{Dean}},\ }\bibfield  {title} {\bibinfo {title} {A {{Gaussian}} theory for
			fluctuations in simple liquids},\ }\href {https://doi.org/10.1063/1.4979659}
	{\bibfield  {journal} {\bibinfo  {journal} {Journal of Chemical Physics}\
		}\textbf {\bibinfo {volume} {146}},\ \bibinfo {pages} {134507} (\bibinfo
		{year} {2017})}\BibitemShut {NoStop}%
	\bibitem [{\citenamefont {Kr{\"u}ger}\ \emph {et~al.}(2018)\citenamefont
		{Kr{\"u}ger}, \citenamefont {Solon}, \citenamefont {D{\'e}mery},
		\citenamefont {Rohwer},\ and\ \citenamefont {Dean}}]{Kruger2018}%
	\BibitemOpen
	\bibfield  {author} {\bibinfo {author} {\bibfnamefont {M.}~\bibnamefont
			{Kr{\"u}ger}}, \bibinfo {author} {\bibfnamefont {A.}~\bibnamefont {Solon}},
		\bibinfo {author} {\bibfnamefont {V.}~\bibnamefont {D{\'e}mery}}, \bibinfo
		{author} {\bibfnamefont {C.~M.}\ \bibnamefont {Rohwer}},\ and\ \bibinfo
		{author} {\bibfnamefont {D.~S.}\ \bibnamefont {Dean}},\ }\bibfield  {title}
	{\bibinfo {title} {Stresses in non-equilibrium fluids: {{Exact}} formulation
			and coarse-grained theory},\ }\href {https://doi.org/10.1063/1.5019424}
	{\bibfield  {journal} {\bibinfo  {journal} {The Journal of Chemical Physics}\
		}\textbf {\bibinfo {volume} {148}},\ \bibinfo {pages} {084503} (\bibinfo
		{year} {2018})}\BibitemShut {NoStop}%
	\bibitem [{\citenamefont {Jin}\ and\ \citenamefont {Reichman}(2024)}]{Jin2024}%
	\BibitemOpen
	\bibfield  {author} {\bibinfo {author} {\bibfnamefont {J.}~\bibnamefont
			{Jin}}\ and\ \bibinfo {author} {\bibfnamefont {D.~R.}\ \bibnamefont
			{Reichman}},\ }\bibfield  {title} {\bibinfo {title} {Perturbative
			{{Expansion}} in {{Reciprocal Space}}: {{Bridging Microscopic}} and
			{{Mesoscopic Descriptions}} of {{Molecular Interactions}}},\ }\href
	{https://doi.org/10.1021/acs.jpcb.3c06048} {\bibfield  {journal} {\bibinfo
			{journal} {The Journal of Physical Chemistry B}\ }\textbf {\bibinfo {volume}
			{128}},\ \bibinfo {pages} {1061} (\bibinfo {year} {2024})}\BibitemShut
	{NoStop}%
	\bibitem [{\citenamefont {Frusawa}(2019{\natexlab{a}})}]{Frusawa2019a}%
	\BibitemOpen
	\bibfield  {author} {\bibinfo {author} {\bibfnamefont {H.}~\bibnamefont
			{Frusawa}},\ }\bibfield  {title} {\bibinfo {title} {Stochastic dynamics and
			thermodynamics around a metastable state based on the linear
			{{Dean}}--{{Kawasaki}} equation},\ }\href
	{https://doi.org/10.1088/1751-8121/aaf65c} {\bibfield  {journal} {\bibinfo
			{journal} {Journal of Physics A: Mathematical and Theoretical}\ }\textbf
		{\bibinfo {volume} {52}},\ \bibinfo {pages} {065003} (\bibinfo {year}
		{2019}{\natexlab{a}})}\BibitemShut {NoStop}%
	\bibitem [{\citenamefont {Frusawa}(2021)}]{Frusawa2021}%
	\BibitemOpen
	\bibfield  {author} {\bibinfo {author} {\bibfnamefont {H.}~\bibnamefont
			{Frusawa}},\ }\bibfield  {title} {\bibinfo {title} {Non-hyperuniform
			metastable states around a disordered hyperuniform state of densely packed
			spheres: Stochastic density functional theory at strong coupling},\ }\href
	{https://doi.org/10.1039/D1SM01052B} {\bibfield  {journal} {\bibinfo
			{journal} {Soft Matter}\ }\textbf {\bibinfo {volume} {17}},\ \bibinfo {pages}
		{8810} (\bibinfo {year} {2021})}\BibitemShut {NoStop}%
	\bibitem [{\citenamefont {Adhikari}\ \emph {et~al.}(2005)\citenamefont
		{Adhikari}, \citenamefont {Stratford}, \citenamefont {Cates},\ and\
		\citenamefont {Wagner}}]{Adhikari2005}%
	\BibitemOpen
	\bibfield  {author} {\bibinfo {author} {\bibfnamefont {R.}~\bibnamefont
			{Adhikari}}, \bibinfo {author} {\bibfnamefont {K.}~\bibnamefont {Stratford}},
		\bibinfo {author} {\bibfnamefont {M.~E.}\ \bibnamefont {Cates}},\ and\
		\bibinfo {author} {\bibfnamefont {A.~J.}\ \bibnamefont {Wagner}},\ }\bibfield
	{title} {\bibinfo {title} {Fluctuating lattice {{Boltzmann}}},\ }\href
	{https://doi.org/10.1209/epl/i2004-10542-5} {\bibfield  {journal} {\bibinfo
			{journal} {Europhysics Letters (EPL)}\ }\textbf {\bibinfo {volume} {71}},\
		\bibinfo {pages} {473} (\bibinfo {year} {2005})}\BibitemShut {NoStop}%
	\bibitem [{\citenamefont {Russo}\ \emph {et~al.}(2021)\citenamefont {Russo},
		\citenamefont {Perez}, \citenamefont {{Dur{\'a}n-Olivencia}}, \citenamefont
		{Yatsyshin}, \citenamefont {Carrillo},\ and\ \citenamefont
		{Kalliadasis}}]{Russo2021}%
	\BibitemOpen
	\bibfield  {author} {\bibinfo {author} {\bibfnamefont {A.}~\bibnamefont
			{Russo}}, \bibinfo {author} {\bibfnamefont {S.~P.}\ \bibnamefont {Perez}},
		\bibinfo {author} {\bibfnamefont {M.~A.}\ \bibnamefont
			{{Dur{\'a}n-Olivencia}}}, \bibinfo {author} {\bibfnamefont {P.}~\bibnamefont
			{Yatsyshin}}, \bibinfo {author} {\bibfnamefont {J.~A.}\ \bibnamefont
			{Carrillo}},\ and\ \bibinfo {author} {\bibfnamefont {S.}~\bibnamefont
			{Kalliadasis}},\ }\bibfield  {title} {\bibinfo {title} {A finite-volume
			method for fluctuating dynamical density functional theory},\ }\href
	{https://doi.org/10.1016/j.jcp.2020.109796} {\bibfield  {journal} {\bibinfo
			{journal} {Journal of Computational Physics}\ }\textbf {\bibinfo {volume}
			{428}},\ \bibinfo {pages} {109796} (\bibinfo {year} {2021})}\BibitemShut
	{NoStop}%
	\bibitem [{\citenamefont {Mendes}\ \emph {et~al.}(2021)\citenamefont {Mendes},
		\citenamefont {Russo}, \citenamefont {Perez},\ and\ \citenamefont
		{Kalliadasis}}]{Mendes2021}%
	\BibitemOpen
	\bibfield  {author} {\bibinfo {author} {\bibfnamefont {J.}~\bibnamefont
			{Mendes}}, \bibinfo {author} {\bibfnamefont {A.}~\bibnamefont {Russo}},
		\bibinfo {author} {\bibfnamefont {S.~P.}\ \bibnamefont {Perez}},\ and\
		\bibinfo {author} {\bibfnamefont {S.}~\bibnamefont {Kalliadasis}},\
	}\bibfield  {title} {\bibinfo {title} {A finite-volume scheme for
			gradient-flow equations with non-homogeneous diffusion},\ }\href
	{https://doi.org/10.1016/j.camwa.2021.02.004} {\bibfield  {journal} {\bibinfo
			{journal} {Computers \& Mathematics with Applications}\ }\textbf {\bibinfo
			{volume} {89}},\ \bibinfo {pages} {150} (\bibinfo {year} {2021})}\BibitemShut
	{NoStop}%
	\bibitem [{\citenamefont {{Mart{\'i}nez-Lera}}\ and\ \citenamefont
		{De~Corato}(2024)}]{Martinez-Lera2024}%
	\BibitemOpen
	\bibfield  {author} {\bibinfo {author} {\bibfnamefont {P.}~\bibnamefont
			{{Mart{\'i}nez-Lera}}}\ and\ \bibinfo {author} {\bibfnamefont
			{M.}~\bibnamefont {De~Corato}},\ }\bibfield  {title} {\bibinfo {title} {A
			finite element method for stochastic diffusion equations using fluctuating
			hydrodynamics},\ }\href {https://doi.org/10.1016/j.jcp.2024.113098}
	{\bibfield  {journal} {\bibinfo  {journal} {Journal of Computational
				Physics}\ }\textbf {\bibinfo {volume} {510}},\ \bibinfo {pages} {113098}
		(\bibinfo {year} {2024})}\BibitemShut {NoStop}%
	\bibitem [{\citenamefont {Magaletti}\ \emph {et~al.}(2022)\citenamefont
		{Magaletti}, \citenamefont {Gallo}, \citenamefont {Perez}, \citenamefont
		{Carrillo},\ and\ \citenamefont {Kalliadasis}}]{Magaletti2022}%
	\BibitemOpen
	\bibfield  {author} {\bibinfo {author} {\bibfnamefont {F.}~\bibnamefont
			{Magaletti}}, \bibinfo {author} {\bibfnamefont {M.}~\bibnamefont {Gallo}},
		\bibinfo {author} {\bibfnamefont {S.~P.}\ \bibnamefont {Perez}}, \bibinfo
		{author} {\bibfnamefont {J.~A.}\ \bibnamefont {Carrillo}},\ and\ \bibinfo
		{author} {\bibfnamefont {S.}~\bibnamefont {Kalliadasis}},\ }\bibfield
	{title} {\bibinfo {title} {A positivity-preserving scheme for fluctuating
			hydrodynamics},\ }\href {https://doi.org/10.1016/j.jcp.2022.111248}
	{\bibfield  {journal} {\bibinfo  {journal} {Journal of Computational
				Physics}\ }\textbf {\bibinfo {volume} {463}},\ \bibinfo {pages} {111248}
		(\bibinfo {year} {2022})}\BibitemShut {NoStop}%
	\bibitem [{\citenamefont {Cornalba}\ and\ \citenamefont
		{Fischer}(2023)}]{Cornalba2023a}%
	\BibitemOpen
	\bibfield  {author} {\bibinfo {author} {\bibfnamefont {F.}~\bibnamefont
			{Cornalba}}\ and\ \bibinfo {author} {\bibfnamefont {J.}~\bibnamefont
			{Fischer}},\ }\bibfield  {title} {\bibinfo {title} {The {{Dean}}--{{Kawasaki
					Equation}} and the {{Structure}} of {{Density Fluctuations}} in {{Systems}}
			of {{Diffusing Particles}}},\ }\href
	{https://doi.org/10.1007/s00205-023-01903-7} {\bibfield  {journal} {\bibinfo
			{journal} {Archive for Rational Mechanics and Analysis}\ }\textbf {\bibinfo
			{volume} {247}},\ \bibinfo {pages} {76} (\bibinfo {year} {2023})}\BibitemShut
	{NoStop}%
	\bibitem [{\citenamefont {Cornalba}\ and\ \citenamefont
		{Shardlow}(2023)}]{Cornalba2023b}%
	\BibitemOpen
	\bibfield  {author} {\bibinfo {author} {\bibfnamefont {F.}~\bibnamefont
			{Cornalba}}\ and\ \bibinfo {author} {\bibfnamefont {T.}~\bibnamefont
			{Shardlow}},\ }\bibfield  {title} {\bibinfo {title} {The regularised inertial
			{{Dean}}--{{Kawasaki}} equation: Discontinuous {{Galerkin}} approximation and
			modelling for low-density regime},\ }\href
	{https://doi.org/10.1051/m2an/2023077} {\bibfield  {journal} {\bibinfo
			{journal} {ESAIM: Mathematical Modelling and Numerical Analysis}\ }\textbf
		{\bibinfo {volume} {57}},\ \bibinfo {pages} {3061} (\bibinfo {year}
		{2023})}\BibitemShut {NoStop}%
	\bibitem [{\citenamefont {Gupta}\ \emph {et~al.}(2011)\citenamefont {Gupta},
		\citenamefont {Das},\ and\ \citenamefont {Barrat}}]{Gupta2011}%
	\BibitemOpen
	\bibfield  {author} {\bibinfo {author} {\bibfnamefont {B.~S.}\ \bibnamefont
			{Gupta}}, \bibinfo {author} {\bibfnamefont {S.~P.}\ \bibnamefont {Das}},\
		and\ \bibinfo {author} {\bibfnamefont {J.-L.}\ \bibnamefont {Barrat}},\
	}\bibfield  {title} {\bibinfo {title} {Time-dependent correlations in a
			supercooled liquid from nonlinear fluctuating hydrodynamics},\ }\href
	{https://doi.org/10.1103/PhysRevE.83.041506} {\bibfield  {journal} {\bibinfo
			{journal} {Physical Review E}\ }\textbf {\bibinfo {volume} {83}},\ \bibinfo
		{pages} {041506} (\bibinfo {year} {2011})}\BibitemShut {NoStop}%
	\bibitem [{\citenamefont {Bidhoodi}\ and\ \citenamefont
		{Das}(2016)}]{Bidhoodi2016}%
	\BibitemOpen
	\bibfield  {author} {\bibinfo {author} {\bibfnamefont {N.}~\bibnamefont
			{Bidhoodi}}\ and\ \bibinfo {author} {\bibfnamefont {S.~P.}\ \bibnamefont
			{Das}},\ }\bibfield  {title} {\bibinfo {title} {Ergodicity and slow diffusion
			in a supercooled liquid},\ }\href
	{https://doi.org/10.1016/j.physa.2015.12.076} {\bibfield  {journal} {\bibinfo
			{journal} {Physica A: Statistical Mechanics and its Applications}\ }\textbf
		{\bibinfo {volume} {449}},\ \bibinfo {pages} {357} (\bibinfo {year}
		{2016})}\BibitemShut {NoStop}%
	\bibitem [{\citenamefont {Ramaswamy}(2010)}]{Ramaswamy2010}%
	\BibitemOpen
	\bibfield  {author} {\bibinfo {author} {\bibfnamefont {S.}~\bibnamefont
			{Ramaswamy}},\ }\bibfield  {title} {\bibinfo {title} {The {{Mechanics}} and
			{{Statistics}} of {{Active Matter}}},\ }\href
	{https://doi.org/10.1146/annurev-conmatphys-070909-104101} {\bibfield
		{journal} {\bibinfo  {journal} {Annual Review of Condensed Matter Physics}\
		}\textbf {\bibinfo {volume} {1}},\ \bibinfo {pages} {323} (\bibinfo {year}
		{2010})}\BibitemShut {NoStop}%
	\bibitem [{\citenamefont {Marchetti}\ \emph {et~al.}(2013)\citenamefont
		{Marchetti}, \citenamefont {Joanny}, \citenamefont {Ramaswamy}, \citenamefont
		{Liverpool}, \citenamefont {Prost}, \citenamefont {Rao},\ and\ \citenamefont
		{Simha}}]{Marchetti2013}%
	\BibitemOpen
	\bibfield  {author} {\bibinfo {author} {\bibfnamefont {M.~C.}\ \bibnamefont
			{Marchetti}}, \bibinfo {author} {\bibfnamefont {J.~F.}\ \bibnamefont
			{Joanny}}, \bibinfo {author} {\bibfnamefont {S.}~\bibnamefont {Ramaswamy}},
		\bibinfo {author} {\bibfnamefont {T.~B.}\ \bibnamefont {Liverpool}}, \bibinfo
		{author} {\bibfnamefont {J.}~\bibnamefont {Prost}}, \bibinfo {author}
		{\bibfnamefont {M.}~\bibnamefont {Rao}},\ and\ \bibinfo {author}
		{\bibfnamefont {R.~A.}\ \bibnamefont {Simha}},\ }\bibfield  {title} {\bibinfo
		{title} {Hydrodynamics of soft active matter},\ }\href
	{https://doi.org/10.1103/RevModPhys.85.1143} {\bibfield  {journal} {\bibinfo
			{journal} {Reviews of Modern Physics}\ }\textbf {\bibinfo {volume} {85}},\
		\bibinfo {pages} {1143} (\bibinfo {year} {2013})}\BibitemShut {NoStop}%
	\bibitem [{\citenamefont {Tailleur}\ and\ \citenamefont
		{Cates}(2008)}]{Tailleur2008}%
	\BibitemOpen
	\bibfield  {author} {\bibinfo {author} {\bibfnamefont {J.}~\bibnamefont
			{Tailleur}}\ and\ \bibinfo {author} {\bibfnamefont {M.}~\bibnamefont
			{Cates}},\ }\bibfield  {title} {\bibinfo {title} {Statistical {{Mechanics}}
			of {{Interacting Run-and-Tumble Bacteria}}},\ }\href
	{https://doi.org/10.1103/PhysRevLett.100.218103} {\bibfield  {journal}
		{\bibinfo  {journal} {Physical Review Letters}\ }\textbf {\bibinfo {volume}
			{100}},\ \bibinfo {pages} {218103} (\bibinfo {year} {2008})}\BibitemShut
	{NoStop}%
	\bibitem [{\citenamefont {{\'O}~Laighl{\'e}is}\ \emph
		{et~al.}(2018)\citenamefont {{\'O}~Laighl{\'e}is}, \citenamefont {Evans},\
		and\ \citenamefont {Blythe}}]{OLaighleis2018}%
	\BibitemOpen
	\bibfield  {author} {\bibinfo {author} {\bibfnamefont {E.}~\bibnamefont
			{{\'O}~Laighl{\'e}is}}, \bibinfo {author} {\bibfnamefont {M.~R.}\
			\bibnamefont {Evans}},\ and\ \bibinfo {author} {\bibfnamefont {R.~A.}\
			\bibnamefont {Blythe}},\ }\bibfield  {title} {\bibinfo {title} {Minimal
			stochastic field equations for one-dimensional flocking},\ }\href
	{https://doi.org/10.1103/PhysRevE.98.062127} {\bibfield  {journal} {\bibinfo
			{journal} {Physical Review E}\ }\textbf {\bibinfo {volume} {98}},\ \bibinfo
		{pages} {062127} (\bibinfo {year} {2018})}\BibitemShut {NoStop}%
	\bibitem [{\citenamefont {Bertin}\ \emph {et~al.}(2013)\citenamefont {Bertin},
		\citenamefont {Chat{\'e}}, \citenamefont {Ginelli}, \citenamefont {Mishra},
		\citenamefont {Peshkov},\ and\ \citenamefont {Ramaswamy}}]{Bertin2013}%
	\BibitemOpen
	\bibfield  {author} {\bibinfo {author} {\bibfnamefont {E.}~\bibnamefont
			{Bertin}}, \bibinfo {author} {\bibfnamefont {H.}~\bibnamefont {Chat{\'e}}},
		\bibinfo {author} {\bibfnamefont {F.}~\bibnamefont {Ginelli}}, \bibinfo
		{author} {\bibfnamefont {S.}~\bibnamefont {Mishra}}, \bibinfo {author}
		{\bibfnamefont {A.}~\bibnamefont {Peshkov}},\ and\ \bibinfo {author}
		{\bibfnamefont {S.}~\bibnamefont {Ramaswamy}},\ }\bibfield  {title} {\bibinfo
		{title} {Mesoscopic theory for fluctuating active nematics},\ }\href
	{https://doi.org/10.1088/1367-2630/15/8/085032} {\bibfield  {journal}
		{\bibinfo  {journal} {New Journal of Physics}\ }\textbf {\bibinfo {volume}
			{15}},\ \bibinfo {pages} {085032} (\bibinfo {year} {2013})}\BibitemShut
	{NoStop}%
	\bibitem [{\citenamefont {Solon}\ \emph {et~al.}(2015)\citenamefont {Solon},
		\citenamefont {Cates},\ and\ \citenamefont {Tailleur}}]{Solon2015}%
	\BibitemOpen
	\bibfield  {author} {\bibinfo {author} {\bibfnamefont {A.~P.}\ \bibnamefont
			{Solon}}, \bibinfo {author} {\bibfnamefont {M.~E.}\ \bibnamefont {Cates}},\
		and\ \bibinfo {author} {\bibfnamefont {J.}~\bibnamefont {Tailleur}},\
	}\bibfield  {title} {\bibinfo {title} {Active brownian particles and
			run-and-tumble particles: {{A}} comparative study},\ }\href
	{https://doi.org/10.1140/epjst/e2015-02457-0} {\bibfield  {journal} {\bibinfo
			{journal} {The European Physical Journal Special Topics}\ }\textbf {\bibinfo
			{volume} {224}},\ \bibinfo {pages} {1231} (\bibinfo {year}
		{2015})}\BibitemShut {NoStop}%
	\bibitem [{\citenamefont {Martin}\ \emph {et~al.}(2021)\citenamefont {Martin},
		\citenamefont {O'Byrne}, \citenamefont {Cates}, \citenamefont {Fodor},
		\citenamefont {Nardini}, \citenamefont {Tailleur},\ and\ \citenamefont
		{Van~Wijland}}]{Martin2021}%
	\BibitemOpen
	\bibfield  {author} {\bibinfo {author} {\bibfnamefont {D.}~\bibnamefont
			{Martin}}, \bibinfo {author} {\bibfnamefont {J.}~\bibnamefont {O'Byrne}},
		\bibinfo {author} {\bibfnamefont {M.~E.}\ \bibnamefont {Cates}}, \bibinfo
		{author} {\bibfnamefont {{\'E}.}~\bibnamefont {Fodor}}, \bibinfo {author}
		{\bibfnamefont {C.}~\bibnamefont {Nardini}}, \bibinfo {author} {\bibfnamefont
			{J.}~\bibnamefont {Tailleur}},\ and\ \bibinfo {author} {\bibfnamefont
			{F.}~\bibnamefont {Van~Wijland}},\ }\bibfield  {title} {\bibinfo {title}
		{Statistical mechanics of active {{Ornstein-Uhlenbeck}} particles},\ }\href
	{https://doi.org/10.1103/PhysRevE.103.032607} {\bibfield  {journal} {\bibinfo
			{journal} {Physical Review E}\ }\textbf {\bibinfo {volume} {103}},\ \bibinfo
		{pages} {032607} (\bibinfo {year} {2021})}\BibitemShut {NoStop}%
	\bibitem [{\citenamefont {Dinelli}\ \emph {et~al.}(2024)\citenamefont
		{Dinelli}, \citenamefont {O'Byrne},\ and\ \citenamefont
		{Tailleur}}]{Dinelli2024}%
	\BibitemOpen
	\bibfield  {author} {\bibinfo {author} {\bibfnamefont {A.}~\bibnamefont
			{Dinelli}}, \bibinfo {author} {\bibfnamefont {J.}~\bibnamefont {O'Byrne}},\
		and\ \bibinfo {author} {\bibfnamefont {J.}~\bibnamefont {Tailleur}},\
	}\bibfield  {title} {\bibinfo {title} {Fluctuating hydrodynamics of active
			particles interacting via chemotaxis and quorum sensing: static and
			dynamics},\ }\href {https://doi.org/10.1088/1751-8121/ad72bc} {\bibfield
		{journal} {\bibinfo  {journal} {J. Phys. A: Math. Theor.}\ }\textbf {\bibinfo
			{volume} {57}},\ \bibinfo {pages} {395002} (\bibinfo {year}
		{2024})}\BibitemShut {NoStop}%
	\bibitem [{\citenamefont {Poncet}\ \emph
		{et~al.}(2021{\natexlab{b}})\citenamefont {Poncet}, \citenamefont
		{B{\'e}nichou}, \citenamefont {D{\'e}mery},\ and\ \citenamefont
		{Nishiguchi}}]{Poncet2021b}%
	\BibitemOpen
	\bibfield  {author} {\bibinfo {author} {\bibfnamefont {A.}~\bibnamefont
			{Poncet}}, \bibinfo {author} {\bibfnamefont {O.}~\bibnamefont
			{B{\'e}nichou}}, \bibinfo {author} {\bibfnamefont {V.}~\bibnamefont
			{D{\'e}mery}},\ and\ \bibinfo {author} {\bibfnamefont {D.}~\bibnamefont
			{Nishiguchi}},\ }\bibfield  {title} {\bibinfo {title} {Pair correlation of
			dilute active {{Brownian}} particles: {{From}} low-activity dipolar
			correction to high-activity algebraic depletion wings},\ }\href
	{https://doi.org/10.1103/PhysRevE.103.012605} {\bibfield  {journal} {\bibinfo
			{journal} {Physical Review E}\ }\textbf {\bibinfo {volume} {103}},\ \bibinfo
		{pages} {012605} (\bibinfo {year} {2021}{\natexlab{b}})}\BibitemShut
	{NoStop}%
	\bibitem [{\citenamefont {Hohenberg}\ and\ \citenamefont
		{Halperin}(1977)}]{Hohenberg1977}%
	\BibitemOpen
	\bibfield  {author} {\bibinfo {author} {\bibfnamefont {P.~C.}\ \bibnamefont
			{Hohenberg}}\ and\ \bibinfo {author} {\bibfnamefont {B.~I.}\ \bibnamefont
			{Halperin}},\ }\bibfield  {title} {\bibinfo {title} {Theory of dynamic
			critical phenomena},\ }\href {https://doi.org/10.1103/RevModPhys.49.435}
	{\bibfield  {journal} {\bibinfo  {journal} {Reviews of Modern Physics}\
		}\textbf {\bibinfo {volume} {49}},\ \bibinfo {pages} {435} (\bibinfo {year}
		{1977})}\BibitemShut {NoStop}%
	\bibitem [{\citenamefont {Chaikin}\ and\ \citenamefont {Lubensky}()}]{Chaikin}%
	\BibitemOpen
	\bibfield  {author} {\bibinfo {author} {\bibfnamefont {P.~M.}\ \bibnamefont
			{Chaikin}}\ and\ \bibinfo {author} {\bibfnamefont {T.~C.}\ \bibnamefont
			{Lubensky}},\ }\href@noop {} {\emph {\bibinfo {title} {Principles of
				Condensed Matter Physics}}}\ (\bibinfo  {publisher} {Cambridge University
		Press})\BibitemShut {NoStop}%
	\bibitem [{\citenamefont {Tjhung}\ \emph {et~al.}(2018)\citenamefont {Tjhung},
		\citenamefont {Nardini},\ and\ \citenamefont {Cates}}]{Tjhung2018}%
	\BibitemOpen
	\bibfield  {author} {\bibinfo {author} {\bibfnamefont {E.}~\bibnamefont
			{Tjhung}}, \bibinfo {author} {\bibfnamefont {C.}~\bibnamefont {Nardini}},\
		and\ \bibinfo {author} {\bibfnamefont {M.~E.}\ \bibnamefont {Cates}},\
	}\bibfield  {title} {\bibinfo {title} {Cluster {{Phases}} and {{Bubbly Phase
					Separation}} in {{Active Fluids}}: {{Reversal}} of the {{Ostwald Process}}},\
	}\href {https://doi.org/10.1103/PhysRevX.8.031080} {\bibfield  {journal}
		{\bibinfo  {journal} {Phys. Rev. X}\ }\textbf {\bibinfo {volume} {8}},\
		\bibinfo {pages} {31080} (\bibinfo {year} {2018})}\BibitemShut {NoStop}%
	\bibitem [{\citenamefont {GrandPre}\ \emph {et~al.}(2021)\citenamefont
		{GrandPre}, \citenamefont {Klymko}, \citenamefont {Mandadapu},\ and\
		\citenamefont {Limmer}}]{GrandPre2021}%
	\BibitemOpen
	\bibfield  {author} {\bibinfo {author} {\bibfnamefont {T.}~\bibnamefont
			{GrandPre}}, \bibinfo {author} {\bibfnamefont {K.}~\bibnamefont {Klymko}},
		\bibinfo {author} {\bibfnamefont {K.~K.}\ \bibnamefont {Mandadapu}},\ and\
		\bibinfo {author} {\bibfnamefont {D.~T.}\ \bibnamefont {Limmer}},\ }\bibfield
	{title} {\bibinfo {title} {Entropy production fluctuations encode collective
			behavior in active matter},\ }\href
	{https://doi.org/10.1103/PhysRevE.103.012613} {\bibfield  {journal} {\bibinfo
			{journal} {Physical Review E}\ }\textbf {\bibinfo {volume} {103}},\ \bibinfo
		{pages} {012613} (\bibinfo {year} {2021})}\BibitemShut {NoStop}%
	\bibitem [{\citenamefont {Tociu}\ \emph {et~al.}(2019)\citenamefont {Tociu},
		\citenamefont {Fodor}, \citenamefont {Nemoto},\ and\ \citenamefont
		{Vaikuntanathan}}]{Tociu2019}%
	\BibitemOpen
	\bibfield  {author} {\bibinfo {author} {\bibfnamefont {L.}~\bibnamefont
			{Tociu}}, \bibinfo {author} {\bibfnamefont {{\'E}.}~\bibnamefont {Fodor}},
		\bibinfo {author} {\bibfnamefont {T.}~\bibnamefont {Nemoto}},\ and\ \bibinfo
		{author} {\bibfnamefont {S.}~\bibnamefont {Vaikuntanathan}},\ }\bibfield
	{title} {\bibinfo {title} {How {{Dissipation Constrains Fluctuations}} in
			{{Nonequilibrium Liquids}}: {{Diffusion}}, {{Structure}}, and {{Biased
					Interactions}}},\ }\href {https://doi.org/10.1103/PhysRevX.9.041026}
	{\bibfield  {journal} {\bibinfo  {journal} {Physical Review X}\ }\textbf
		{\bibinfo {volume} {9}},\ \bibinfo {pages} {41026} (\bibinfo {year}
		{2019})}\BibitemShut {NoStop}%
	\bibitem [{\citenamefont {Fodor}\ \emph {et~al.}(2020)\citenamefont {Fodor},
		\citenamefont {Nemoto},\ and\ \citenamefont {Vaikuntanathan}}]{Fodor2020}%
	\BibitemOpen
	\bibfield  {author} {\bibinfo {author} {\bibfnamefont {{\'E}.}~\bibnamefont
			{Fodor}}, \bibinfo {author} {\bibfnamefont {T.}~\bibnamefont {Nemoto}},\ and\
		\bibinfo {author} {\bibfnamefont {S.}~\bibnamefont {Vaikuntanathan}},\
	}\bibfield  {title} {\bibinfo {title} {Dissipation controls transport and
			phase transitions in active fluids: Mobility, diffusion and biased
			ensembles},\ }\href {https://doi.org/10.1088/1367-2630/ab6353} {\bibfield
		{journal} {\bibinfo  {journal} {New Journal of Physics}\ }\textbf {\bibinfo
			{volume} {22}},\ \bibinfo {pages} {013052} (\bibinfo {year}
		{2020})}\BibitemShut {NoStop}%
	\bibitem [{\citenamefont {Rassolov}\ \emph {et~al.}(2022)\citenamefont
		{Rassolov}, \citenamefont {Tociu}, \citenamefont {Fodor},\ and\ \citenamefont
		{Vaikuntanathan}}]{Rassolov2022}%
	\BibitemOpen
	\bibfield  {author} {\bibinfo {author} {\bibfnamefont {G.}~\bibnamefont
			{Rassolov}}, \bibinfo {author} {\bibfnamefont {L.}~\bibnamefont {Tociu}},
		\bibinfo {author} {\bibfnamefont {{\'E}.}~\bibnamefont {Fodor}},\ and\
		\bibinfo {author} {\bibfnamefont {S.}~\bibnamefont {Vaikuntanathan}},\
	}\bibfield  {title} {\bibinfo {title} {From predicting to learning
			dissipation from pair correlations of active liquids},\ }\href
	{https://doi.org/10.1063/5.0097863} {\bibfield  {journal} {\bibinfo
			{journal} {The Journal of Chemical Physics}\ }\textbf {\bibinfo {volume}
			{157}},\ \bibinfo {pages} {054901} (\bibinfo {year} {2022})}\BibitemShut
	{NoStop}%
	\bibitem [{\citenamefont {Tociu}\ \emph {et~al.}(2022)\citenamefont {Tociu},
		\citenamefont {Rassolov}, \citenamefont {Fodor},\ and\ \citenamefont
		{Vaikuntanathan}}]{Tociu2022}%
	\BibitemOpen
	\bibfield  {author} {\bibinfo {author} {\bibfnamefont {L.}~\bibnamefont
			{Tociu}}, \bibinfo {author} {\bibfnamefont {G.}~\bibnamefont {Rassolov}},
		\bibinfo {author} {\bibfnamefont {{\'E}.}~\bibnamefont {Fodor}},\ and\
		\bibinfo {author} {\bibfnamefont {S.}~\bibnamefont {Vaikuntanathan}},\
	}\bibfield  {title} {\bibinfo {title} {Mean-field theory for the structure of
			strongly interacting active liquids},\ }\href
	{https://doi.org/10.1063/5.0096710} {\bibfield  {journal} {\bibinfo
			{journal} {The Journal of Chemical Physics}\ }\textbf {\bibinfo {volume}
			{157}},\ \bibinfo {pages} {014902} (\bibinfo {year} {2022})}\BibitemShut
	{NoStop}%
	\bibitem [{\citenamefont {Kuroda}\ and\ \citenamefont
		{Miyazaki}(2023)}]{Kuroda2023}%
	\BibitemOpen
	\bibfield  {author} {\bibinfo {author} {\bibfnamefont {Y.}~\bibnamefont
			{Kuroda}}\ and\ \bibinfo {author} {\bibfnamefont {K.}~\bibnamefont
			{Miyazaki}},\ }\bibfield  {title} {\bibinfo {title} {Microscopic theory for
			hyperuniformity in two-dimensional chiral active fluid},\ }\href
	{https://doi.org/10.1088/1742-5468/ad0639} {\bibfield  {journal} {\bibinfo
			{journal} {Journal of Statistical Mechanics: Theory and Experiment}\ }\textbf
		{\bibinfo {volume} {2023}},\ \bibinfo {pages} {103203} (\bibinfo {year}
		{2023})}\BibitemShut {NoStop}%
	\bibitem [{\citenamefont {Damman}\ \emph {et~al.}(2024)\citenamefont {Damman},
		\citenamefont {Démery}, \citenamefont {Palumbo},\ and\ \citenamefont
		{Thomas}}]{Damman2024}%
	\BibitemOpen
	\bibfield  {author} {\bibinfo {author} {\bibfnamefont {P.}~\bibnamefont
			{Damman}}, \bibinfo {author} {\bibfnamefont {V.}~\bibnamefont {Démery}},
		\bibinfo {author} {\bibfnamefont {G.}~\bibnamefont {Palumbo}},\ and\ \bibinfo
		{author} {\bibfnamefont {Q.}~\bibnamefont {Thomas}},\ }\href
	{https://arxiv.org/abs/2406.11616} {\bibinfo {title} {Algebraic depletion
			interactions in two-temperature mixtures}} (\bibinfo {year} {2024}),\ \Eprint
	{https://arxiv.org/abs/2406.11616} {arXiv:2406.11616} \BibitemShut {NoStop}%
	\bibitem [{\citenamefont {Grosberg}\ and\ \citenamefont
		{Joanny}(2015)}]{Grosberg2015}%
	\BibitemOpen
	\bibfield  {author} {\bibinfo {author} {\bibfnamefont {A.~Y.}\ \bibnamefont
			{Grosberg}}\ and\ \bibinfo {author} {\bibfnamefont {J.-F.}\ \bibnamefont
			{Joanny}},\ }\bibfield  {title} {\bibinfo {title} {Nonequilibrium statistical
			mechanics of mixtures of particles in contact with different thermostats},\
	}\href {https://doi.org/10.1103/PhysRevE.92.032118} {\bibfield  {journal}
		{\bibinfo  {journal} {Physical Review E}\ }\textbf {\bibinfo {volume} {92}},\
		\bibinfo {pages} {032118} (\bibinfo {year} {2015})}\BibitemShut {NoStop}%
	\bibitem [{\citenamefont {Ghimenti}\ \emph {et~al.}(2024)\citenamefont
		{Ghimenti}, \citenamefont {Berthier}, \citenamefont {Szamel},\ and\
		\citenamefont {van Wijland}}]{Ghimenti2024}%
	\BibitemOpen
	\bibfield  {author} {\bibinfo {author} {\bibfnamefont {F.}~\bibnamefont
			{Ghimenti}}, \bibinfo {author} {\bibfnamefont {L.}~\bibnamefont {Berthier}},
		\bibinfo {author} {\bibfnamefont {G.}~\bibnamefont {Szamel}},\ and\ \bibinfo
		{author} {\bibfnamefont {F.}~\bibnamefont {van Wijland}},\ }\bibfield
	{title} {\bibinfo {title} {Irreversible {Boltzmann} samplers in dense
			liquids: {Weak}-coupling approximation and mode-coupling theory},\ }\href
	{https://doi.org/10.1103/PhysRevE.110.034604} {\bibfield  {journal} {\bibinfo
			{journal} {Phys. Rev. E}\ }\textbf {\bibinfo {volume} {110}},\ \bibinfo
		{pages} {034604} (\bibinfo {year} {2024})}\BibitemShut {NoStop}%
	\bibitem [{\citenamefont {Dinelli}\ \emph {et~al.}(2023)\citenamefont
		{Dinelli}, \citenamefont {O'Byrne}, \citenamefont {Curatolo}, \citenamefont
		{Zhao}, \citenamefont {Sollich},\ and\ \citenamefont
		{Tailleur}}]{Dinelli2023}%
	\BibitemOpen
	\bibfield  {author} {\bibinfo {author} {\bibfnamefont {A.}~\bibnamefont
			{Dinelli}}, \bibinfo {author} {\bibfnamefont {J.}~\bibnamefont {O'Byrne}},
		\bibinfo {author} {\bibfnamefont {A.}~\bibnamefont {Curatolo}}, \bibinfo
		{author} {\bibfnamefont {Y.}~\bibnamefont {Zhao}}, \bibinfo {author}
		{\bibfnamefont {P.}~\bibnamefont {Sollich}},\ and\ \bibinfo {author}
		{\bibfnamefont {J.}~\bibnamefont {Tailleur}},\ }\bibfield  {title} {\bibinfo
		{title} {Non-reciprocity across scales in active mixtures},\ }\href
	{https://doi.org/10.1038/s41467-023-42713-5} {\bibfield  {journal} {\bibinfo
			{journal} {Nature Communications}\ }\textbf {\bibinfo {volume} {14}},\
		\bibinfo {pages} {7035} (\bibinfo {year} {2023})}\BibitemShut {NoStop}%
	\bibitem [{\citenamefont {Chavanis}(2010)}]{Chavanis2010}%
	\BibitemOpen
	\bibfield  {author} {\bibinfo {author} {\bibfnamefont {P.-H.}\ \bibnamefont
			{Chavanis}},\ }\bibfield  {title} {\bibinfo {title} {A stochastic
			{{Keller}}--{{Segel}} model of chemotaxis},\ }\href
	{https://doi.org/10.1016/j.cnsns.2008.09.002} {\bibfield  {journal} {\bibinfo
			{journal} {Communications in Nonlinear Science and Numerical Simulation}\
		}\textbf {\bibinfo {volume} {15}},\ \bibinfo {pages} {60} (\bibinfo {year}
		{2010})}\BibitemShut {NoStop}%
	\bibitem [{\citenamefont {Chavanis}\ and\ \citenamefont
		{Delfini}(2014)}]{Chavanis2014}%
	\BibitemOpen
	\bibfield  {author} {\bibinfo {author} {\bibfnamefont {P.~H.}\ \bibnamefont
			{Chavanis}}\ and\ \bibinfo {author} {\bibfnamefont {L.}~\bibnamefont
			{Delfini}},\ }\bibfield  {title} {\bibinfo {title} {Random transitions
			described by the stochastic {{Smoluchowski-Poisson}} system and by the
			stochastic {{Keller-Segel}} model},\ }\href
	{https://doi.org/10.1103/PhysRevE.89.032139} {\bibfield  {journal} {\bibinfo
			{journal} {Physical Review E}\ }\textbf {\bibinfo {volume} {89}},\ \bibinfo
		{pages} {032139} (\bibinfo {year} {2014})}\BibitemShut {NoStop}%
	\bibitem [{\citenamefont {Forster}\ \emph {et~al.}(1977)\citenamefont
		{Forster}, \citenamefont {Nelson},\ and\ \citenamefont
		{Stephen}}]{Forster1977}%
	\BibitemOpen
	\bibfield  {author} {\bibinfo {author} {\bibfnamefont {D.}~\bibnamefont
			{Forster}}, \bibinfo {author} {\bibfnamefont {D.~R.}\ \bibnamefont
			{Nelson}},\ and\ \bibinfo {author} {\bibfnamefont {J.}~\bibnamefont
			{Stephen}},\ }\bibfield  {title} {\bibinfo {title} {Large-distance and
			long-time properties of a randomly stirred fluid},\ }\href
	{https://doi.org/10.1016/0003-4916(61)90056-2} {\bibfield  {journal}
		{\bibinfo  {journal} {Phys. Rev. A}\ }\textbf {\bibinfo {volume} {16}},\
		\bibinfo {pages} {732} (\bibinfo {year} {1977})}\BibitemShut {NoStop}%
	\bibitem [{\citenamefont {Medina}\ \emph {et~al.}(1989)\citenamefont {Medina},
		\citenamefont {Hwa}, \citenamefont {Kardar},\ and\ \citenamefont
		{Zhang}}]{Medina1989}%
	\BibitemOpen
	\bibfield  {author} {\bibinfo {author} {\bibfnamefont {E.}~\bibnamefont
			{Medina}}, \bibinfo {author} {\bibfnamefont {T.}~\bibnamefont {Hwa}},
		\bibinfo {author} {\bibfnamefont {M.}~\bibnamefont {Kardar}},\ and\ \bibinfo
		{author} {\bibfnamefont {Y.~C.}\ \bibnamefont {Zhang}},\ }\bibfield  {title}
	{\bibinfo {title} {Burgers equation with correlated noise:
			{{Renormalization-group}} analysis and applications to directed polymers and
			interface growth},\ }\href {https://doi.org/10.1103/PhysRevA.39.3053}
	{\bibfield  {journal} {\bibinfo  {journal} {Physical Review A}\ }\textbf
		{\bibinfo {volume} {39}},\ \bibinfo {pages} {3053} (\bibinfo {year}
		{1989})}\BibitemShut {NoStop}%
	\bibitem [{\citenamefont {Gelimson}\ and\ \citenamefont
		{Golestanian}(2015)}]{Gelimson2015}%
	\BibitemOpen
	\bibfield  {author} {\bibinfo {author} {\bibfnamefont {A.}~\bibnamefont
			{Gelimson}}\ and\ \bibinfo {author} {\bibfnamefont {R.}~\bibnamefont
			{Golestanian}},\ }\bibfield  {title} {\bibinfo {title} {Collective
			{{Dynamics}} of {{Dividing Chemotactic Cells}}},\ }\href
	{https://doi.org/10.1103/PhysRevLett.114.028101} {\bibfield  {journal}
		{\bibinfo  {journal} {Physical Review Letters}\ }\textbf {\bibinfo {volume}
			{114}},\ \bibinfo {pages} {028101} (\bibinfo {year} {2015})}\BibitemShut
	{NoStop}%
	\bibitem [{\citenamefont {Mahdisoltani}\ \emph {et~al.}(2021)\citenamefont
		{Mahdisoltani}, \citenamefont {Zinati}, \citenamefont {Duclut}, \citenamefont
		{Gambassi},\ and\ \citenamefont {Golestanian}}]{Mahdisoltani2021}%
	\BibitemOpen
	\bibfield  {author} {\bibinfo {author} {\bibfnamefont {S.}~\bibnamefont
			{Mahdisoltani}}, \bibinfo {author} {\bibfnamefont {R.~B.~A.}\ \bibnamefont
			{Zinati}}, \bibinfo {author} {\bibfnamefont {C.}~\bibnamefont {Duclut}},
		\bibinfo {author} {\bibfnamefont {A.}~\bibnamefont {Gambassi}},\ and\
		\bibinfo {author} {\bibfnamefont {R.}~\bibnamefont {Golestanian}},\
	}\bibfield  {title} {\bibinfo {title} {Nonequilibrium polarity-induced
			chemotaxis: {{Emergent Galilean}} symmetry and exact scaling exponents},\
	}\href {https://doi.org/10.1103/PhysRevResearch.3.013100} {\bibfield
		{journal} {\bibinfo  {journal} {Physical Review Research}\ }\textbf {\bibinfo
			{volume} {3}},\ \bibinfo {pages} {013100} (\bibinfo {year}
		{2021})}\BibitemShut {NoStop}%
	\bibitem [{\citenamefont {Ben Al{\`i}~Zinati}\ \emph
		{et~al.}(2021)\citenamefont {Ben Al{\`i}~Zinati}, \citenamefont {Duclut},
		\citenamefont {Mahdisoltani}, \citenamefont {Gambassi},\ and\ \citenamefont
		{Golestanian}}]{BenAliZinati2021}%
	\BibitemOpen
	\bibfield  {author} {\bibinfo {author} {\bibfnamefont {R.}~\bibnamefont {Ben
				Al{\`i}~Zinati}}, \bibinfo {author} {\bibfnamefont {C.}~\bibnamefont
			{Duclut}}, \bibinfo {author} {\bibfnamefont {S.}~\bibnamefont
			{Mahdisoltani}}, \bibinfo {author} {\bibfnamefont {A.}~\bibnamefont
			{Gambassi}},\ and\ \bibinfo {author} {\bibfnamefont {R.}~\bibnamefont
			{Golestanian}},\ }\bibfield  {title} {\bibinfo {title} {Stochastic dynamics
			of chemotactic colonies with logistic growth},\ }\href
	{https://doi.org/10.1209/0295-5075/ac48c9} {\bibfield  {journal} {\bibinfo
			{journal} {Europhysics Letters}\ }\textbf {\bibinfo {volume} {136}},\
		\bibinfo {pages} {50003} (\bibinfo {year} {2021})}\BibitemShut {NoStop}%
	\bibitem [{\citenamefont {Mahdisoltani}\ and\ \citenamefont
		{Golestanian}(2023)}]{Mahdisoltani2023}%
	\BibitemOpen
	\bibfield  {author} {\bibinfo {author} {\bibfnamefont {S.}~\bibnamefont
			{Mahdisoltani}}\ and\ \bibinfo {author} {\bibfnamefont {R.}~\bibnamefont
			{Golestanian}},\ }\bibfield  {title} {\bibinfo {title} {Nonequilibrium
			phenomena in driven and active {{Coulomb}} field theories},\ }\href
	{https://doi.org/10.1016/j.physa.2022.127947} {\bibfield  {journal} {\bibinfo
			{journal} {Physica A: Statistical Mechanics and its Applications}\ }\textbf
		{\bibinfo {volume} {631}},\ \bibinfo {pages} {127947} (\bibinfo {year}
		{2023})}\BibitemShut {NoStop}%
	\bibitem [{\citenamefont {Van Der~Kolk}\ \emph {et~al.}(2023)\citenamefont {Van
			Der~Kolk}, \citenamefont {Ra{\ss}hofer}, \citenamefont {Swiderski},
		\citenamefont {Haldar}, \citenamefont {Basu},\ and\ \citenamefont
		{Frey}}]{VanDerKolk2023}%
	\BibitemOpen
	\bibfield  {author} {\bibinfo {author} {\bibfnamefont {J.}~\bibnamefont {Van
				Der~Kolk}}, \bibinfo {author} {\bibfnamefont {F.}~\bibnamefont
			{Ra{\ss}hofer}}, \bibinfo {author} {\bibfnamefont {R.}~\bibnamefont
			{Swiderski}}, \bibinfo {author} {\bibfnamefont {A.}~\bibnamefont {Haldar}},
		\bibinfo {author} {\bibfnamefont {A.}~\bibnamefont {Basu}},\ and\ \bibinfo
		{author} {\bibfnamefont {E.}~\bibnamefont {Frey}},\ }\bibfield  {title}
	{\bibinfo {title} {Anomalous {{Collective Dynamics}} of {{Autochemotactic
					Populations}}},\ }\href {https://doi.org/10.1103/PhysRevLett.131.088201}
	{\bibfield  {journal} {\bibinfo  {journal} {Physical Review Letters}\
		}\textbf {\bibinfo {volume} {131}},\ \bibinfo {pages} {088201} (\bibinfo
		{year} {2023})}\BibitemShut {NoStop}%
	\bibitem [{\citenamefont {Samanta}\ and\ \citenamefont
		{Thirumalai}(2019)}]{Samanta2019}%
	\BibitemOpen
	\bibfield  {author} {\bibinfo {author} {\bibfnamefont {H.~S.}\ \bibnamefont
			{Samanta}}\ and\ \bibinfo {author} {\bibfnamefont {D.}~\bibnamefont
			{Thirumalai}},\ }\bibfield  {title} {\bibinfo {title} {Origin of
			superdiffusive behavior in a class of nonequilibrium systems},\ }\href
	{https://doi.org/10.1103/PhysRevE.99.032401} {\bibfield  {journal} {\bibinfo
			{journal} {Physical Review E}\ }\textbf {\bibinfo {volume} {99}},\ \bibinfo
		{pages} {032401} (\bibinfo {year} {2019})}\BibitemShut {NoStop}%
	\bibitem [{\citenamefont {Samanta}(2020)}]{Samanta2020}%
	\BibitemOpen
	\bibfield  {author} {\bibinfo {author} {\bibfnamefont {H.~S.}\ \bibnamefont
			{Samanta}},\ }\bibfield  {title} {\bibinfo {title} {Interstitial flows
			regulate collective cell migration heterogeneity through adhesion},\ }\href
	{https://doi.org/10.1103/PhysRevResearch.2.013048} {\bibfield  {journal}
		{\bibinfo  {journal} {Physical Review Research}\ }\textbf {\bibinfo {volume}
			{2}},\ \bibinfo {pages} {013048} (\bibinfo {year} {2020})}\BibitemShut
	{NoStop}%
	\bibitem [{\citenamefont {Kjellander}(2019)}]{Kjellander2019a}%
	\BibitemOpen
	\bibfield  {author} {\bibinfo {author} {\bibfnamefont {R.}~\bibnamefont
			{Kjellander}},\ }\href {https://doi.org/10.1201/9780429194368} {\emph
		{\bibinfo {title} {Statistical {{Mechanics}} of {{Liquids}} and
				{{Solutions}}: {{Intermolecular Forces}}, {{Structure}} and {{Surface
						Interactions}}}}},\ \bibinfo {edition} {1st}\ ed.\ (\bibinfo  {publisher}
	{CRC Press},\ \bibinfo {year} {2019})\BibitemShut {NoStop}%
	\bibitem [{\citenamefont {Bonneau}\ \emph {et~al.}(2023)\citenamefont
		{Bonneau}, \citenamefont {D{\'e}mery},\ and\ \citenamefont
		{Rapha{\"e}l}}]{Bonneau2023}%
	\BibitemOpen
	\bibfield  {author} {\bibinfo {author} {\bibfnamefont {H.}~\bibnamefont
			{Bonneau}}, \bibinfo {author} {\bibfnamefont {V.}~\bibnamefont
			{D{\'e}mery}},\ and\ \bibinfo {author} {\bibfnamefont {{\'E}.}~\bibnamefont
			{Rapha{\"e}l}},\ }\bibfield  {title} {\bibinfo {title} {Temporal response of
			the conductivity of electrolytes},\ }\href
	{https://doi.org/10.1088/1742-5468/acdced} {\bibfield  {journal} {\bibinfo
			{journal} {J. Stat. Mech.}\ }\textbf {\bibinfo {volume} {2023}},\ \bibinfo
		{pages} {073205} (\bibinfo {year} {2023})}\BibitemShut {NoStop}%
	\bibitem [{\citenamefont {Bonneau}\ \emph {et~al.}(2025)\citenamefont
		{Bonneau}, \citenamefont {D{\'e}mery},\ and\ \citenamefont
		{Rapha{\"e}l}}]{Bonneau2025}%
	\BibitemOpen
	\bibfield  {author} {\bibinfo {author} {\bibfnamefont {H.}~\bibnamefont
			{Bonneau}}, \bibinfo {author} {\bibfnamefont {V.}~\bibnamefont
			{D{\'e}mery}},\ and\ \bibinfo {author} {\bibfnamefont {E.}~\bibnamefont
			{Rapha{\"e}l}},\ }\bibfield  {title} {\bibinfo {title} {Stationary and
			transient correlations in driven electrolytes},\ }\href
	{https://doi.org/10.1088/1742-5468/adb4ce} {\bibfield  {journal} {\bibinfo
			{journal} {J. Stat. Mech.}\ }\textbf {\bibinfo {volume} {2025}},\ \bibinfo
		{pages} {033201} (\bibinfo {year} {2025})}\BibitemShut {NoStop}%
	\bibitem [{\citenamefont {Berthoumieux}\ \emph {et~al.}(2024)\citenamefont
		{Berthoumieux}, \citenamefont {D{\'e}mery},\ and\ \citenamefont
		{Maggs}}]{Berthoumieux2024}%
	\BibitemOpen
	\bibfield  {author} {\bibinfo {author} {\bibfnamefont {H.}~\bibnamefont
			{Berthoumieux}}, \bibinfo {author} {\bibfnamefont {V.}~\bibnamefont
			{D{\'e}mery}},\ and\ \bibinfo {author} {\bibfnamefont {A.~C.}\ \bibnamefont
			{Maggs}},\ }\href@noop {} {\bibinfo {title} {Non-monotonic conductivity of
			aqueous electrolytes: Beyond the first {{Wien}} effect}} (\bibinfo {year}
	{2024})\BibitemShut {NoStop}%
	\bibitem [{\citenamefont {Frusawa}(2019{\natexlab{b}})}]{Frusawa2019}%
	\BibitemOpen
	\bibfield  {author} {\bibinfo {author} {\bibfnamefont {H.}~\bibnamefont
			{Frusawa}},\ }\bibfield  {title} {\bibinfo {title} {Transverse {{Density
					Fluctuations}} around the {{Ground State Distribution}} of {{Counterions}}
			near {{One Charged Plate}}: {{Stochastic Density Functional View}}},\ }\href
	{https://doi.org/10.3390/e22010034} {\bibfield  {journal} {\bibinfo
			{journal} {Entropy}\ }\textbf {\bibinfo {volume} {22}},\ \bibinfo {pages}
		{34} (\bibinfo {year} {2019}{\natexlab{b}})}\BibitemShut {NoStop}%
	\bibitem [{\citenamefont {Hoang~Ngoc}\ \emph {et~al.}(2023)\citenamefont
		{Hoang~Ngoc}, \citenamefont {Rotenberg},\ and\ \citenamefont
		{Marbach}}]{HoangNgoc2023}%
	\BibitemOpen
	\bibfield  {author} {\bibinfo {author} {\bibfnamefont {M.-T.}\ \bibnamefont
			{Hoang~Ngoc}}, \bibinfo {author} {\bibfnamefont {B.}~\bibnamefont
			{Rotenberg}},\ and\ \bibinfo {author} {\bibfnamefont {S.}~\bibnamefont
			{Marbach}},\ }\bibfield  {title} {\bibinfo {title} {Ionic fluctuations in
			finite volumes: Fractional noise and hyperuniformity},\ }\href
	{https://doi.org//10.1039/D3FD00031A} {\bibfield  {journal} {\bibinfo
			{journal} {Faraday Discussions}\ }\textbf {\bibinfo {volume} {246}},\
		\bibinfo {pages} {225} (\bibinfo {year} {2023})}\BibitemShut {NoStop}%
	\bibitem [{\citenamefont {Okamoto}(2022)}]{Okamoto2022}%
	\BibitemOpen
	\bibfield  {author} {\bibinfo {author} {\bibfnamefont {R.}~\bibnamefont
			{Okamoto}},\ }\bibfield  {title} {\bibinfo {title} {Fluctuating hydrodynamics
			of dilute electrolyte solutions: Systematic perturbation calculation of
			effective transport coefficients governing large-scale dynamics},\ }\href
	{https://doi.org/10.1088/1742-5468/ac8c8d} {\bibfield  {journal} {\bibinfo
			{journal} {Journal of Statistical Mechanics: Theory and Experiment}\ }\textbf
		{\bibinfo {volume} {2022}},\ \bibinfo {pages} {093203} (\bibinfo {year}
		{2022})}\BibitemShut {NoStop}%
	\bibitem [{\citenamefont {Poitevin}\ \emph {et~al.}(2016)\citenamefont
		{Poitevin}, \citenamefont {Delarue},\ and\ \citenamefont
		{Orland}}]{Poitevin2016}%
	\BibitemOpen
	\bibfield  {author} {\bibinfo {author} {\bibfnamefont {F.}~\bibnamefont
			{Poitevin}}, \bibinfo {author} {\bibfnamefont {M.}~\bibnamefont {Delarue}},\
		and\ \bibinfo {author} {\bibfnamefont {H.}~\bibnamefont {Orland}},\
	}\bibfield  {title} {\bibinfo {title} {Beyond {{Poisson}}--{{Boltzmann}}:
			{{Numerical Sampling}} of {{Charge Density Fluctuations}}},\ }\href
	{https://doi.org/10.1021/acs.jpcb.6b02650} {\bibfield  {journal} {\bibinfo
			{journal} {The Journal of Physical Chemistry B}\ }\textbf {\bibinfo {volume}
			{120}},\ \bibinfo {pages} {6270} (\bibinfo {year} {2016})}\BibitemShut
	{NoStop}%
	\bibitem [{\citenamefont {Avni}\ \emph
		{et~al.}(2022{\natexlab{a}})\citenamefont {Avni}, \citenamefont {Adar},
		\citenamefont {Andelman},\ and\ \citenamefont {Orland}}]{Avni2022a}%
	\BibitemOpen
	\bibfield  {author} {\bibinfo {author} {\bibfnamefont {Y.}~\bibnamefont
			{Avni}}, \bibinfo {author} {\bibfnamefont {R.~M.}\ \bibnamefont {Adar}},
		\bibinfo {author} {\bibfnamefont {D.}~\bibnamefont {Andelman}},\ and\
		\bibinfo {author} {\bibfnamefont {H.}~\bibnamefont {Orland}},\ }\bibfield
	{title} {\bibinfo {title} {Conductivity of {{Concentrated Electrolytes}}},\
	}\href {https://doi.org/10.1103/PhysRevLett.128.098002} {\bibfield  {journal}
		{\bibinfo  {journal} {Physical Review Letters}\ }\textbf {\bibinfo {volume}
			{128}},\ \bibinfo {pages} {098002} (\bibinfo {year}
		{2022}{\natexlab{a}})}\BibitemShut {NoStop}%
	\bibitem [{\citenamefont {Avni}\ \emph
		{et~al.}(2022{\natexlab{b}})\citenamefont {Avni}, \citenamefont {Andelman},\
		and\ \citenamefont {Orland}}]{Avni2022}%
	\BibitemOpen
	\bibfield  {author} {\bibinfo {author} {\bibfnamefont {Y.}~\bibnamefont
			{Avni}}, \bibinfo {author} {\bibfnamefont {D.}~\bibnamefont {Andelman}},\
		and\ \bibinfo {author} {\bibfnamefont {H.}~\bibnamefont {Orland}},\
	}\bibfield  {title} {\bibinfo {title} {Conductance of concentrated
			electrolytes: Multivalency and the {{Wien}} effect},\ }\href
	{https://doi.org/10.1063/5.0111645} {\bibfield  {journal} {\bibinfo
			{journal} {The Journal of Chemical Physics}\ }\textbf {\bibinfo {volume}
			{157}},\ \bibinfo {pages} {154502} (\bibinfo {year}
		{2022}{\natexlab{b}})}\BibitemShut {NoStop}%
	\bibitem [{\citenamefont {Robin}(2024)}]{Robin2024}%
	\BibitemOpen
	\bibfield  {author} {\bibinfo {author} {\bibfnamefont {P.}~\bibnamefont
			{Robin}},\ }\bibfield  {title} {\bibinfo {title} {Correlation-induced viscous
			dissipation in concentrated electrolytes},\ }\href
	{https://doi.org/10.1063/5.0188215} {\bibfield  {journal} {\bibinfo
			{journal} {The Journal of Chemical Physics}\ }\textbf {\bibinfo {volume}
			{160}},\ \bibinfo {pages} {064503} (\bibinfo {year} {2024})}\BibitemShut
	{NoStop}%
	\bibitem [{\citenamefont {Bernard}\ \emph {et~al.}(2023)\citenamefont
		{Bernard}, \citenamefont {Jardat}, \citenamefont {Rotenberg},\ and\
		\citenamefont {Illien}}]{Bernard2023}%
	\BibitemOpen
	\bibfield  {author} {\bibinfo {author} {\bibfnamefont {O.}~\bibnamefont
			{Bernard}}, \bibinfo {author} {\bibfnamefont {M.}~\bibnamefont {Jardat}},
		\bibinfo {author} {\bibfnamefont {B.}~\bibnamefont {Rotenberg}},\ and\
		\bibinfo {author} {\bibfnamefont {P.}~\bibnamefont {Illien}},\ }\bibfield
	{title} {\bibinfo {title} {On analytical theories for conductivity and
			self-diffusion in concentrated electrolytes},\ }\href
	{https://doi.org/10.1063/5.0165533} {\bibfield  {journal} {\bibinfo
			{journal} {The Journal of Chemical Physics}\ }\textbf {\bibinfo {volume}
			{159}},\ \bibinfo {pages} {164105} (\bibinfo {year} {2023})}\BibitemShut
	{NoStop}%
	\bibitem [{\citenamefont {Frusawa}(2022)}]{Frusawa2022}%
	\BibitemOpen
	\bibfield  {author} {\bibinfo {author} {\bibfnamefont {H.}~\bibnamefont
			{Frusawa}},\ }\bibfield  {title} {\bibinfo {title} {Electric-field-induced
			oscillations in ionic fluids: A unified formulation of modified
			{{Poisson}}--{{Nernst}}--{{Planck}} models and its relevance to correlation
			function analysis},\ }\href {https://doi.org/10.1039/D1SM01811F} {\bibfield
		{journal} {\bibinfo  {journal} {Soft Matter}\ }\textbf {\bibinfo {volume}
			{18}},\ \bibinfo {pages} {4280} (\bibinfo {year} {2022})}\BibitemShut
	{NoStop}%
	\bibitem [{\citenamefont {Kardar}\ and\ \citenamefont
		{Golestanian}(1999)}]{Kardar1999}%
	\BibitemOpen
	\bibfield  {author} {\bibinfo {author} {\bibfnamefont {M.}~\bibnamefont
			{Kardar}}\ and\ \bibinfo {author} {\bibfnamefont {R.}~\bibnamefont
			{Golestanian}},\ }\bibfield  {title} {\bibinfo {title} {The "friction" of
			vacuum, and other fluctuation-induced forces},\ }\href
	{https://doi.org/10.1103/RevModPhys.71.1233} {\bibfield  {journal} {\bibinfo
			{journal} {Rev. Mod. Phys.}\ }\textbf {\bibinfo {volume} {71}},\ \bibinfo
		{pages} {1233} (\bibinfo {year} {1999})}\BibitemShut {NoStop}%
	\bibitem [{\citenamefont {Lu}\ \emph {et~al.}(2015)\citenamefont {Lu},
		\citenamefont {Dean},\ and\ \citenamefont {Podgornik}}]{Lu2015}%
	\BibitemOpen
	\bibfield  {author} {\bibinfo {author} {\bibfnamefont {B.-S.}\ \bibnamefont
			{Lu}}, \bibinfo {author} {\bibfnamefont {D.~S.}\ \bibnamefont {Dean}},\ and\
		\bibinfo {author} {\bibfnamefont {R.}~\bibnamefont {Podgornik}},\ }\bibfield
	{title} {\bibinfo {title} {Out-of-equilibrium thermal {{Casimir}} effect
			between {{Brownian}} conducting plates},\ }\href
	{https://doi.org/10.1209/0295-5075/112/20001} {\bibfield  {journal} {\bibinfo
			{journal} {EPL (Europhysics Letters)}\ }\textbf {\bibinfo {volume} {112}},\
		\bibinfo {pages} {20001} (\bibinfo {year} {2015})}\BibitemShut {NoStop}%
	\bibitem [{\citenamefont {Dean}\ \emph {et~al.}(2016)\citenamefont {Dean},
		\citenamefont {Lu}, \citenamefont {Maggs},\ and\ \citenamefont
		{Podgornik}}]{Dean2016}%
	\BibitemOpen
	\bibfield  {author} {\bibinfo {author} {\bibfnamefont {D.~S.}\ \bibnamefont
			{Dean}}, \bibinfo {author} {\bibfnamefont {B.-S.}\ \bibnamefont {Lu}},
		\bibinfo {author} {\bibfnamefont {A.~C.}\ \bibnamefont {Maggs}},\ and\
		\bibinfo {author} {\bibfnamefont {R.}~\bibnamefont {Podgornik}},\ }\bibfield
	{title} {\bibinfo {title} {Nonequilibrium {{Tuning}} of the {{Thermal Casimir
					Effect}}},\ }\href {https://doi.org/10.1103/PhysRevLett.116.240602}
	{\bibfield  {journal} {\bibinfo  {journal} {Physical Review Letters}\
		}\textbf {\bibinfo {volume} {116}},\ \bibinfo {pages} {240602} (\bibinfo
		{year} {2016})}\BibitemShut {NoStop}%
	\bibitem [{\citenamefont {Mahdisoltani}\ and\ \citenamefont
		{Golestanian}(2021{\natexlab{a}})}]{Mahdisoltani2021a}%
	\BibitemOpen
	\bibfield  {author} {\bibinfo {author} {\bibfnamefont {S.}~\bibnamefont
			{Mahdisoltani}}\ and\ \bibinfo {author} {\bibfnamefont {R.}~\bibnamefont
			{Golestanian}},\ }\bibfield  {title} {\bibinfo {title} {Long-{{Range
					Fluctuation-Induced Forces}} in {{Driven Electrolytes}}},\ }\href
	{https://doi.org/10.1103/PhysRevLett.126.158002} {\bibfield  {journal}
		{\bibinfo  {journal} {Physical Review Letters}\ }\textbf {\bibinfo {volume}
			{126}},\ \bibinfo {pages} {158002} (\bibinfo {year}
		{2021}{\natexlab{a}})}\BibitemShut {NoStop}%
	\bibitem [{\citenamefont {Mahdisoltani}\ and\ \citenamefont
		{Golestanian}(2021{\natexlab{b}})}]{Mahdisoltani2021b}%
	\BibitemOpen
	\bibfield  {author} {\bibinfo {author} {\bibfnamefont {S.}~\bibnamefont
			{Mahdisoltani}}\ and\ \bibinfo {author} {\bibfnamefont {R.}~\bibnamefont
			{Golestanian}},\ }\bibfield  {title} {\bibinfo {title} {Transient
			fluctuation-induced forces in driven electrolytes after an electric field
			quench},\ }\href {https://doi.org/10.1088/1367-2630/ac0f1a} {\bibfield
		{journal} {\bibinfo  {journal} {New Journal of Physics}\ }\textbf {\bibinfo
			{volume} {23}},\ \bibinfo {pages} {073034} (\bibinfo {year}
		{2021}{\natexlab{b}})}\BibitemShut {NoStop}%
	\bibitem [{\citenamefont {Du}\ \emph {et~al.}(2024)\citenamefont {Du},
		\citenamefont {Dean}, \citenamefont {Miao},\ and\ \citenamefont
		{Podgornik}}]{Du2024}%
	\BibitemOpen
	\bibfield  {author} {\bibinfo {author} {\bibfnamefont {G.}~\bibnamefont
			{Du}}, \bibinfo {author} {\bibfnamefont {D.~S.}\ \bibnamefont {Dean}},
		\bibinfo {author} {\bibfnamefont {B.}~\bibnamefont {Miao}},\ and\ \bibinfo
		{author} {\bibfnamefont {R.}~\bibnamefont {Podgornik}},\ }\href
	{https://arxiv.org/abs/2404.06028} {\bibinfo {title} {Correlation decoupling
			of {C}asimir interaction in an electrolyte driven by external electric
			fields}} (\bibinfo {year} {2024}),\ \Eprint
	{https://arxiv.org/abs/2404.06028} {arXiv:2404.06028} \BibitemShut {NoStop}%
	\bibitem [{\citenamefont {Du}\ \emph {et~al.}(2025)\citenamefont {Du},
		\citenamefont {Dean}, \citenamefont {Miao},\ and\ \citenamefont
		{Podgornik}}]{Du2025}%
	\BibitemOpen
	\bibfield  {author} {\bibinfo {author} {\bibfnamefont {G.}~\bibnamefont
			{Du}}, \bibinfo {author} {\bibfnamefont {D.~S.}\ \bibnamefont {Dean}},
		\bibinfo {author} {\bibfnamefont {B.}~\bibnamefont {Miao}},\ and\ \bibinfo
		{author} {\bibfnamefont {R.}~\bibnamefont {Podgornik}},\ }\bibfield  {title}
	{\bibinfo {title} {Repulsive thermal van der {{Waals}} interaction in
			multispecies asymmetric electrolytes driven by external electric fields},\
	}\href {https://doi.org/10.1103/PhysRevE.111.044108} {\bibfield  {journal}
		{\bibinfo  {journal} {Physical Review E}\ }\textbf {\bibinfo {volume}
			{111}},\ \bibinfo {pages} {044108} (\bibinfo {year} {2025})}\BibitemShut
	{NoStop}%
	\bibitem [{\citenamefont {D{\'e}mery}\ and\ \citenamefont
		{Dean}(2011)}]{Demery2011}%
	\BibitemOpen
	\bibfield  {author} {\bibinfo {author} {\bibfnamefont {V.}~\bibnamefont
			{D{\'e}mery}}\ and\ \bibinfo {author} {\bibfnamefont {D.~S.}\ \bibnamefont
			{Dean}},\ }\bibfield  {title} {\bibinfo {title} {Perturbative path-integral
			study of active- and passive-tracer diffusion in fluctuating fields},\ }\href
	{https://doi.org/10.1103/PhysRevE.84.011148} {\bibfield  {journal} {\bibinfo
			{journal} {Physical Review E}\ }\textbf {\bibinfo {volume} {84}},\ \bibinfo
		{pages} {011148} (\bibinfo {year} {2011})}\BibitemShut {NoStop}%
	\bibitem [{\citenamefont {D{\'e}mery}(2015)}]{Demery2015}%
	\BibitemOpen
	\bibfield  {author} {\bibinfo {author} {\bibfnamefont {V.}~\bibnamefont
			{D{\'e}mery}},\ }\bibfield  {title} {\bibinfo {title} {Mean-field
			microrheology of a very soft colloidal suspension: {{Inertia}} induces shear
			thickening},\ }\href {https://doi.org/10.1103/PhysRevE.91.062301} {\bibfield
		{journal} {\bibinfo  {journal} {Physical Review E}\ }\textbf {\bibinfo
			{volume} {91}},\ \bibinfo {pages} {062301} (\bibinfo {year}
		{2015})}\BibitemShut {NoStop}%
	\bibitem [{\citenamefont {D{\'e}mery}\ and\ \citenamefont
		{Fodor}(2019)}]{Demery2019}%
	\BibitemOpen
	\bibfield  {author} {\bibinfo {author} {\bibfnamefont {V.}~\bibnamefont
			{D{\'e}mery}}\ and\ \bibinfo {author} {\bibfnamefont {{\'E}.}~\bibnamefont
			{Fodor}},\ }\bibfield  {title} {\bibinfo {title} {Driven probe under harmonic
			confinement in a colloidal bath},\ }\href
	{https://doi.org/10.1088/1742-5468/ab02e9} {\bibfield  {journal} {\bibinfo
			{journal} {Journal of Statistical Mechanics: Theory and Experiment}\ }\textbf
		{\bibinfo {volume} {2019}},\ \bibinfo {pages} {033202} (\bibinfo {year}
		{2019})}\BibitemShut {NoStop}%
	\bibitem [{\citenamefont {Martin}\ \emph {et~al.}(2018)\citenamefont {Martin},
		\citenamefont {Nardini}, \citenamefont {Cates},\ and\ \citenamefont
		{Fodor}}]{Martin2018}%
	\BibitemOpen
	\bibfield  {author} {\bibinfo {author} {\bibfnamefont {D.}~\bibnamefont
			{Martin}}, \bibinfo {author} {\bibfnamefont {C.}~\bibnamefont {Nardini}},
		\bibinfo {author} {\bibfnamefont {M.~E.}\ \bibnamefont {Cates}},\ and\
		\bibinfo {author} {\bibfnamefont {{\'E}.}~\bibnamefont {Fodor}},\ }\bibfield
	{title} {\bibinfo {title} {Extracting maximum power from active colloidal
			heat engines},\ }\href {https://doi.org/10.1209/0295-5075/121/60005}
	{\bibfield  {journal} {\bibinfo  {journal} {EPL (Europhysics Letters)}\
		}\textbf {\bibinfo {volume} {121}},\ \bibinfo {pages} {60005} (\bibinfo
		{year} {2018})}\BibitemShut {NoStop}%
	\bibitem [{\citenamefont {Feng}\ and\ \citenamefont {Hou}(2023)}]{Feng2023}%
	\BibitemOpen
	\bibfield  {author} {\bibinfo {author} {\bibfnamefont {M.}~\bibnamefont
			{Feng}}\ and\ \bibinfo {author} {\bibfnamefont {Z.}~\bibnamefont {Hou}},\
	}\bibfield  {title} {\bibinfo {title} {Unraveling on kinesin acceleration in
			intracellular environments: {{A}} theory for active bath},\ }\href
	{https://doi.org/10.1103/PhysRevResearch.5.013206} {\bibfield  {journal}
		{\bibinfo  {journal} {Physical Review Research}\ }\textbf {\bibinfo {volume}
			{5}},\ \bibinfo {pages} {013206} (\bibinfo {year} {2023})}\BibitemShut
	{NoStop}%
	\bibitem [{\citenamefont {Wang}\ \emph {et~al.}(2023)\citenamefont {Wang},
		\citenamefont {Dean}, \citenamefont {Marbach},\ and\ \citenamefont
		{Zakine}}]{Wang2023}%
	\BibitemOpen
	\bibfield  {author} {\bibinfo {author} {\bibfnamefont {Y.}~\bibnamefont
			{Wang}}, \bibinfo {author} {\bibfnamefont {D.}~\bibnamefont {Dean}}, \bibinfo
		{author} {\bibfnamefont {S.}~\bibnamefont {Marbach}},\ and\ \bibinfo {author}
		{\bibfnamefont {R.}~\bibnamefont {Zakine}},\ }\bibfield  {title} {\bibinfo
		{title} {Interactions enhance dispersion in fluctuating channels via emergent
			flows},\ }\href {https://doi.org/10.1017/jfm.2023.640} {\bibfield  {journal}
		{\bibinfo  {journal} {Journal of Fluid Mechanics}\ }\textbf {\bibinfo
			{volume} {972}},\ \bibinfo {pages} {A8} (\bibinfo {year} {2023})}\BibitemShut
	{NoStop}%
	\bibitem [{\citenamefont {Muzzeddu}\ \emph {et~al.}(2025)\citenamefont
		{Muzzeddu}, \citenamefont {Kalz}, \citenamefont {Gambassi}, \citenamefont
		{Sharma},\ and\ \citenamefont {Metzler}}]{Muzzeddu2025a}%
	\BibitemOpen
	\bibfield  {author} {\bibinfo {author} {\bibfnamefont {P.~L.}\ \bibnamefont
			{Muzzeddu}}, \bibinfo {author} {\bibfnamefont {E.}~\bibnamefont {Kalz}},
		\bibinfo {author} {\bibfnamefont {A.}~\bibnamefont {Gambassi}}, \bibinfo
		{author} {\bibfnamefont {A.}~\bibnamefont {Sharma}},\ and\ \bibinfo {author}
		{\bibfnamefont {R.}~\bibnamefont {Metzler}},\ }\bibfield  {title} {\bibinfo
		{title} {Self-diffusion anomalies of an odd tracer in soft-core media},\
	}\href {https://doi.org/10.1088/1367-2630/adbdea} {\bibfield  {journal}
		{\bibinfo  {journal} {New Journal of Physics}\ }\textbf {\bibinfo {volume}
			{27}},\ \bibinfo {pages} {033025} (\bibinfo {year} {2025})}\BibitemShut
	{NoStop}%
	\bibitem [{\citenamefont {Venturelli}\ \emph
		{et~al.}(2025{\natexlab{a}})\citenamefont {Venturelli}, \citenamefont
		{Illien}, \citenamefont {Grabsch},\ and\ \citenamefont
		{B{\'e}nichou}}]{Venturelli2025}%
	\BibitemOpen
	\bibfield  {author} {\bibinfo {author} {\bibfnamefont {D.}~\bibnamefont
			{Venturelli}}, \bibinfo {author} {\bibfnamefont {P.}~\bibnamefont {Illien}},
		\bibinfo {author} {\bibfnamefont {A.}~\bibnamefont {Grabsch}},\ and\ \bibinfo
		{author} {\bibfnamefont {O.}~\bibnamefont {B{\'e}nichou}},\ }\href
	{https://doi.org/10.48550/arXiv.2411.09326} {\bibinfo {title} {Universal
			scale-free decay of tracer-bath correlations in \$d\$-dimensional interacting
			particle systems}} (\bibinfo {year} {2025}{\natexlab{a}}),\ \Eprint
	{https://arxiv.org/abs/2411.09326} {arXiv:2411.09326} \BibitemShut {NoStop}%
	\bibitem [{\citenamefont {Venturelli}\ \emph
		{et~al.}(2025{\natexlab{b}})\citenamefont {Venturelli}, \citenamefont
		{Illien}, \citenamefont {Grabsch},\ and\ \citenamefont
		{B{\'e}nichou}}]{Venturelli2025a}%
	\BibitemOpen
	\bibfield  {author} {\bibinfo {author} {\bibfnamefont {D.}~\bibnamefont
			{Venturelli}}, \bibinfo {author} {\bibfnamefont {P.}~\bibnamefont {Illien}},
		\bibinfo {author} {\bibfnamefont {A.}~\bibnamefont {Grabsch}},\ and\ \bibinfo
		{author} {\bibfnamefont {O.}~\bibnamefont {B{\'e}nichou}},\ }\href
	{https://doi.org/10.48550/arXiv.2502.16951} {\bibinfo {title} {Dynamics of
			soft interacting particles on a comb}} (\bibinfo {year}
	{2025}{\natexlab{b}}),\ \Eprint {https://arxiv.org/abs/2502.16951}
	{arXiv:2502.16951} \BibitemShut {NoStop}%
	\bibitem [{\citenamefont {Ooshida}\ \emph {et~al.}(2011)\citenamefont
		{Ooshida}, \citenamefont {Goto}, \citenamefont {Matsumoto}, \citenamefont
		{Nakahara},\ and\ \citenamefont {Otsuki}}]{Ooshida2011}%
	\BibitemOpen
	\bibfield  {author} {\bibinfo {author} {\bibfnamefont {T.}~\bibnamefont
			{Ooshida}}, \bibinfo {author} {\bibfnamefont {S.}~\bibnamefont {Goto}},
		\bibinfo {author} {\bibfnamefont {T.}~\bibnamefont {Matsumoto}}, \bibinfo
		{author} {\bibfnamefont {A.}~\bibnamefont {Nakahara}},\ and\ \bibinfo
		{author} {\bibfnamefont {M.}~\bibnamefont {Otsuki}},\ }\bibfield  {title}
	{\bibinfo {title} {Continuum {{Theory}} of {{Single-File Diffusion}} in
			{{Terms}} of {{Label Variable}}},\ }\href
	{https://doi.org/10.1143/JPSJ.80.074007} {\bibfield  {journal} {\bibinfo
			{journal} {Journal of the Physical Society of Japan}\ }\textbf {\bibinfo
			{volume} {80}},\ \bibinfo {pages} {074007} (\bibinfo {year}
		{2011})}\BibitemShut {NoStop}%
	\bibitem [{\citenamefont {Ooshida}\ \emph {et~al.}(2013)\citenamefont
		{Ooshida}, \citenamefont {Goto}, \citenamefont {Matsumoto}, \citenamefont
		{Nakahara},\ and\ \citenamefont {Otsuki}}]{Ooshida2013}%
	\BibitemOpen
	\bibfield  {author} {\bibinfo {author} {\bibfnamefont {T.}~\bibnamefont
			{Ooshida}}, \bibinfo {author} {\bibfnamefont {S.}~\bibnamefont {Goto}},
		\bibinfo {author} {\bibfnamefont {T.}~\bibnamefont {Matsumoto}}, \bibinfo
		{author} {\bibfnamefont {A.}~\bibnamefont {Nakahara}},\ and\ \bibinfo
		{author} {\bibfnamefont {M.}~\bibnamefont {Otsuki}},\ }\bibfield  {title}
	{\bibinfo {title} {Analytical calculation of four-point correlations for a
			simple model of cages involving numerous particles},\ }\href
	{https://doi.org/10.1103/PhysRevE.88.062108} {\bibfield  {journal} {\bibinfo
			{journal} {Physical Review E}\ }\textbf {\bibinfo {volume} {88}},\ \bibinfo
		{pages} {062108} (\bibinfo {year} {2013})}\BibitemShut {NoStop}%
	\bibitem [{\citenamefont {Ooshida}\ \emph {et~al.}(2015)\citenamefont
		{Ooshida}, \citenamefont {Goto}, \citenamefont {Matsumoto},\ and\
		\citenamefont {Otsuki}}]{Ooshida2015}%
	\BibitemOpen
	\bibfield  {author} {\bibinfo {author} {\bibfnamefont {T.}~\bibnamefont
			{Ooshida}}, \bibinfo {author} {\bibfnamefont {S.}~\bibnamefont {Goto}},
		\bibinfo {author} {\bibfnamefont {T.}~\bibnamefont {Matsumoto}},\ and\
		\bibinfo {author} {\bibfnamefont {M.}~\bibnamefont {Otsuki}},\ }\bibfield
	{title} {\bibinfo {title} {Displacement correlation as an indicator of
			collective motion in one-dimensional and quasi-one-dimensional systems of
			repulsive {{Brownian}} particles},\ }\href
	{https://doi.org/10.1142/S0217984915502218} {\bibfield  {journal} {\bibinfo
			{journal} {Modern Physics Letters B}\ }\textbf {\bibinfo {volume} {29}},\
		\bibinfo {pages} {1550221} (\bibinfo {year} {2015})}\BibitemShut {NoStop}%
	\bibitem [{\citenamefont {Ooshida}\ and\ \citenamefont
		{Otsuki}(2018)}]{Ooshida2018}%
	\BibitemOpen
	\bibfield  {author} {\bibinfo {author} {\bibfnamefont {T.}~\bibnamefont
			{Ooshida}}\ and\ \bibinfo {author} {\bibfnamefont {M.}~\bibnamefont
			{Otsuki}},\ }\bibfield  {title} {\bibinfo {title} {Two-tag correlations and
			nonequilibrium fluctuation--response relation in ageing single-file
			diffusion},\ }\href {https://doi.org/10.1088/1361-648X/aad4cc} {\bibfield
		{journal} {\bibinfo  {journal} {Journal of Physics: Condensed Matter}\
		}\textbf {\bibinfo {volume} {30}},\ \bibinfo {pages} {374001} (\bibinfo
		{year} {2018})}\BibitemShut {NoStop}%
	\bibitem [{\citenamefont {Abel}\ \emph {et~al.}(2009)\citenamefont {Abel},
		\citenamefont {Steve~Tse},\ and\ \citenamefont {Andersenb}}]{Abel2009}%
	\BibitemOpen
	\bibfield  {author} {\bibinfo {author} {\bibfnamefont {S.~M.}\ \bibnamefont
			{Abel}}, \bibinfo {author} {\bibfnamefont {Y.-L.}\ \bibnamefont
			{Steve~Tse}},\ and\ \bibinfo {author} {\bibfnamefont {H.~C.}\ \bibnamefont
			{Andersenb}},\ }\bibfield  {title} {\bibinfo {title} {Kinetic theories of
			dynamics and persistent caging in a one-dimensional lattice gas},\ }\href
	{https://doi.org/10.1073/pnas.0901693106} {\bibfield  {journal} {\bibinfo
			{journal} {Proceedings of the National Academy of Sciences}\ }\textbf
		{\bibinfo {volume} {106}},\ \bibinfo {pages} {15142} (\bibinfo {year}
		{2009})}\BibitemShut {NoStop}%
	\bibitem [{\citenamefont {Ooshida}\ \emph
		{et~al.}(2016{\natexlab{a}})\citenamefont {Ooshida}, \citenamefont {Goto},
		\citenamefont {Matsumoto},\ and\ \citenamefont {Otsuki}}]{Ooshida2016}%
	\BibitemOpen
	\bibfield  {author} {\bibinfo {author} {\bibfnamefont {T.}~\bibnamefont
			{Ooshida}}, \bibinfo {author} {\bibfnamefont {S.}~\bibnamefont {Goto}},
		\bibinfo {author} {\bibfnamefont {T.}~\bibnamefont {Matsumoto}},\ and\
		\bibinfo {author} {\bibfnamefont {M.}~\bibnamefont {Otsuki}},\ }\bibfield
	{title} {\bibinfo {title} {Calculation of displacement correlation tensor
			indicating vortical cooperative motion in two-dimensional colloidal
			liquids},\ }\href {https://doi.org/10.1103/PhysRevE.94.022125} {\bibfield
		{journal} {\bibinfo  {journal} {Physical Review E}\ }\textbf {\bibinfo
			{volume} {94}},\ \bibinfo {pages} {022125} (\bibinfo {year}
		{2016}{\natexlab{a}})}\BibitemShut {NoStop}%
	\bibitem [{\citenamefont {Ooshida}\ \emph
		{et~al.}(2016{\natexlab{b}})\citenamefont {Ooshida}, \citenamefont {Goto},
		\citenamefont {Matsumoto},\ and\ \citenamefont {Otsuki}}]{Ooshida2016a}%
	\BibitemOpen
	\bibfield  {author} {\bibinfo {author} {\bibfnamefont {T.}~\bibnamefont
			{Ooshida}}, \bibinfo {author} {\bibfnamefont {S.}~\bibnamefont {Goto}},
		\bibinfo {author} {\bibfnamefont {T.}~\bibnamefont {Matsumoto}},\ and\
		\bibinfo {author} {\bibfnamefont {M.}~\bibnamefont {Otsuki}},\ }\bibfield
	{title} {\bibinfo {title} {Insights from {{Single-File Diffusion}} into
			{{Cooperativity}} in {{Higher Dimensions}}},\ }\href
	{https://doi.org/10.1142/S1793048015400019} {\bibfield  {journal} {\bibinfo
			{journal} {Biophysical Reviews and Letters}\ }\textbf {\bibinfo {volume}
			{11}},\ \bibinfo {pages} {9} (\bibinfo {year}
		{2016}{\natexlab{b}})}\BibitemShut {NoStop}%
	\bibitem [{\citenamefont {Touzo}\ \emph {et~al.}(2023)\citenamefont {Touzo},
		\citenamefont {Le~Doussal},\ and\ \citenamefont {Schehr}}]{Touzo2023}%
	\BibitemOpen
	\bibfield  {author} {\bibinfo {author} {\bibfnamefont {L.}~\bibnamefont
			{Touzo}}, \bibinfo {author} {\bibfnamefont {P.}~\bibnamefont {Le~Doussal}},\
		and\ \bibinfo {author} {\bibfnamefont {G.}~\bibnamefont {Schehr}},\
	}\bibfield  {title} {\bibinfo {title} {Interacting, running and tumbling:
			{{The}} active {{Dyson Brownian}} motion},\ }\href
	{https://doi.org/10.1209/0295-5075/acdabb} {\bibfield  {journal} {\bibinfo
			{journal} {Europhysics Letters}\ }\textbf {\bibinfo {volume} {142}},\
		\bibinfo {pages} {61004} (\bibinfo {year} {2023})}\BibitemShut {NoStop}%
	\bibitem [{\citenamefont {Le~Doussal}(2022)}]{LeDoussal2022}%
	\BibitemOpen
	\bibfield  {author} {\bibinfo {author} {\bibfnamefont {P.}~\bibnamefont
			{Le~Doussal}},\ }\bibfield  {title} {\bibinfo {title} {Ranked diffusion,
			delta {{Bose}} gas, and {{Burgers}} equation},\ }\href
	{https://doi.org/10.1103/PhysRevE.105.L012103} {\bibfield  {journal}
		{\bibinfo  {journal} {Physical Review E}\ }\textbf {\bibinfo {volume}
			{105}},\ \bibinfo {pages} {L012103} (\bibinfo {year} {2022})}\BibitemShut
	{NoStop}%
	\bibitem [{\citenamefont {Flack}\ \emph {et~al.}(2023)\citenamefont {Flack},
		\citenamefont {Le~Doussal}, \citenamefont {Majumdar},\ and\ \citenamefont
		{Schehr}}]{Flack2023}%
	\BibitemOpen
	\bibfield  {author} {\bibinfo {author} {\bibfnamefont {A.}~\bibnamefont
			{Flack}}, \bibinfo {author} {\bibfnamefont {P.}~\bibnamefont {Le~Doussal}},
		\bibinfo {author} {\bibfnamefont {S.~N.}\ \bibnamefont {Majumdar}},\ and\
		\bibinfo {author} {\bibfnamefont {G.}~\bibnamefont {Schehr}},\ }\bibfield
	{title} {\bibinfo {title} {Out-of-equilibrium dynamics of repulsive ranked
			diffusions: {{The}} expanding crystal},\ }\href
	{https://doi.org/10.1103/PhysRevE.107.064105} {\bibfield  {journal} {\bibinfo
			{journal} {Physical Review E}\ }\textbf {\bibinfo {volume} {107}},\ \bibinfo
		{pages} {064105} (\bibinfo {year} {2023})}\BibitemShut {NoStop}%
	\bibitem [{\citenamefont {Dandekar}\ \emph {et~al.}(2023)\citenamefont
		{Dandekar}, \citenamefont {Krapivsky},\ and\ \citenamefont
		{Mallick}}]{Dandekar2023}%
	\BibitemOpen
	\bibfield  {author} {\bibinfo {author} {\bibfnamefont {R.}~\bibnamefont
			{Dandekar}}, \bibinfo {author} {\bibfnamefont {P.~L.}\ \bibnamefont
			{Krapivsky}},\ and\ \bibinfo {author} {\bibfnamefont {K.}~\bibnamefont
			{Mallick}},\ }\bibfield  {title} {\bibinfo {title} {Dynamical fluctuations in
			the {{Riesz}} gas},\ }\href {https://doi.org/10.1103/PhysRevE.107.044129}
	{\bibfield  {journal} {\bibinfo  {journal} {Physical Review E}\ }\textbf
		{\bibinfo {volume} {107}},\ \bibinfo {pages} {044129} (\bibinfo {year}
		{2023})}\BibitemShut {NoStop}%
	\bibitem [{\citenamefont {Dandekar}\ \emph {et~al.}(2024)\citenamefont
		{Dandekar}, \citenamefont {Krapivsky},\ and\ \citenamefont
		{Mallick}}]{Dandekar2024}%
	\BibitemOpen
	\bibfield  {author} {\bibinfo {author} {\bibfnamefont {R.}~\bibnamefont
			{Dandekar}}, \bibinfo {author} {\bibfnamefont {P.~L.}\ \bibnamefont
			{Krapivsky}},\ and\ \bibinfo {author} {\bibfnamefont {K.}~\bibnamefont
			{Mallick}},\ }\href {https://arxiv.org/abs/2409.06881} {\bibinfo {title}
		{Current fluctuations in the dyson gas}} (\bibinfo {year} {2024}),\ \Eprint
	{https://arxiv.org/abs/2409.06881} {arXiv:2409.06881} \BibitemShut {NoStop}%
	\bibitem [{\citenamefont {Rotskoff}\ and\ \citenamefont
		{Vanden-Eijnden}(2022)}]{Rotskoff2022}%
	\BibitemOpen
	\bibfield  {author} {\bibinfo {author} {\bibfnamefont {G.}~\bibnamefont
			{Rotskoff}}\ and\ \bibinfo {author} {\bibfnamefont {E.}~\bibnamefont
			{Vanden-Eijnden}},\ }\bibfield  {title} {\bibinfo {title} {Trainability and
			{{Accuracy}} of {{Artificial Neural Networks}}: {{An Interacting Particle
					System Approach}}},\ }\href {https://doi.org/10.1002/cpa.22074} {\bibfield
		{journal} {\bibinfo  {journal} {Communications on Pure and Applied
				Mathematics}\ }\textbf {\bibinfo {volume} {75}},\ \bibinfo {pages} {1889}
		(\bibinfo {year} {2022})}\BibitemShut {NoStop}%
	\bibitem [{\citenamefont {Lutsko}(2012)}]{Lutsko2012}%
	\BibitemOpen
	\bibfield  {author} {\bibinfo {author} {\bibfnamefont {J.~F.}\ \bibnamefont
			{Lutsko}},\ }\bibfield  {title} {\bibinfo {title} {A dynamical theory of
			nucleation for colloids and macromolecules},\ }\href
	{https://doi.org/10.1063/1.3677191} {\bibfield  {journal} {\bibinfo
			{journal} {The Journal of Chemical Physics}\ }\textbf {\bibinfo {volume}
			{136}},\ \bibinfo {pages} {034509} (\bibinfo {year} {2012})}\BibitemShut
	{NoStop}%
	\bibitem [{\citenamefont {Liu}\ \emph {et~al.}(2024)\citenamefont {Liu},
		\citenamefont {Sprittles},\ and\ \citenamefont {Grafke}}]{Liu2024}%
	\BibitemOpen
	\bibfield  {author} {\bibinfo {author} {\bibfnamefont {J.}~\bibnamefont
			{Liu}}, \bibinfo {author} {\bibfnamefont {J.~E.}\ \bibnamefont {Sprittles}},\
		and\ \bibinfo {author} {\bibfnamefont {T.}~\bibnamefont {Grafke}},\ }\href
	{https://arxiv.org/abs/2405.13490} {\bibinfo {title} {Mean first passage
			times and eyring-kramers formula for fluctuating hydrodynamics}} (\bibinfo
	{year} {2024}),\ \Eprint {https://arxiv.org/abs/2405.13490}
	{arXiv:2405.13490} \BibitemShut {NoStop}%
\end{thebibliography}


\end{document}